\documentclass[useAMS,usenatbib]{mn2e}

\usepackage{tabularx}
\usepackage{graphicx}
\usepackage{txfonts}
\usepackage{longtable}
\usepackage{hhline}
\usepackage{arydshln}
\usepackage{multirow}
\usepackage{lscape}
\usepackage{array}
\usepackage{rotating}

\newcommand{\hersc}{{\it Herschel}}
\newcommand{\lab}{LABOCA}

\newcommand{\spitz}{{\it  Spitzer}}

\newcommand{\iras}{{\it IRAS}}

\newcommand{\lsun}{$L_\odot$}
\newcommand{\msun}{$M_\odot$}
\newcommand{\zsun}{$Z_\odot$}
\newcommand{\mic}{$\mu$m}

\newcolumntype{R}[1]{>{\raggedleft\arraybackslash }b{#1}}
\newcolumntype{L}[1]{>{\raggedright\arraybackslash }b{#1}}
\newcolumntype{C}[1]{>{\centering\arraybackslash }b{#1}}

\newlength{\pointwidth}
\begin{document}

  \title[Dust properties and physical conditions in N11]{The dust properties and physical conditions of the interstellar medium in the LMC massive star forming complex N11 }

\author[Galametz et al.]
{\parbox{\textwidth}{M. Galametz$^{1}$\thanks{e-mail: maud.galametz@eso.org}, 
S. Hony$^{2}$, 
M. Albrecht$^{3}$,
F. Galliano$^{4}$,
D. Cormier$^{2}$,
V. Lebouteiller$^{4}$,
M.Y. Lee$^{4}$,
S. C. Madden$^{4}$,
A. Bolatto$^{5}$,
C. Bot$^{6}$,
A. Hughes$^{7}$,
F. Israel$^{8}$,
M. Meixner$^{9,10}$,
J. M. Oliviera$^{11}$,
D. Paradis$^{7,12}$,
E. Pellegrini$^{2,13}$,
J. Roman-Duval$^{9}$,
M. Rubio$^{14}$,
M. Sewi{\l}o$^{15,16}$,
Y. Fukui$^{17}$,
A. Kawamura$^{17}$,
T. Onishi$^{18}$
}\vspace{0.5cm}\\
\parbox{\textwidth}{
$^{1}$ European Southern Observatory, Karl-Schwarzschild-Str. 2, D-85748 Garching-bei-M\"unchen, Germany \\
$^{2}$ Universit\"at Heidelberg, Zentrum f\"ur Astronomie, Institut f\"ur Theoretische Astrophysik, Albert-Ueberle-Str. 2, 69120 Heidelberg, Germany \\
$^{3}$ Argelander-Institut f\"ur Astronomie, Auf dem H\"ugel 71, D-53121 Bonn, Germany \\
$^{4}$ Laboratoire AIM, CEA, Universit\'{e} Paris Diderot, IRFU/Service d'Astrophysique, Bat. 709, 91191 Gif-sur-Yvette, France \\
$^{5}$ Department of Astronomy and Laboratory for Millimeter-wave Astronomy, University of Maryland, College Park, MD 20742, USA \\
$^{6}$ Observatoire Astronomique de Strasbourg, Universit\'{e} de Strasbourg, UMR 7550, 11 rue de l'Universit\'{e}, 67000, Strasbourg, France \\
$^{7}$ CNRS; IRAP; 9 Av. du Colonel Roche, BP 44346, 31028, Toulouse, Cedex 4, France \\
$^{8}$ Sterrewacht Leiden, Leiden University, P.O. Box 9513, NL-2300 RA Leiden, The Netherlands \\ 
$^{9}$ Space Telescope Science Institute, 3700 San Martin Drive, Baltimore, MD 21218 \\
$^{10}$ The Johns Hopkins University, Department of Physics and Astronomy, 366 Bloomberg Center, 3400 N. Charles Street, Baltimore, MD 21218, USA \\
$^{11}$ Lennard-Jones Laboratories, School of Physical \& Geographical Sciences, Keele University, Staffordshire ST5 5BG, UK \\
$^{12}$ Universit\'{e} de Toulouse; UPS-OMP; IRAP; Toulouse, France \\
$^{13}$ Department of Physics \& Astronomy, Mail Drop 111, University of Toledo, 2801 West Bancroft Street, Toledo, OH 43606, USA \\
$^{14}$ Departamento de Astronomia, Universidad de Chile, Casilla 36-D, Santiago, Chile \\
$^{15}$ Exoplanets and Stellar Astrophysics Laboratory, Code 667, NASA Goddard Space Flight Center, Greenbelt, MD 20771, USA\\
$^{16}$ Oak Ridge Associated Universities (ORAU), Oak Ridge, TN 37831, USA\\
$^{17}$ Department of Physics, Nagoya University, Chikusa-ku, Magoya 464-8602, Japan \\
$^{18}$ Department of Physical Science, Osaka Prefecture University, Gakuen 1-1, Sakai, Osaka 599-8531, Japan \\
}
}

\maketitle{}

 
\begin{abstract}

We combine \spitz\ and \hersc\ data of the star-forming region N11 in the Large Magellanic Cloud to 
produce detailed maps of the dust properties in the complex and study their variations with the ISM 
conditions. We also compare APEX/LABOCA 870\mic\ observations with our model predictions in order 
to decompose the 870 \mic\ emission into dust and non-dust (free-free emission and CO(3-2) line) 
contributions. We find that in N11, the 870 \mic\ can be fully accounted for by these 3 components. 
The dust surface density map of N11 is combined with H\,{\sc i} and CO observations to study local variations 
in the gas-to-dust mass ratios. Our analysis leads to values lower than those expected from the LMC low-metallicity 
as well as to a decrease of the gas-to-dust mass ratio with the dust surface density. 
We explore potential hypotheses that could explain the low `observed' gas-to-dust mass ratios (variations in the X$_{\rm CO}$ factor, 
presence of CO-dark gas or of optically thick H\,{\sc i} or variations in the dust abundance in the dense regions). We finally decompose the 
local SEDs using a Principal Component Analysis (i.e. with no {\it a priori} assumption on the dust composition in the complex). Our results lead to a promising
decomposition of the local SEDs in various dust components (hot, warm, cold) coherent with that expected for the region. Further analysis on a larger sample 
of galaxies will follow in order to understand how unique this decomposition is or how it evolves from one environment to another.

\end{abstract}
  


\section{Introduction}

Measuring the different gas and dust reservoirs and their dependence on the physical conditions in the interstellar medium (ISM) is
fundamental to further our understanding of star formation in galaxies and understand their chemical evolution. 
However, the various methods used to trace these reservoirs have a number of fundamental flaws. For instance, CO observations
are often used to indirectly trace the molecular hydrogen H$_{\rm 2}$ in galaxies. Departures from the well-constrained 
Galactic conversion factor between the CO line intensity and the H$_{\rm 2}$ mass - the so-called X$_{\rm CO}$ factor - 
are expected on local (region to region) or global (galaxy to galaxy) scales. However, these variations are still poorly understood. Dust 
emission in the IR-to-submillimeter regime is often used as a complementary tracer of the gas reservoirs. This technique, however, also requires 
assumptions on both the dust composition and the gas-to-dust mass ratio (GDR). Both quantities also vary from one environment 
to another. 

All these effects are even less constrained at lower metallicities. In a dust-poor environment for instance, UV photons penetrate 
more easily in the ISM, creating large reservoirs of gas where H$_{\rm 2}$ is efficiently self-shielded but CO is photo-dissociated, 
thus not properly tracing H$_{\rm 2}$. The conversion factor X$_{\rm CO}$ thus strongly depends on the dust content and the ISM 
morphology \citep[we refer to][for a review on X$_{\rm CO}$]{Bolatto2013}. In these metal-poor environments, dust appears as a 
less biased tracer of the gaseous phase. However, metallicity also affects the dust properties, whether we are talking about the size distribution 
of the dust grains \citep{Galliano2005} or their composition. Submillimeter (submm) observations with the 
{\it Herschel Space Observatory} or from the ground have in particular helped us characterize the variations of the cold dust 
properties (dust emissivity, submm excess) with the physical conditions of the ISM \citep{RemyRuyer2014,RemyRuyer2015}. \\

Further studies targeting a wide range of physical conditions, with a particular focus on lower metallicity environments, 
are necessary to understand the dust and the gas reservoirs and study the influence of standard assumptions 
on their apparent relations. This paper is part of a series to study the ISM components in the nearby Large Magellanic Cloud (LMC) at small 
spatial scales. The LMC is a prime extragalactic laboratory to perform a detailed study of low metallicity ISM \citep[Z$_{\rm LMC}$ 
= 1/2 \zsun;][]{Dufour1982}. Its proximity \citep[50 kpc;][]{Schaefer2008} and its almost face-on orientation \citep[23-37$^{\circ}$;][]{Subramanian2012} enable us 
to study its star forming complexes at a resolution of 10 pc (the best resolution currently available for external galaxies). 
 This analysis aims to constrain the dust and gas properties along a large number of sight-lines toward N11, the 
 second brightest H\,{\sc ii} region in the LMC. This paper has three main goals. Our first goal is to model the local infrared 
 (IR) to submm dust Spectral Energy Distributions (SED) across the N11 complex in order to provide maps of the dust 
 parameters (excitation, dust column density) and study their dependence on the ISM physical conditions. The second 
 goal is to relate the dust reservoir to the atomic and molecular gas reservoirs in order to study the local variations in 
GDR and understand their origin. The methodologies we discuss will enable us to gauge the effects of the various assumptions 
on the derived GDR observed. The last goal is to explore a Principal Component Analysis (PCA) of the local SEDs as an alternative and possibly 
promising method to investigate the ``resolved" dust populations in galaxies with no {\it a priori} assumption on the dust composition. 

We describe the N11 star-forming complex, the \spitz, \hersc\ and \lab\ observations and the correlations between these 
different bands in $\S$2. We present the SED modeling technique we apply on resolved scales in $\S$3 and describe the 
dust properties we obtain (dust temperatures, mean stellar radiation field intensities) in $\S$4. We also analyze the origin 
of the 870 \mic\ emission in this section. We compare the dust and gas reservoirs and discuss the implications of the low 
gas-to-dust we observe in $\S$5. We finally provide the results of the Principal Component Analysis we perform on the N11 
local SEDs in $\S$6.

\begin{figure*}
    \centering
\includegraphics[width=12cm]{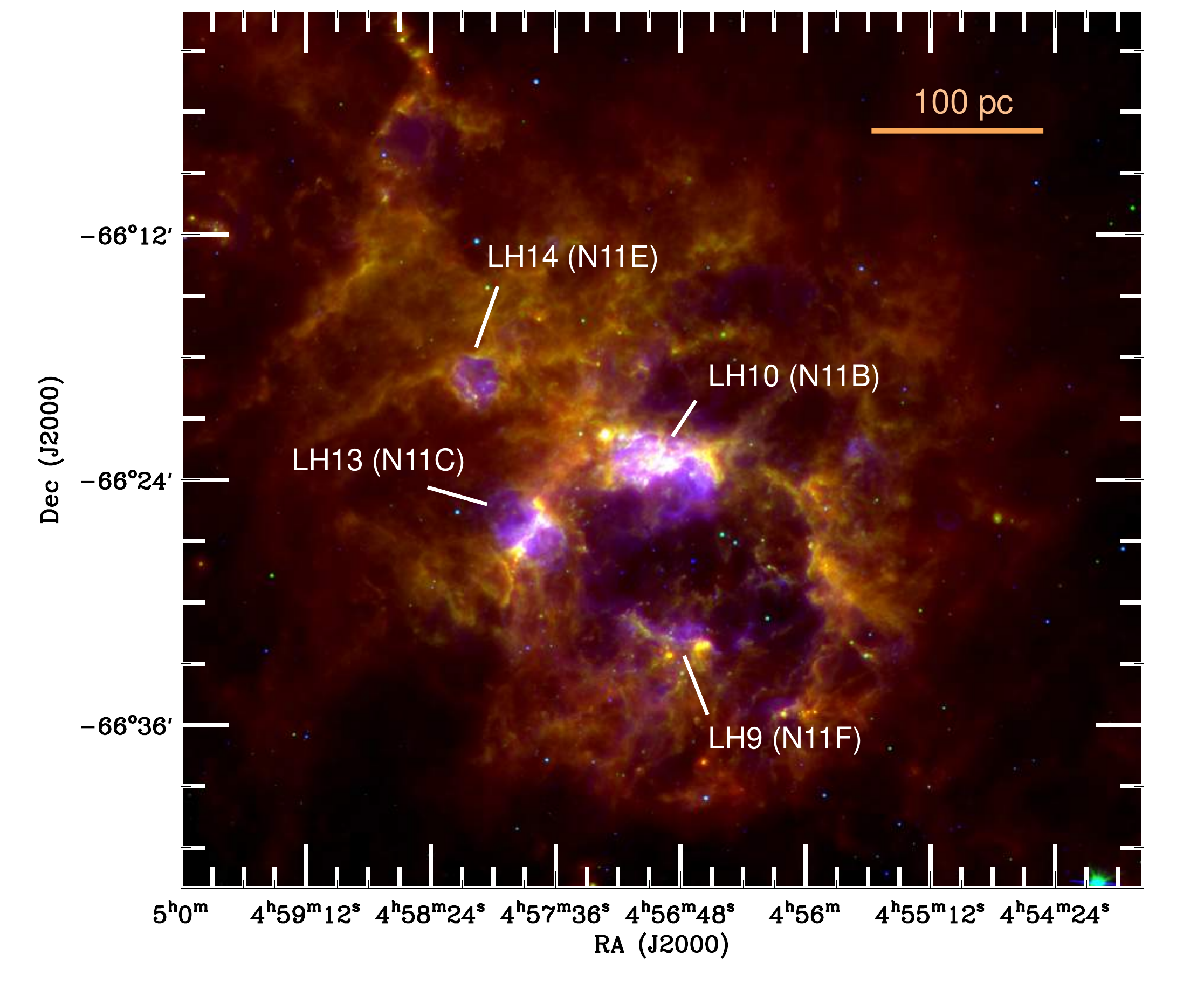}     \\    
     \caption{Color-composition of N11 with H$\alpha$ in blue, IRAC 8 \mic\ in green and SPIRE 250 \mic\ in red. The 4 main OB associations are indicated (from 
     north to south: LH14 in the nebula N11E, LH10 in N11B, LH13 in N11C and LH9 at the center of the N11 cavity in the nebula N11F). 
     The H$\alpha$ image was taken as part of the Magellanic Cloud Emission Line Survey \citep[MCELS;][]{Smith1998}. For all the maps in this paper, north is up, 
     east is left and coordinates (RA, Dec) are given in equinox J2000.}
    \label{N11_Ha_I4_S250}
\end{figure*}

\section{Observations}

\subsection{The N11 complex}

On the north-west edge of the galaxy, N11 is the second brightest H\,{\sc ii} region in the LMC \citep[after 30 Doradus;][]{Kennicutt_Hodge_1986}. It exhibits several 
prominent secondary H\,{\sc ii} regions along its periphery as well as dense filamentary ISM structures, as traced by the prominent dust and CO emission. It is characterized 
by an evacuated central cavity with an inner diameter of 170 pc. Four star clusters are the main sources of the ionization in the complex (see \citet{Lucke1970} or 
\citet{Bica2008} for studies of the stellar associations in the whole LMC). These clusters are indicated in Fig.~\ref{N11_Ha_I4_S250}. The rich OB association LH9 
is located at the center of the cavity. The age of this cluster was estimated to be $\sim$7 Myr \citep{Mokiem2007}. The star cluster LH10 \citep[$\sim$3 Myr old,][]{Mokiem2007}, 
is located in the north-east region of the cavity and is embedded in the N11B nebula. The OB association LH13 and its corresponding nebula (N11C) are located on 
the eastern edge of the super-bubble. The main exciting source of N11C is the compact star cluster Sk-66$^{\circ}$41 \citep[$<$5 Myr old;][]{Heydari2000}. Finally, the 
OB association LH14 (in N11E) is located in the northeastern filament. Its main exciting source is the Sk-66$^{\circ}$43 star cluster. The region also harbors a few massive 
stars (see \citet{Heydari1987} for a detail on its stellar content). The cluster ages and the initial mass functions of the OB associations suggest a sequential star formation 
in N11, with a star formation in the peripheral molecular clouds triggered by the central LH9 association \citep[][among others]{Rosado1996,Hatano2006}. N11 is also 
associated with a giant molecular complex.  \citet{Israel2003_1} and \citet{Herrera2013} have suggested that N11 is a shell compounded of discrete CO clouds (rather 
than bright clouds bathing in continuous inter cloud CO emission) and estimate that more than 50 $\%$ of the CO emission resides in discrete molecular clumps. \\

To construct our local IR to submm SEDs in N11, we use data from the {\it Spitzer Space Telescope} \citep{Werner2004} and the {\it Herschel Space Observatory} 
\citep{Pilbratt2010}. We complement the submm coverage with observations obtained on the Large APEX Bolometer Camera (LABOCA) instrument at 870\mic.

\subsection{Spitzer IRAC and MIPS}

The LMC has been observed with \spitz\ as part of the SAGE project \citep[][Surveying the Agents of a Galaxy's Evolution;]{Meixner2006} and the \spitz\ IRAC 
\citep[InfraRed Array Camera;][]{Fazio2004} and MIPS \citep[Multiband Imaging Photometer;][]{Rieke2004} and data have been reduced by the SAGE consortium. 
IRAC observed at 3.6, 4.5, 5.8, and 8 \mic\ (full width half maximum (FWHM) of its point spread function (PSF) $<$2\arcsec). The calibration errors of IRAC maps 
are 2$\%$ \citep{Reach2005}. MIPS observed at 24, 70 and 160 \mic\ (PSF FWHMs of 6\arcsec, 18\arcsec\ and 40\arcsec\ respectively). The MIPS 160 \mic\ map 
of N11 is not used in the following analysis because of its lower resolution (40\arcsec) than to PACS 160 \mic\ but the data is used for the calibration of the \hersc/PACS 
160 \mic\ map (see $\S$2.4). The respective calibration errors in the MIPS 24 and 70 \mic\ observations are 4 and 5$\%$ \citep{Engelbracht2007,Gordon2007}. We 
refer to \citet{Meixner2006} for a detailed description of the various steps of the data reduction. Additional steps have been added to the SAGE MIPS data reduction 
pipeline by \citet{Gordon2014} (see $\S$2.5).

\begin{figure*}
    \centering

    \includegraphics[width=18cm]{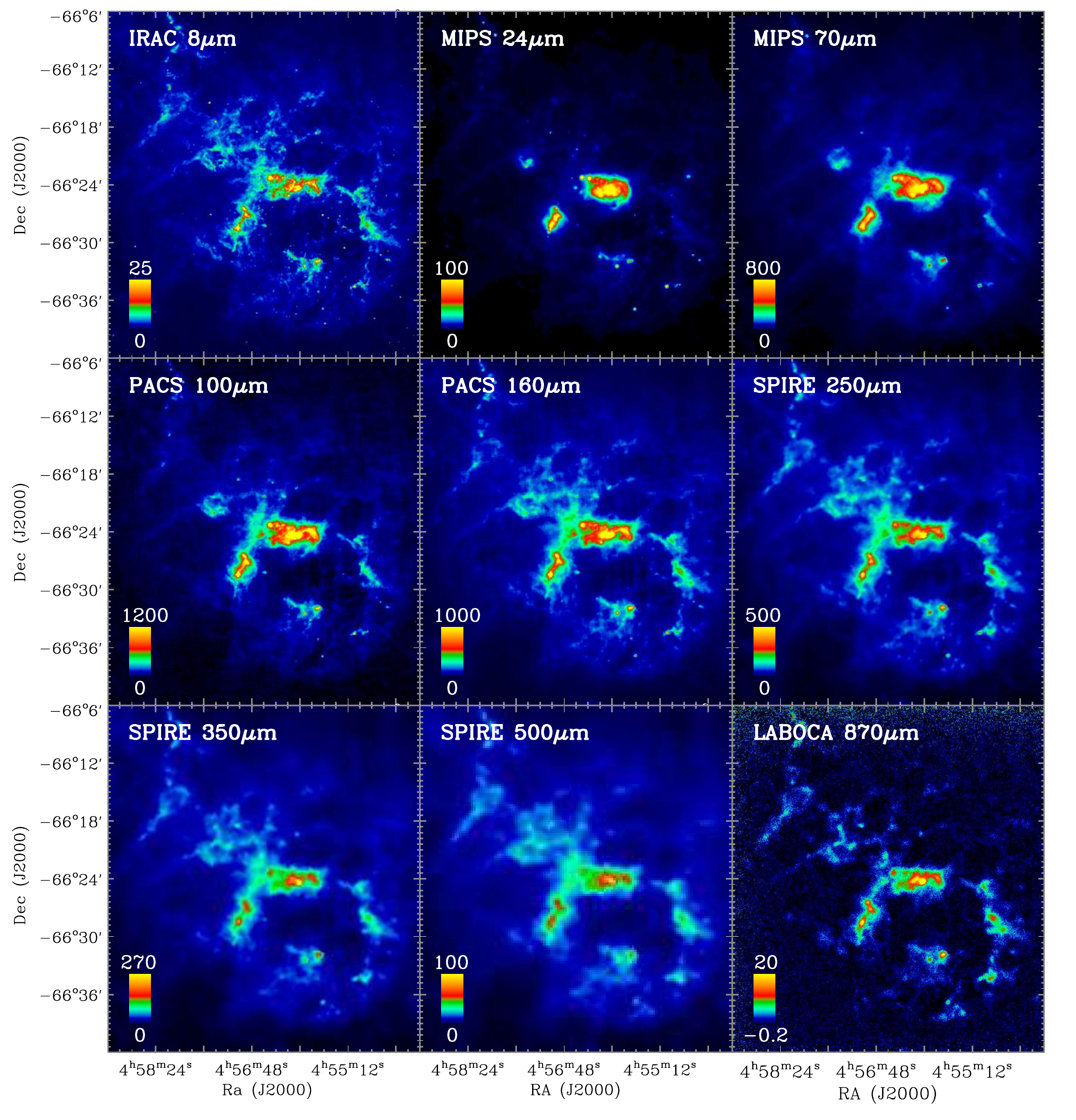}
     \caption{The IR/submm emission of the N11 star-forming complex from 8 to 870 \mic\ (original resolution). The color bars indicate the intensity scale in MJy~sr$^{-1}$. 
     Square root scaling has been applied in order to enhance the fainter structures.}
    \label{Spitzer_Herschel_LABOCA_maps}
\end{figure*}

\subsection{Herschel PACS and SPIRE}

The LMC has been observed with {\it Herschel} as part of the successor project of SAGE, HERITAGE \citep[HERschel Inventory of The Agents of Galaxy's Evolution; ][]{Meixner2010,Meixner2013}. 
The \hersc\ PACS \citep[Photodetector Array Camera and Spectrometer;][]{Poglitsch2010} and SPIRE \citep[Spectral and Photometric Imaging Receiver;][]{Griffin2010} 
and data have been reduced by the HERITAGE consortium. The LMC was mapped at 100 and 160 \mic\ with PACS (PSF FWHMs of $\sim$7\farcs7 and $\sim$12\arcsec\ 
respectively) and 250, 350 and 500 \mic\ with SPIRE (PSF FWHMs of 18\arcsec, 25\arcsec and 36\arcsec\ respectively). The SPIRE 500 \mic\ map possesses the lowest 
resolution (FWHM: 36\arcsec) of the IR-to-submm dataset. Details of the \hersc\ data reduction can be found in \citet{Meixner2013}. We particularly refer the reader to the 
$\S$3.9 that describes the cross-calibration applied between the two PACS maps and the \iras\ 100 \mic\ \citep{Schwering1989} and the MIPS 160 \mic\ maps \citep{Meixner2006} 
in order to correct for the drifting baseline of the PACS bolometers. The PACS instrument has an absolute uncertainty of $\sim$5$\%$ \citep[the accuracy is mostly limited 
by the uncertainty of the celestial standard models used to derive the absolute calibration;][]{Balog2014} to which we linearly add an additional 5$\%$ to account for uncertainties in 
the total beam area. The SPIRE instrument has an absolute calibration uncertainty of 5$\%$ to which we linearly add an additional 4$\%$ to account for the uncertainty in the total 
beam area \citep{Griffin2013}. Additional steps have been added to the HERITAGE PACS and SPIRE data reduction pipelines by \citet{Gordon2014} (see $\S$2.5).

\begin{table*}
\caption{Spearman rank correlation coefficients between the IR/submm luminosities from 8 to 870 \mic\ (log $\nu$L$_{\rm \nu}$).}
\label{Correlation_coefficients}
 \centering
 \begin{tabular}{cccccccccc}
\hline
\hline
&\\
Band & 8 \mic\ & 24 \mic\ & 70 \mic\ & 100 \mic\ & 160 \mic\ & 250 \mic\ & 350 \mic\ & 500 \mic\ \\
&\\
\hline
&\\
24\mic\   &       0.85 &  \\
70\mic\   &       0.86 &       0.93  \\
100\mic\   &       0.93 &       0.93 &       0.96 \\
160\mic\   &       0.96 &       0.87 &       0.90 &       0.96 \\
250\mic\   &       0.97 &       0.85 &       0.86 &       0.94 &       0.97  \\
350\mic\   &       0.95 &       0.83 &       0.82 &       0.91 &       0.96 &       1.00  \\
500\mic\   &       0.94 &       0.81 &       0.80 &       0.89 &       0.94 &       0.99 &       1.00  \\
870\mic\   &       0.73 &       0.68 &       0.63 &       0.69 &       0.73 &       0.77 &       0.78 &       0.79 \\
&\\
\hline
\end{tabular}
 \end{table*} 

\subsection{LABOCA observations and data reduction}

LABOCA is a submm bolometer array installed on the APEX (Atacama Pathfinder EXperiment) telescope in North Chile. Its under-sampled field of view is 11\farcm4 and its PSF 
FWHM is 19\farcs5, thus a resolution equivalent to that of the SPIRE 250 \mic\ instrument onboard \hersc. N11 was observed at 870 \mic\ with \lab\ in December 2008 and April, 
July and September 2009 (Program ID: O-081.F-9329A-2008 - PI: Hony). A raster of spiral patterns was used to obtain a regularly sampled map. Data are reduced with BoA 
(BOlometer Array Analysis Software)\footnote{http://www.apex-telescope.org/bolometer/laboca/boa/}. Every scan is reduced individually. They are first calibrated using the 
observations of planets (Mars, Uranus, Venus, Jupiter and Neptune) as well as secondary calibrators (PMNJ0450-8100, PMNJ0210-5101, PMNJ0303-6211, PKS0537-441, 
PKS0506-61, CW-Leo, Carina, V883-ORI, N2071IR and VY-CMa). Zenith opacities are obtained using a linear combination of the opacity determined via skydips and that 
computed from the precipitable water vapor\footnote{The tabulated sky opacities for our observations can be retrieved at http://www.apex-telescope.org/bolometer/laboca/calibration/.}. 
We also remove dead or noisy channels, subtract the correlated noise induced by the coupling of amplifier boxes and cables of the detectors. Stationary points and data taken 
at fast scanning velocity or above an acceleration threshold are removed from the time-ordered data stream. Our reduction procedure then includes steps of median noise removal, 
baseline correction (order 1) and despiking. The reduced scans are then combined into a final map in BoA. As noticed in \citet{Galametz2013a}, the steps of median noise removal 
or baseline correction are responsible for the over-subtraction of faint extended emission around the bright structures. In order to recover (most of) this extended emission we apply 
and additional iterative process to the data treatment.
We use the reduced map as a ``source model" and isolate pixels above a given signal-to-noise. This source model is masked or subtracted from the time-ordered data stream before 
the median noise removal, baseline correction or despiking steps of the following iteration, then added back in. We repeat the process until the process converges\footnote{In more 
detail, we used one blind reduction, then 3 steps where pixels at a 1.5-$\sigma$ level are masked from the time-ordered data. 5 steps where data at a 1.5-$\sigma$ level are 
subtracted from the time-ordered data are then applied until the process converges.}. Final rms and signal-to-noise maps are then generated. The average rms across the field 
is $\sim$8.4 mJy~beam$^{-1}$.  Figure~\ref{Spitzer_Herschel_LABOCA_maps} shows the final LABOCA map. We can see that our iterative data reduction helps us to significantly 
recover and resolve low surface brightnesses around the main structure of the complex.

\subsection{Preparation of the IR/submm data for the analysis}

We use the SAGE and HERITAGE MIPS, PACS and SPIRE maps of the LMC reprocessed by \citet{Gordon2014} in the following analysis. They include an additional step of 
foreground subtraction in order to remove the contamination by Milky Way (MW) cirrus dust. The morphology of the MW dust contamination has been predicted using the integrated 
velocity H\,{\sc i} gas map along the LMC line of sight and using the \citet{Desert1990} model to convert the H\,{\sc i} column density into expected contamination in the PACS and 
SPIRE bands. The image background is also estimated using a surface polynomial interpolation of the external regions of the LMC and subtracted from each image. All the maps 
are convolved to the resolution of SPIRE 500 \mic\ (FWHM: 36\arcsec) using the convolution kernels developed by \citet{Aniano2011}\footnote{Available at http://www.astro.princeton.edu/$\sim$ganiano/Kernels.html}. 
We refer to \citet{Gordon2014} for further details on these additional steps. We convolve the IRAC and LABOCA maps using the same convolution kernel library. For these maps, 
we estimate the background from each image by masking the emission linked with the complex, fitting the distribution of the remaining pixels with a Gaussian and using the peak 
value as a background estimate. Pixels of the final maps are 14\arcsec, which corresponds to 3.4pc at the distance of the LMC.

\subsection{Qualitative description of the dust emission}

Figure~\ref{Spitzer_Herschel_LABOCA_maps} first shows the IRAC 8 \mic, MIPS 24 and 70 \mic, the two PACS and three SPIRE maps at their respective original resolutions. 
The emission in the 8 \mic\ band observed with the IRAC instrument is primarily coming from polycyclic aromatic hydrocarbons (PAH). PAHs are large organic molecules thought to be responsible 
of the broad emission features often detected in the near- to mid-IR bands. They are present everywhere across the complex. The 24 \mic\ emission is a tracer of the hottest dust 
populations. Primarily  associated with H\,{\sc ii} regions, it is often used (by itself or combined with other tracers) as a calibrator of the star formation rate \citep[SFR,][]{Calzetti2007} 
in nearby objects. We observe that the MIPS 24 \mic\ emission is more compact than that at longer wavelengths and peaks in the H\,{\sc ii} regions of the N11 complex. The MIPS 
70 \mic\ band is essential to properly constrain the Wien side of the FIR SEDs and traces the warm dust in the complex. Part of the 70 \mic\ emission could also be associated with 
a very small grain (VSG) population. Studies at 70 \mic\ in the LMC have indeed revealed a population of VSG ($<$10nm; so larger than PAH molecules) probably produced through 
erosion processes of larger grains in the diffuse medium \citep{Lisenfeld2001,Bernard2008}. Erosion processes in magellanic-type galaxies has also been discussed in \citet{Galliano2003,Galliano2005}. 
The emission above 100 \mic\ is mostly produced by a mixture of equilibrium big ($>$25nm) silicate and carbonaceous grains \citep[see][among others]{Draine_Li_2001}. The 
PACS 100 and 160 \mic\ observations are associated with the cool to warm dust reservoirs (20-40K) in the complex: the observations enable us to sample the peak of the dust 
thermal emission across the field. Residual stripping can be observed in these maps. The SPIRE 250 to 500 \mic\ observations are associated with the coldest dust emission 
($<$20K): they will help us constrain the local dust masses as well as investigate potential emissivity variations of the dust grains in N11.

Using the \spitz, \hersc\ and LABOCA maps convolved to a common 36\arcsec\ resolution ($\S$2.4), we calculate the Spearman rank correlation coefficients between the various 
bands (log scale) for ISM elements that fulfill a 2-$\sigma$ detection criterion in the \hersc\ bands. Those coefficients are tabulated in Table~\ref{Correlation_coefficients} and 
highlight the high correlation between the various wavelengths. Correlation coefficients between the 8 \mic\ map and bands longward of 8 \mic\ are similar when the 8 \mic\ map 
is first corrected from stellar continuum (same to the second decimal place). The 8 \mic\ is more strongly correlated with the emission of cold dust traced by the SPIRE bands 
than with the tracers of hot and warm dust. This could be due to the fact that PAHs (that the 8 \mic\ emission mostly traces) are emitted at the surface of the molecular clouds 
where the shielded dust remains cold. The surface of a cloud (PAH emission) and its interior (cold dust emission) are expected to be probed by the same beam at the spatial 
scale on which we perform our analysis (10 pc). Emission from PAHs traced by the 8 \mic\ emission is less tightly correlated with star forming regions traced by hot dust tracers 
such as 24 \mic\ for instance \citep[as shown by][among others]{Calzetti2007_2}. The \lab\ emission at 870 \mic\ is as strongly correlated with the cold dust tracers as with the PAH emission. The correlation between the \lab\ emission and the other bands improves when we restrict the analysis to elements that fulfill a 10-$\sigma$ detection criterion. This indicates that the lower correlation with the \lab\ emission at 870 \mic is mostly linked with missing diffuse emission across the 870 \mic\ map.

\section{Dust SED modeling of the N11 complex}

\subsection{The method}

We select the \citet{Galliano2011} `AC model' in order to interpret the dust spectral energy distribution (SED) in each resolved element of the N11 structure. 

The SED modeling uses the optical properties of amorphous carbon \citep{Zubko1996} in lieu of the properties of graphite, which are more commonly 
employed to represent the carbonaceous component of the interstellar dust grains. This is motivated by two recent results.
First, studies of the dust emission in the LMC by \citet{Meixner2010} and \citet{Galliano2011} have shown that standard grains often lead to gas-to-dust 
mass ratios inconsistent with the elemental abundances, suggesting that LMC dust grains have a different (larger) intrinsic submm opacity compared to 
models which assume graphitic properties for the carbonaceous grains. Second, the comparison of the optical extinction estimated along the lines-of-sight 
towards a large number of quasi-stellar objects with the Galactic diffuse emission as measured by {\it Planck} \citep{PlanckCollaboration2014_Aniano}. In the latter 
analysis, the Draine \& Li (2007) model, which uses graphitic grains, over-predicts the dust column-densities by a factor of $\sim$2.
Our choice of amorphous carbon material (which would lead to lower masses than graphite) thus echoes these new discoveries currently leading to a general revision 
of the Galactic and extragalactic grain emissivity by the ISM community. \\
 
A detailed description of the modeling technique can be found in \citet{Galliano2011}. We state the most relevant aspects here. 
The \citet{Galliano2011} model assumes that the distribution of starlight intensities per unit dust mass can be approximated by a power-law as proposed by \citet{Dale2001}. 
We can thus derive the various parameters of the radiation field intensity distribution: the index $\alpha$ of the distribution (which characterizes the fraction of dust exposed to 
a given intensity) and the minimum and maximum heating intensity U$_{\rm min}$ and U$_{\rm max}$ (with U=1 corresponding to the intensity of the solar neighborhood, i.e. 
2.2$\times$10$^{5}$ W~m$^{-2}$).
The model assumes that the sources of IR/submm emission comes from the photospheres of old stars and dust grains composed of ionized and neutral PAHs and carbon and 
silicate grains. The old stellar contribution is modeled using a pre-synthesized library of spectra \citep[obtained with the stellar evolution model PEGASE][]{Fioc_Rocca_1997}. 
The old stellar mass parameter is only introduced in this analysis in order to estimate the stellar contribution to the MIR bands. The model provides estimates of the fraction of 
PAHs to the total dust mass ratio. Since the ionized PAH-to-neutral PAH ratio (f$_{\rm PAH+}$) is poorly constrained by the broadband fluxes, we choose to fix this value to 0.5; 
we will discuss the caveats of this approximation in $\S$4.3.2. \\

\begin{figure}
    \centering
     \hspace{-25pt}
     \includegraphics[height=5.8cm]{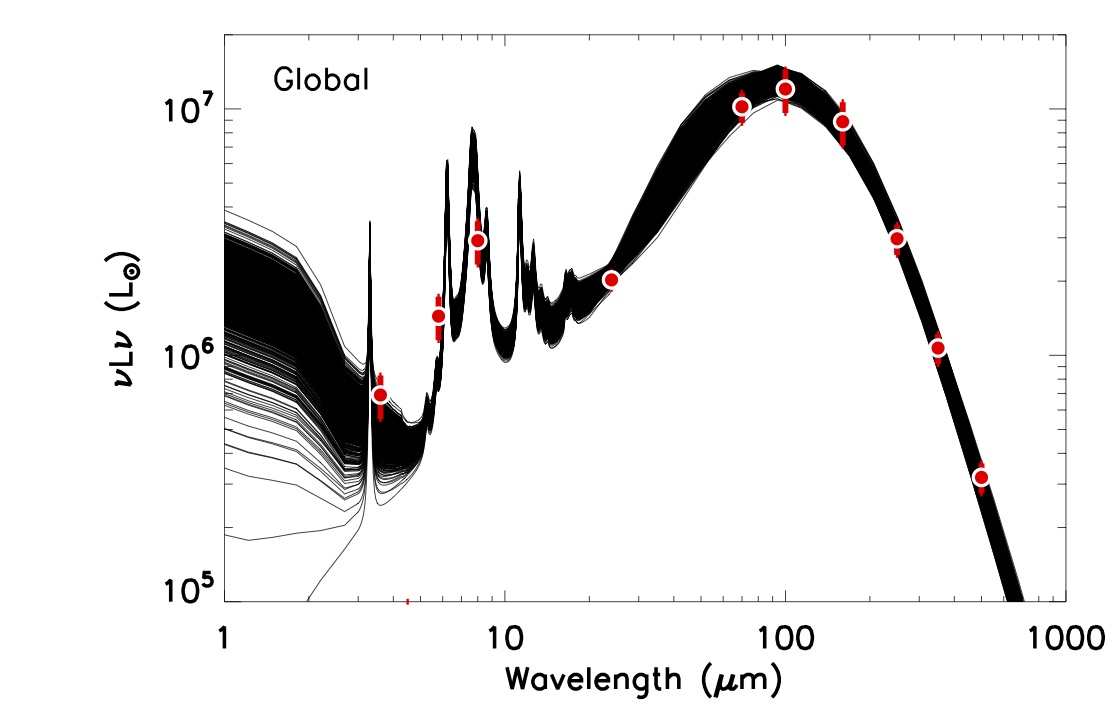}  \\
     \hspace{-25pt}
     \includegraphics[height=5.8cm]{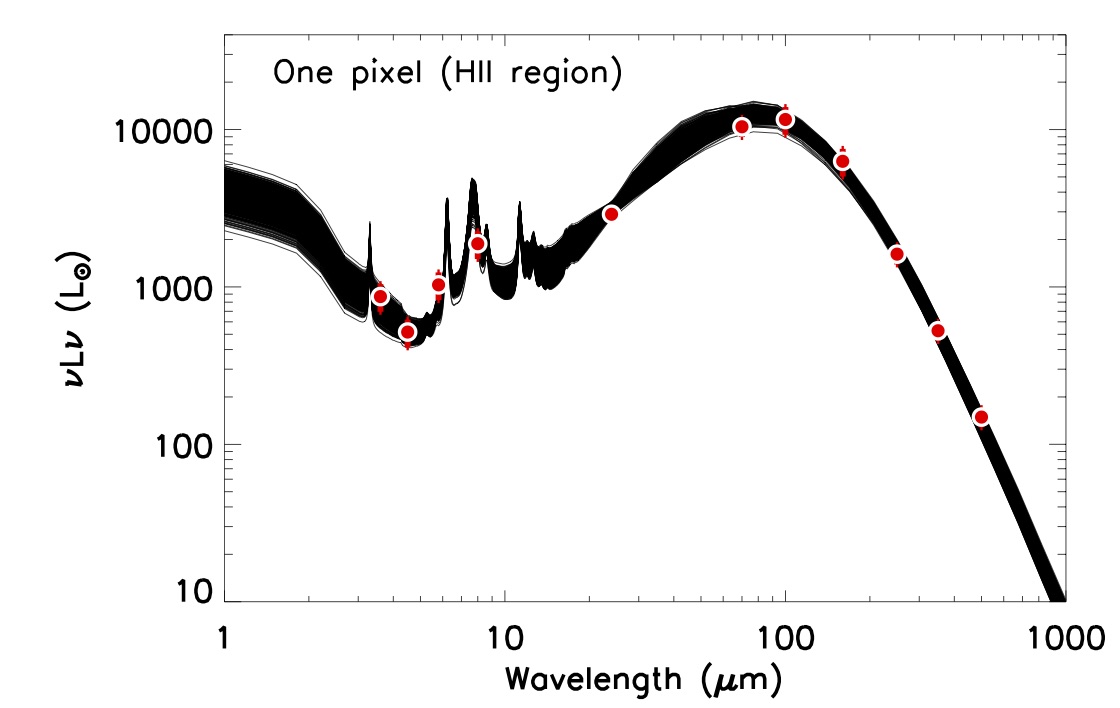}  \\
     \hspace{-25pt}
     \includegraphics[height=5.8cm]{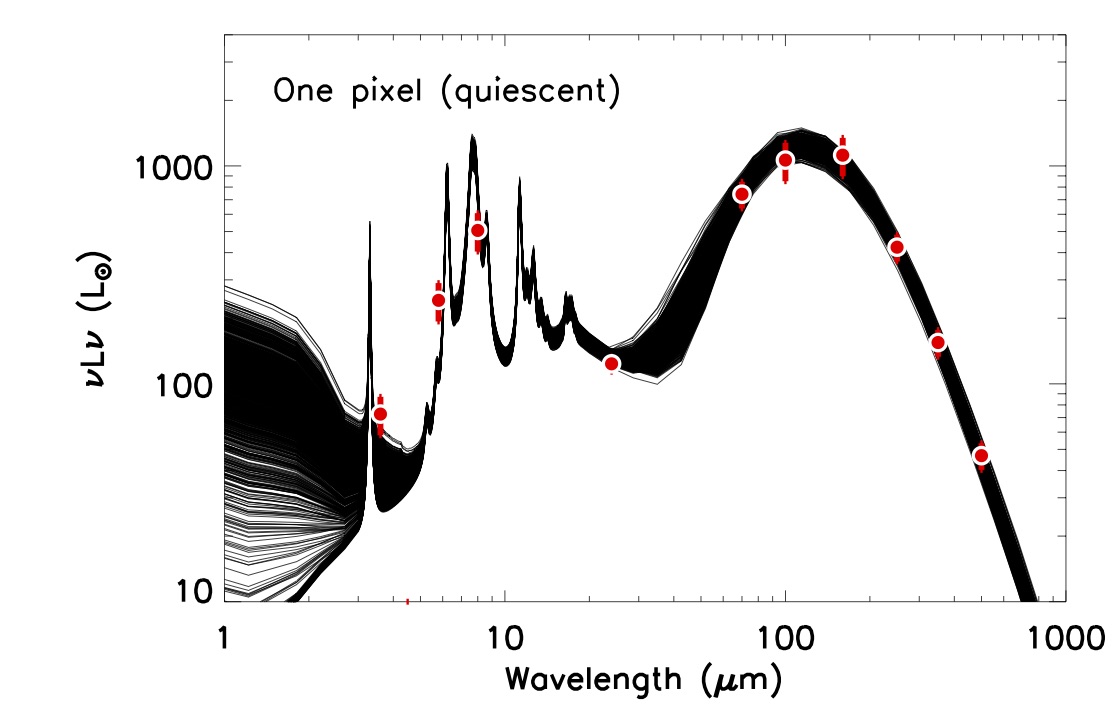}  \\
     \caption{The global SED and 2 local SEDs (from one ISM element in N11B and one element from the more quiescent ISM) modeled with the \citet{Galliano2011} `AC' model. 
     The various lines show the different realizations of the Monte Carlo technique we use to determine the parameter uncertainties (1000 realizations for each of these 3 cases). 
     \hersc\ measurements are overlaid in red. }
    \label{N11_GlobalSED}
\end{figure}

\vspace{-5pt}
\noindent The free parameters of our model are thus:

\vspace{-5pt}
\begin{itemize}
\item the total mass of dust (M$_{\rm dust}$),
\item the PAH-to-dust mass ratio (f$_{\rm PAH}$),
\item the index of the intensity distribution ($\alpha$),
\item  the minimum heating intensity (U$_{\rm min}$),
\item the range of starlight intensities (U),
\item the mass of old stars (M$_{\rm oldstars}$).
\end{itemize}

We apply the model to the \spitz+\hersc\ dataset (3.6 to 500 \mic) and convolve it with the instrumental spectral responses of the different cameras in order to derive the 
expected photometry. The fit is performed using a Levenberg-Marquardt least-squares procedure and uncertainties on flux measurements are taken into account to weight 
the data during the fitting (1 / uncertainty$^2$ weighting). The \citet{Galliano2011} model can help us to quantify the total infrared luminosities (L$_{\rm TIR}$) across the region. 
The model also predicts flux densities at longer wavelengths than the SPIRE 500 \mic\ constraint. We will, in particular, use predictions at 870 \mic\ in order to decompose 
the observed 870 \mic\ emission into its various (thermal dust and non-dust) components in $\S$4.5. \\

Modified blackbodies (MBB) models are commonly used in the literature to obtain average dust temperatures. In order to relate the temperatures derived using this method 
with the radiation 
field intensity derived from the more complex \citet{Galliano2011} fitting procedure, we fit the 24-to-500 \mic\ data using a two-temperature model, i.e. of the form: L$_{\nu}$ = 
A$_{\rm warm}$~$\lambda$$^{-2}$ B$_{\nu}$($\lambda$, T$_{\rm warm}$) + A$_{\rm c}$~$\lambda$$^{- \beta_{cold}}$ B$_{\nu}$($\lambda$, T$_{\rm cold}$). In this equation, 
B$_{\rm \nu}$ is the Planck function, T$_{\rm warm}$ and T$_{\rm cold}$ are the temperature of the warm and cold components, $\beta$$_{\rm cold}$ is the emissivity index of 
the cold dust component and A$_{\rm warm}$ and A$_{\rm cold}$ are scaling coefficients. We follow the standard approximation of the opacity in the \citet{Li_Draine_2001} 
dust models for the warm dust component (emissivity index of the warm dust fixed to 2). We fix the emissivity index of the cold component $\beta$$_{\rm cold}$ to the average 
value derived for the LMC in \citet{Planck_collabo_2011_MagellanicClouds}, i.e. 1.5. Fixing $\beta$ allows us to minimize the degeneracies between the dust temperature 
and the emissivity index linked with the mathematical form of the model we use and limit the biases resulting from this degeneracy \citep[][]{Shetty2009,Galametz2012}. 
Potential variations in the grain emissivity across the N11 complex (so using a free $\beta$$_{\rm cold}$) are discussed in $\S$5.3.4. 


\subsection{Deriving the parameter maps and median local SEDs} 

We run the SED fitting procedures for elements with a 2-$\sigma$ detection in all the \hersc\ bands. We apply a Monte Carlo technique to generate for every ISM element 
30 local SEDs by randomly varying the fluxes within their error bars with a normal distribution around the nominal value. A significant part of the uncertainties in SPIRE bands 
is correlated. To be conservative in the parameters we derive, especially on the dust mass estimates directly affected by variations in the SPIRE fluxes, we decide to link the 
variations of the three SPIRE measurements consistently during the Monte Carlo procedure. Figure~\ref{N11_GlobalSED} shows an example of Monte Carlo realizations of 
the \citet{Galliano2011} `AC' modeling procedure in three cases: if we consider the N11 complex as one single ISM element (top), for one ISM element in N11B (middle) 
and for a more quiescent ISM element (bottom). We finally use these various local Monte Carlo realizations to create a median map for each parameter and a median 
SED for each 14\arcsec\ $\times$ 14\arcsec\ ISM elements. The standard deviations are also providing the uncertainties on each of the parameter derived from the modeling. 
The parameter maps are discussed in $\S$4.

\subsection{Residuals from the Galliano et al (2011) fitting procedure}

In order to assess the ability of the \citet{Galliano2011} procedure to reproduce the observations, we analyze the residuals to the fitting procedure at i = 8, 24, 70, 100, 
160, 250, 350 and 500 \mic. In order to obtain a synthetic photometry (L$_{\nu}^{modeled}$(i)) to which observed fluxes (L$_{\nu}^{observed}$(i)) can be compared directly, 
we integrate the median modeled SEDs obtained on local scales in each instrumental filter. Recall that all these observed fluxes are included as constraints in the fitting 
process. The relative residuals r$_{\rm i}$ determined in this study are defined as:

\begin{equation}
r_{\rm i} = \frac{L_{\rm \nu}^{observed} (i) - L_{\rm i}^{modeled}(i)}{L_{\rm \nu}^{observed}(i)}
\end{equation}

\begin{figure*}
    \centering
\includegraphics[width=18cm]{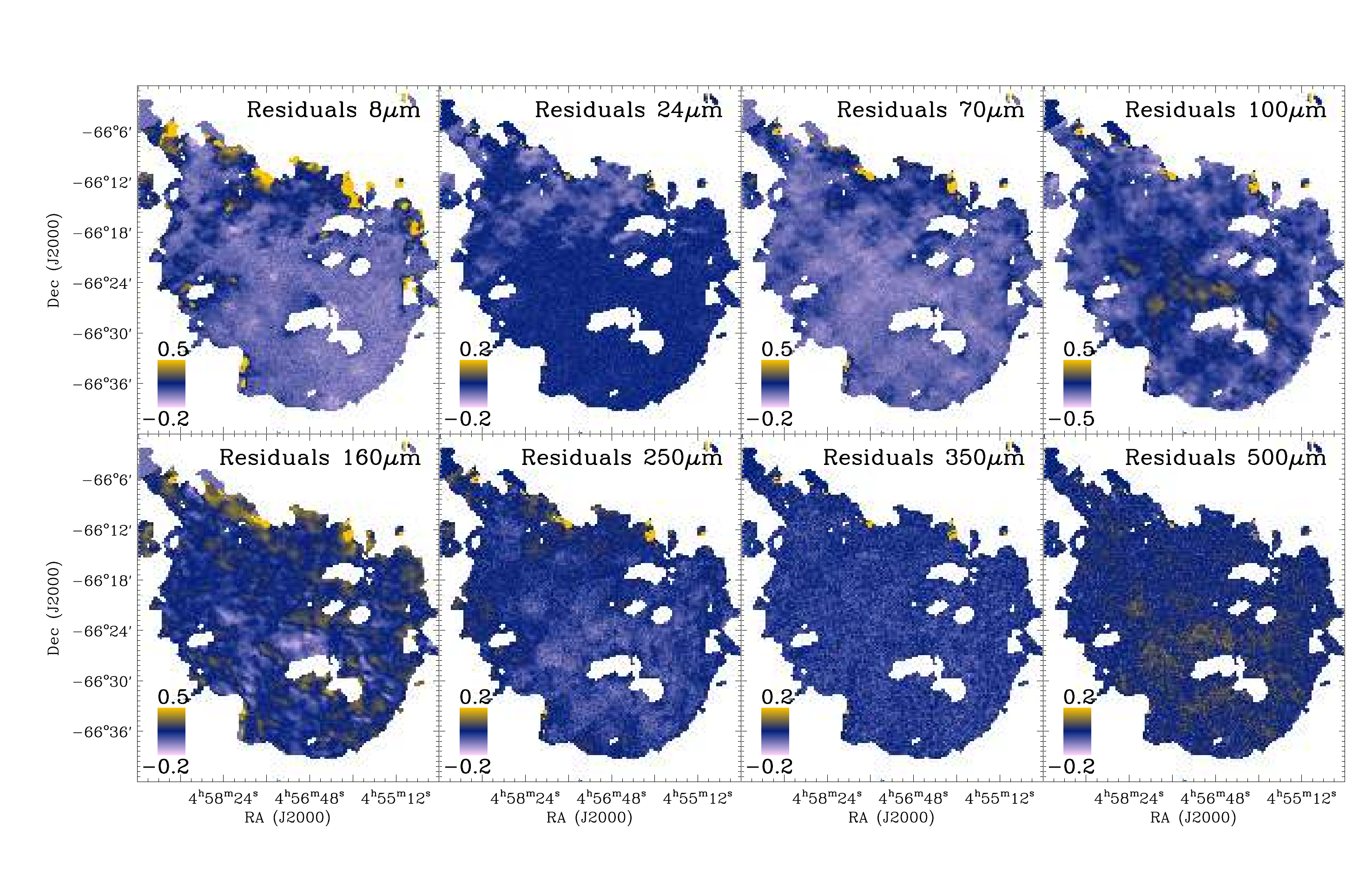}     \\      
 \includegraphics[width=18.5cm]{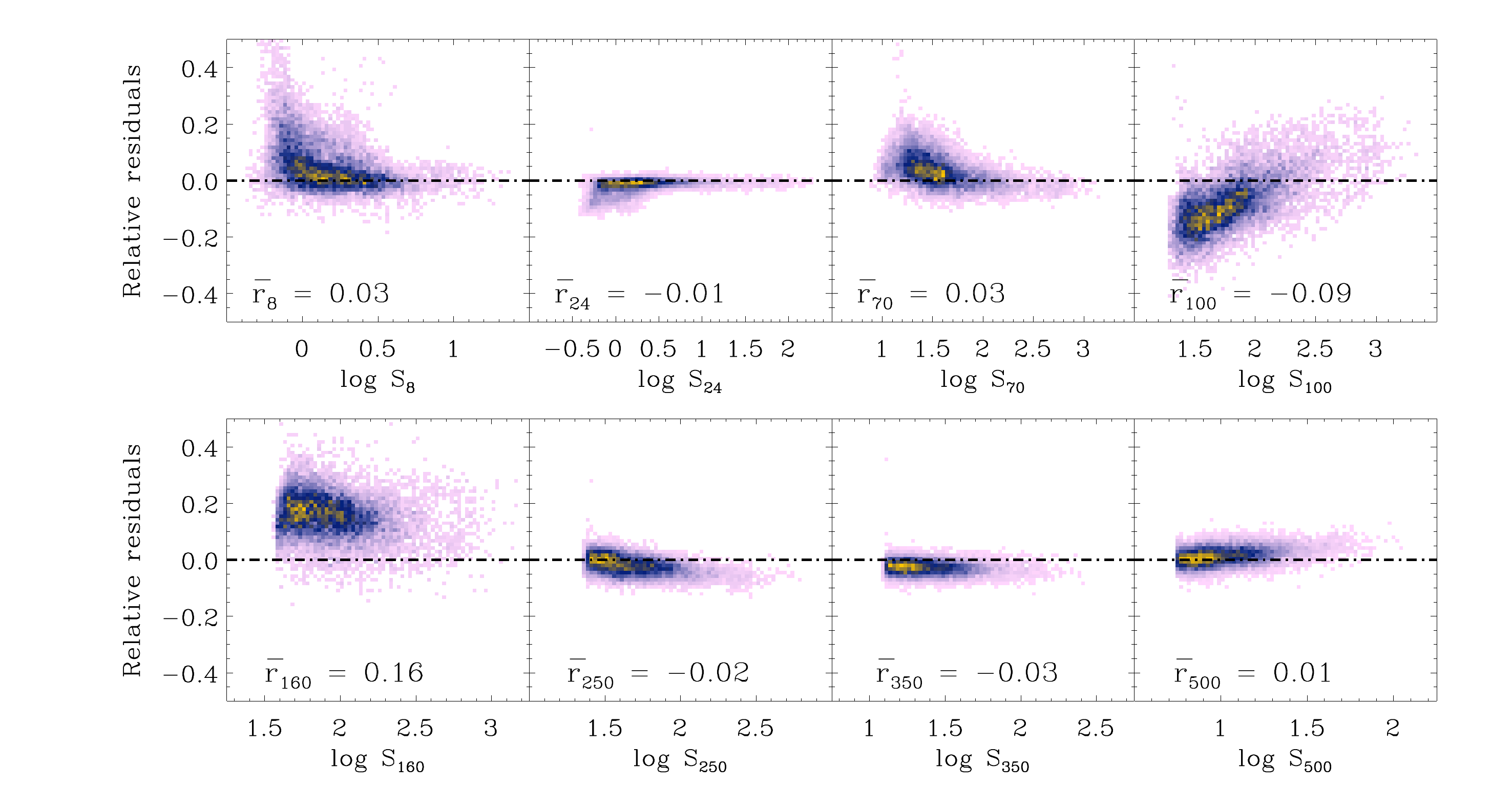}  \\
     \caption{{\it Top -} Distribution of the relative residuals r$_{\rm 8}$, r$_{\rm 24}$, r$_{\rm 70}$, r$_{\rm 100}$, r$_{\rm 160}$, r$_{\rm 250}$, r$_{\rm 350}$ and 
     r$_{\rm 500}$ to the \citet{Galliano2011} fitting procedure across the N11 complex. The relative residuals are defined as (observed flux - modeled flux) / (observed flux). 
     A residual of 0.2 thus indicates that the model underestimates the observed luminosity by 20\%. {\it Bottom -} Relative residuals to the \citet{Galliano2011} fitting 
     procedure as a function of the respective observed fluxes S$_{\rm i}$. S$_{\rm i}$ is expressed in MJy/sr. In this plot, colors scale with the density of points. The 
     median value of the residuals is indicated at the bottom of each plot.}
    \label{N11_allresiduals_maps}
\end{figure*}

Figure~\ref{N11_allresiduals_maps} (top) compares the spatial distribution of the relative residuals of r$_{\rm 8}$, r$_{\rm 24}$, r$_{\rm 70}$, r$_{\rm 100}$, r$_{\rm 160}$, 
r$_{\rm 250}$, r$_{\rm 350}$ and r$_{\rm 500}$ across the N11 complex while Figure~\ref{N11_allresiduals_maps} (bottom) shows these residuals as a function of the 
respective flux densities. Relative residuals are quite small. Except for the 160 \mic\ band, the median values are close to 0, so consistent with a reliable fitting of the 
observational constraints.
The lowest residuals appear in the 24 \mic\ and the 3 SPIRE (250, 350 and 500 \mic) bands, with standard deviations lower than 0.03. The 70 \mic\ residual map shows 
that our modeling procedure slightly underestimates the 70 \mic\ observed flux in the diffuse regions of N11. This could be linked with emission from VSGs produced 
through processes of erosion of larger grains in the diffuse medium where they are less shielded from the interstellar radiation fields \citep{Bot2004,Bernard2008,Paradis2009_1}. 
Residuals in the 100 \mic\ band seems to be dependent on the ISM element surface brightness (overestimation at low surface brightnesses, underestimation at high surface brightnesses). 

The largest residuals are observed in the PACS 160 \mic\ band, with an underestimation of the observed fluxes by 16\% throughout the complex. A non-negligible 
fraction of these residuals could originate from the strong [C\,{\sc ii}] line emission at 157 \mic\ emitting near the peak of the PACS 160 \mic\ filter sensitivity. [C\,{\sc ii}] 
emission is indeed detected in every LMC regions mapped with PACS. We estimate that a C\,{\sc ii}-to-TIR ratio of 2-3\% would be sufficient to explain most of the 
residual we observe at 160 \mic. This ratio is high but consistent with the values estimated in the N11 complex by \citet{Israel2011}.

Part of the discrepancies at 100 and 160 \mic\ could finally be due to a combination of both observational and modeling effects. First, calibration uncertainties in the 
PACS maps at low surface brightnesses as well as uncertainties in background/foreground estimates have a significant impact on the fluxes in the most diffuse regions. 
We remind the reader that 160 \mic\ is where the Galactic cirrus peaks while the LMC SED seems to be flatter than that of the Milky Way in the submm regime. We 
also remind that the modeling procedure we are using assumes AC, thus have a fixed slope of $\sim$1.7 while the \citet{Planck_collabo_2011_MagellanicClouds} 
results lead to effective emissivities closer to $\sim$1.5 on average for the LMC. The residuals we observe could thus also be a sign of different optical properties.

\section{The dust properties in N11}

In this section, we examine local IR-to-submm SEDs across the complex. We also analyze the distributions of the modeling parameters derived from the two SED 
modeling procedures and their correlations. For the rest of the paper, observed maps and parameter maps will be displayed using two different color tables.

\subsection{Local SED variations}

\begin{figure}
    \centering
         \vspace{10pt}
     \begin{tabular}{c}
     \hspace{-25pt}
     \includegraphics[height=9cm]{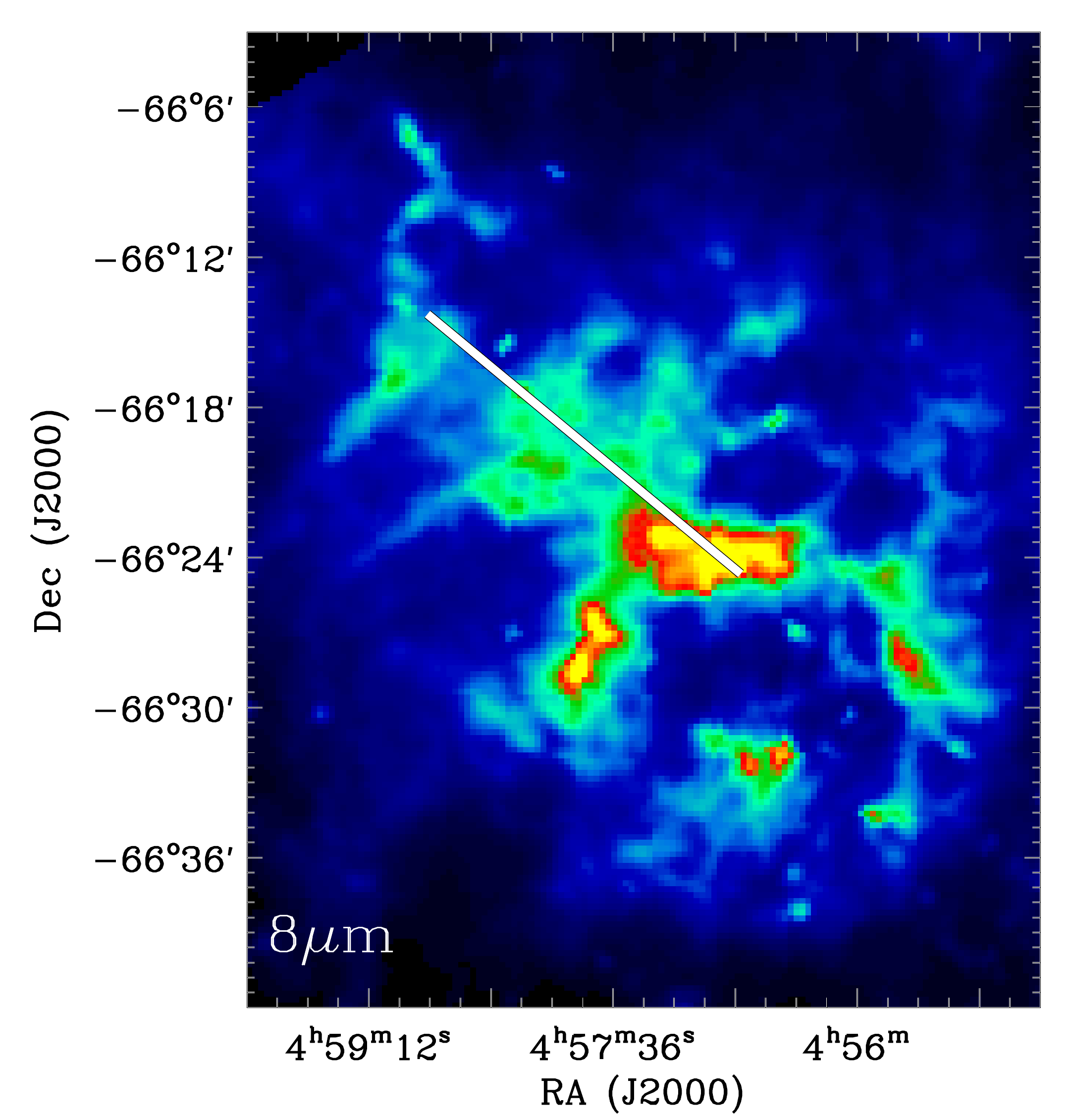}  \\
     \hspace{-30pt}
     \vspace{10pt}
     \includegraphics[height=6cm]{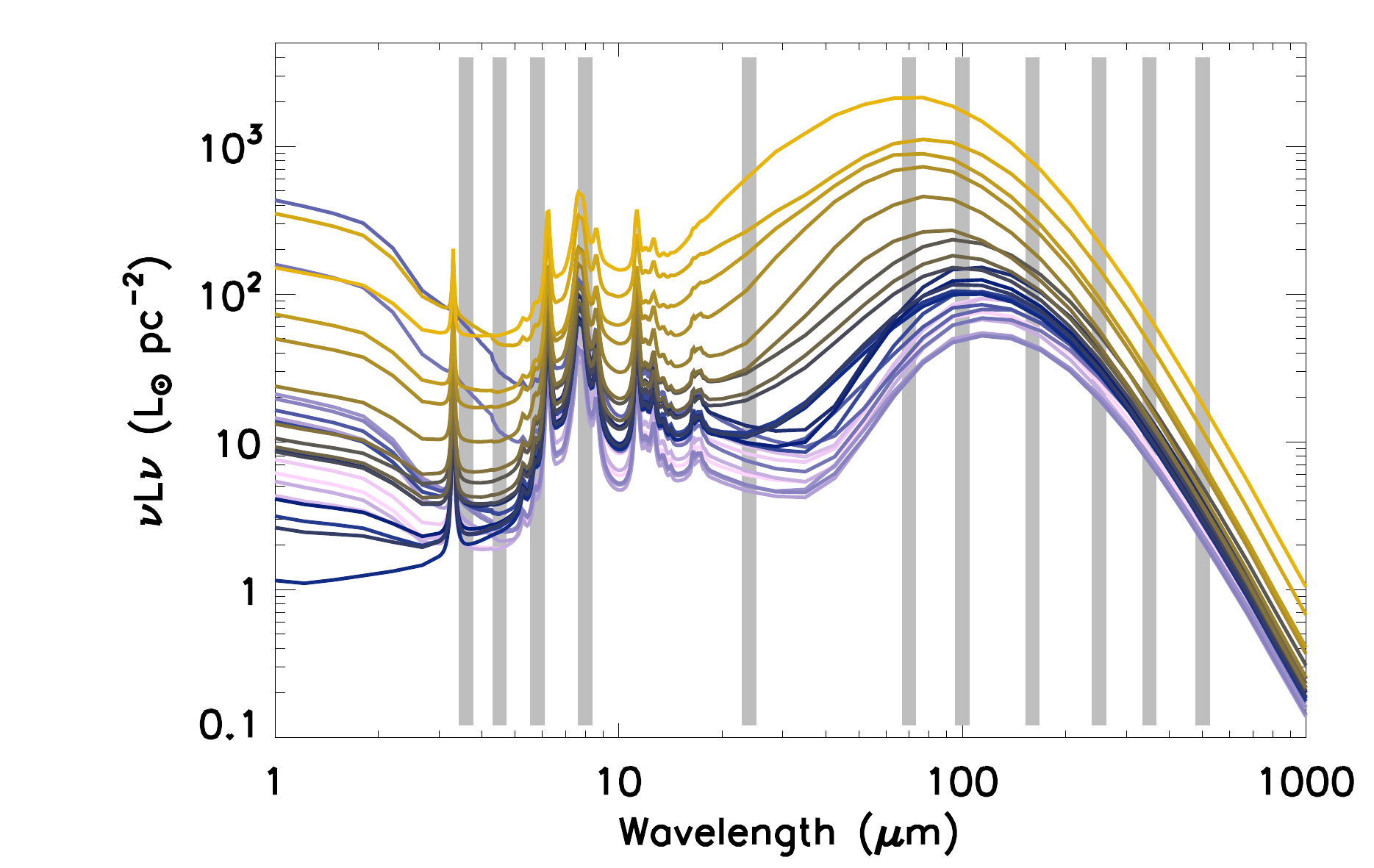}\\
     \end{tabular}\    
     \caption{{\it Top: }The N11 complex observed in the 8 \mic\ band. The image has been convolved to the working resolution of 36\arcsec. The white line indicates the 
     locations of the ISM elements whose individual SEDs are plotted in the figure below. {\it Bottom:} Local SEDs across the N11 complex. Colors indicate the position 
     along the above white line, from light purple in the diffuse ISM (north-east) to yellow in the bright star forming region N11B. The vertical grey lines indicate the 
     position of the \spitz\ and \hersc\ bands.}
    \label{N11_LocalSEDs_vector}
\end{figure}

Figure~\ref{N11_LocalSEDs_vector} (bottom panel) gathers a collection of local IR-to-submm SEDs across the N11 region.
These SEDs are extracted from ISM elements located along the north-eastern filament down to the northern edge of the N11 ring (the location of the selected elements 
is indicated with a white line in the top panel of Fig~\ref{N11_LocalSEDs_vector}). We can see how the local SED varies from star-forming regions to more quiescent 
regions. Bright star forming regions (in yellow) show a wider range of temperatures (broader SED) and a lower PAH fraction (weaker features at 8 \mic) while more 
quiescent regions (in purple) show colder temperatures and a much narrower range of dust temperatures. 

\begin{figure*}
    \centering
    	\vspace{20pt}
     \includegraphics[width=18cm, height=15cm]{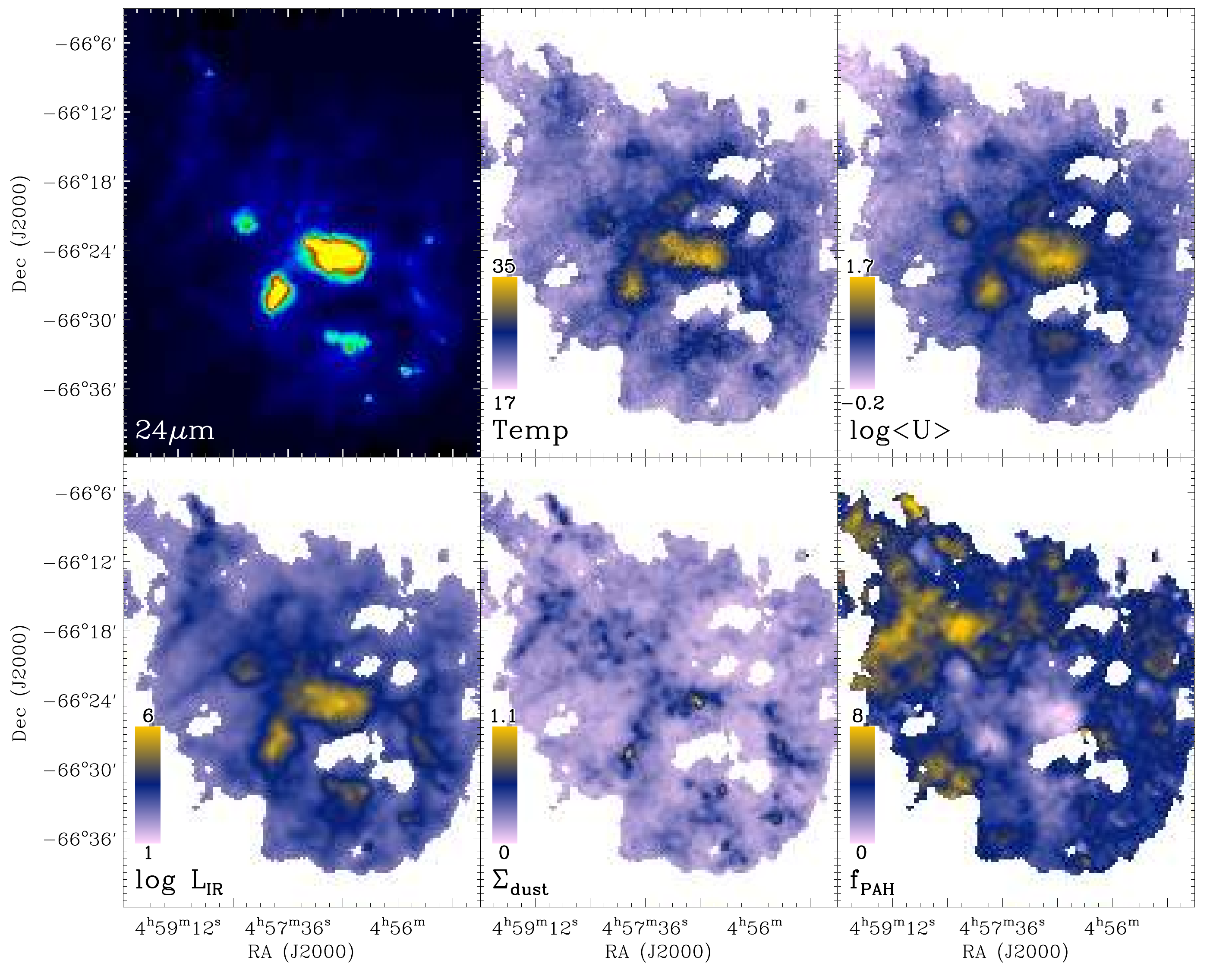}     \\    
     \caption{Maps of the parameters derived from our two SED modeling techniques. 
     The 24 \mic\ map (convolved to a resolution of 36\arcsec) is shown for reference in the upper left panel (square root scaling). 
     {\it Top middle panel:} Cold dust temperature in Kelvins. 
     {\it Top right panel:} Mean starlight heating intensity, with U=1 corresponding to the intensity of the solar neighborhood (log scale). 
     {\it Bottom left panel:} Far-infrared luminosity in \lsun~pc$^{-2}$ (log scale). 
     {\it Bottom middle panel:} Dust surface density in \msun~pc$^{-2}$. All the parameter maps are obtained using the \citet{Galliano2011} modeling technique (with the choice 
     of amorphous carbon grains to model the carbon dust) except the cold dust temperature map that is obtained using the two-temperature modified blackbody fitting technique.   }
     {\it Bottom right panel:} PAH fraction to the total dust mass (in $\%$). 
    \label{SED_results}
\end{figure*}

From our local modeling of the IR SEDs, we can derive a map of the infrared luminosity L$_{\rm IR}$. We integrate the models (in a $\nu$-f$\nu$ space) from 8 \mic\ 
to 1100 \mic. The stellar contribution to the near-IR emission was estimated during the SED modeling process. Even if minor in that wavelength range, this contribution 
is removed from the total L$_{\rm IR}$ to only take the emission from dust grains into account. Figure~\ref{SED_results} (bottom left panel) shows the L$_{\rm IR}$ 
map (in units of \lsun). We can observe that the distribution of L$_{\rm IR}$ is much more extended than the regions where high-mass star formation is taking place, 
with a significant contribution arising from more quiescent regions (regions of weaker 24 \mic\ emission). This extended distribution is due to the fact that part of the 
dust emission is not directly related to star formation occurring in the same beam. A fraction of the dust heating is in fact related to the older stellar populations or to ionizing 
photons leaking out the H\,{\sc ii} regions due to the porous ISM in N11 \citep[see][]{Lebouteiller2012}.

\subsection{Radiation field and dust temperatures}

Figure~\ref{SED_results} (upper middle panel) shows the average cold temperature maps obtained using the MBB model. Temperatures vary significantly across the 
N11 complex and range from 32.5 K in the N11B nebula (associated with the LH10 stellar cluster) down to 17.7 K in the diffuse ISM (with a median of 20.5 K for regions 
detected at a 2 but not 3-$\sigma$ level). The median temperature of the modeled regions is 21.5{\small $\pm$2.1} K. This value is similar to that derived if we keep the 
emissivity index $\beta$ free in the modeling (21.4{\small $\pm$3.4} K). For comparison, \citet{Herrera2013} estimated a mean temperature of $\sim$20 K in the 
region of N11 from the LMC temperature map of \citet{Planck_collabo_2011_MagellanicClouds}, close to what we obtain. The N11 region is representative of the 
average dust temperatures found in the LMC \citep{Bernard2008}. A comparison between N11 and the N158-N159-N160 star forming complex previously studied 
by \citet{Galametz2013a} shows that N11 is colder on average (median temperature of 26.9 K for the N159 region). The coldest temperatures in the N159 complex 
were estimated in the N159 South region ($\sim$22 K), a significant reservoir of dust and molecular gas but with no ongoing massive star formation detected. The 
temperatures of the cold dust grains in the diffuse ISM in N11 are similar to those in N159 South. We finally note that the temperature of N11 when considered as 
one single ISM element (Fig.~\ref{N11_GlobalSED} top) is 23.1{\small $\pm$1.1} K (at the higher end of the uncertainties of the median derived locally). This 
highlights again that determining the dust temperatures on local scales is essential to constrain the coldest phases of the dust.

From the \citet{Galliano2011} SED modeling results, we locally derive three parameters characterizing the distribution of the radiation field intensities, namely the index 
of the simple power law assumed for the distribution $\alpha$ and the minimum and maximum values of the intensities, U$_{\rm min}$ and U$_{\rm max}$ respectively. 
These 3 parameters are combined to derive a map of the mass-weighted mean starlight intensity $<$U$>$ \citep[see][eq. 11]{Galliano2011} shown in Fig.~\ref{SED_results} 
(upper right panel). Recall that values are normalized to those of the solar neighborhood (with U = 1 corresponding to an intensity of 2.2 $\times$ 10$^{-5}$ W m$^{-2}$). 
The dust temperature of the cold grains (derived from the MBB modeling) and the mean radiation field intensity $<$U$>$ that heat those grains (derived from the `AC model') 
have very similar distribution, as expected\footnote{$<$U$>$ is integrated over the whole range of radiation field intensity, and thus includes the contribution from hot regions. 
The temperature T$_{\rm c}$ derived from the MBB modeling is not constrained by $\lambda$$<$100, and thus does not include the hot phases}. While we analyze the local 
and median values of these two parameters in this section, their correlation is studied in more details in $\S$4.4. The median intensity $<$U$>$ of the N11 complex is 2.3 (2.8 
if we restrict the median calculation to regions with a 3-$\sigma$ detection in the \hersc\ bands). Peaks in the $<$U$>$ distribution are observed along the N11 shell, in 
particular in the N11B nebula where it reaches a maximum of 31.8 times the solar neighborhood value. This value is similar to that estimated in the LMC/N158 region in 
\citet{Galametz2013a}. N158 is a H\,{\sc ii} region where two OB associations were detected \citep{Lucke1970}. In N158, the southern association hosts two young stellar 
populations of 2 and 3-6 Myr \citep{Testor1998}. These cluster ages are close to those expected from the intermediate-mass Herbig Ae/Be population detected in the N11B 
nebula by \citet{Barba2003}. The two regions are thus very similar in terms of evolutionary stage.

\subsection{The dust distribution}

\subsubsection{Total dust masses}

Figure~\ref{SED_results} (bottom middle panel) shows the surface density map $\Sigma$$_{\rm dust}$ (units of \msun~pc$^{-2}$) obtained with the `AC model'. The dust 
distribution appears to be very structured and clumpy, with major reservoirs in N11B as well as in the N11C nebula located on the eastern rim. Secondary dust clumps are 
located along the N11 shell. The median of the dust surface density across N11 is 0.22 \msun~pc$^{-2}$. The total dust mass derived for the complex is 3.3{\small $\pm$0.6} 
$\times$ 10$^{4}$ \msun. Uncertainties are the direct sum of the individual uncertainties derived from our Monte-Carlo realizations. The main peaks in the dust distribution 
coincide with the molecular clouds catalogued by \citet[][identification on the SEST CO(2-1) observations at a 23\arcsec\ resolution]{Herrera2013} as shown later in 
Fig.~\ref{HI_CO_Dust} (bottom panel). We find that only 10$\%$ of the total dust mass ($\sim$2.7{\small $\pm$0.5} $\times$ 10$^{3}$ \msun) resides in these individual clumps. 

Using the same dataset than this analysis and a single temperature blackbody modified by a broken power-law emissivity (BEMBB), \citet{Gordon2014} produced a dust 
mass map of the whole LMC. They obtain a total dust mass that is a factor of 4-5 lower than values derived from standard dust models like the \citet{Draine_Li_2007} models. 
We convolve our $\Sigma$$_{\rm dust}$ mass map to their 56\arcsec\ working resolution to position our dust estimates (for the same area) in that dust mass range. We find 
that the mass estimated for the whole N11 region in \citet{Gordon2014} is 2.5 times lower than the dust mass we derive (on average 2.4 times lower for $\Sigma$$_{\rm dust}$ 
$<$ 0.2 \msun~pc$^{-2}$ and 2.6 times lower for $\Sigma$$_{\rm dust}$ $>$ 0.2 \msun~pc$^{-2}$). The dust masses we estimate with the `AC model' are moreover 2.5 times 
lower than those obtained if we use standard graphite in lieu of amorphous carbon to model the carbonaceous grains (as already shown in \citet{Galliano2011} and \citet{Galametz2013a}). 
Our dust masses thus reside in between those derived by the BEMBB model and a standard `graphite' dust model. These results highlight how the choice of dust composition 
can dramatically influence the derived dust masses.

Finally, many studies have shown that total dust masses are usually underestimated when derived globally rather than locally \citep[][among others]{Galliano2011, Galametz2012}. 
On global scales, the SED modeling technique is poorly disentangling between warm / cold / very cold dust at submm wavelengths - due to the combined effects of a poor spatial 
resolution and a small number of submm constraints - and the coldest phases of dust can be diluted in warmer regions. 
This `resolution effect' is not a physical effect but a methodological bias linked with the non-linearity of the SED modeling procedures. To test this effect in N11, we model the 
complex as a single ISM element. We obtain a total dust mass of 2.8$\times$10$^4$ \msun, thus $\sim$15$\%$ lower than the `resolved' dust mass. This estimate is at the 
lower limit of our dust mass uncertainty range.

\begin{figure*}
    \centering
  \includegraphics[width=18cm]{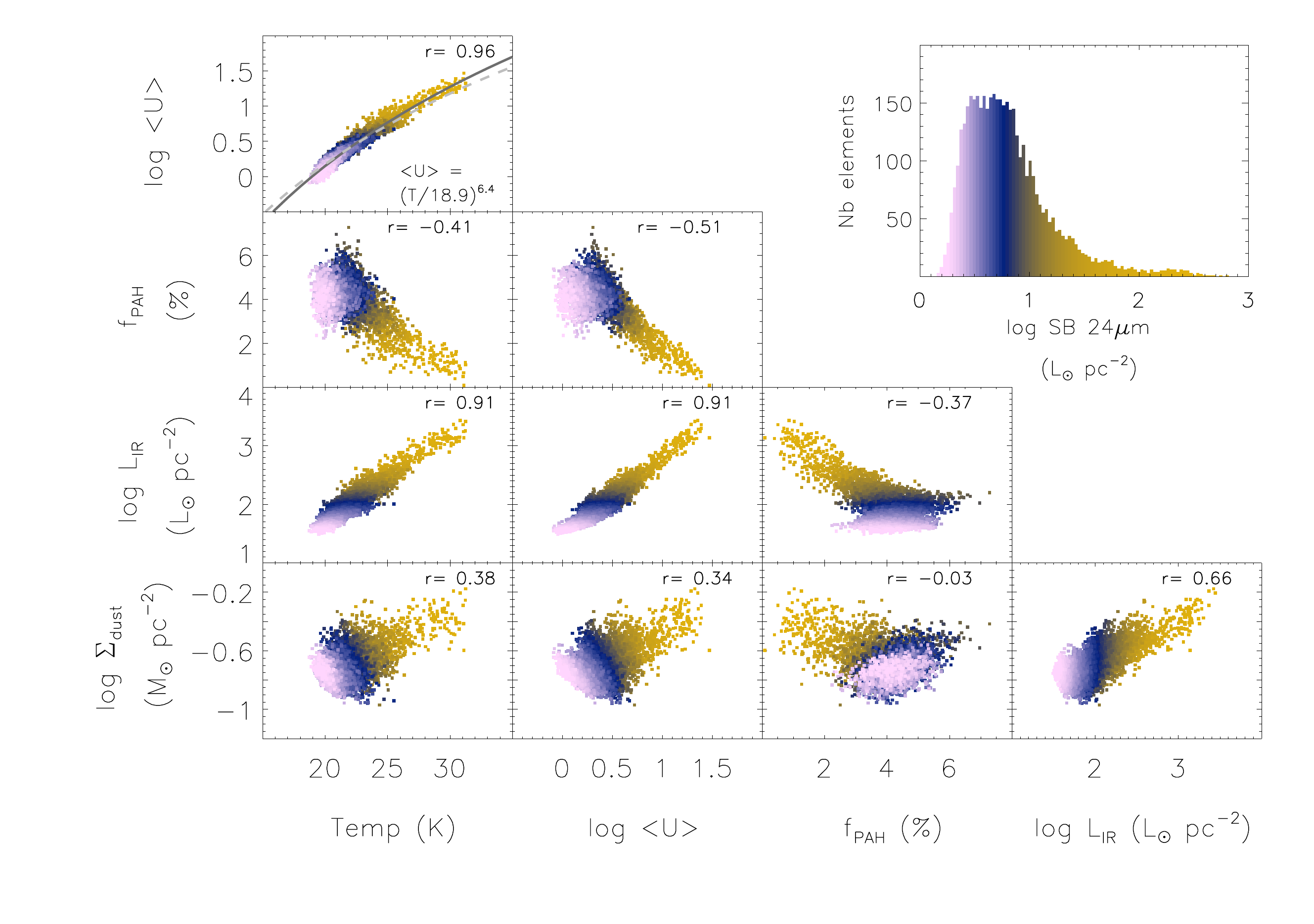}     \\   
  \vspace{-20pt} 
     \caption{Correlation plots between various model parameters: the dust temperature (Temp) in K, the mean starlight heating intensity normalized to the intensity of the solar 
     neighborhood ($<$U$>$), the PAH-to-total dust mass in percent (f$_{\rm PAH}$), the IR luminosity in \lsun\ (L$_{\rm IR}$) and the dust surface density in \msun~pc$^{-2}$ 
     ($\Sigma$$_{\rm dust}$). All parameters are obtained using the \citet{Galliano2011} modeling technique except the cold dust temperature that is obtained using the 
     two-temperature modified blackbody fitting technique. The Spearman rank correlation coefficients are added to each plot. The ISM elements are colored as a 
     function of their MIPS 24 \mic\ surface brightness. The inset plot shows the 24 \mic\ surface brightness histogram and provides the color scale for the plots 
     (from yellow for higher 24 \mic\ surface brightnesses to light purple for lower 24 \mic\ surface brightnesses). The grey curve on the top panel indicates the best-fit 
     curve. The equation is provided in the panel. The dashed grey curve indicates the fit with a coefficient fixed to 5.7 ($<$U$>$ = (T/18.6)$^{5.7}$).}
    \label{ModelsResults_correlations}
\end{figure*}

\subsubsection{PAH fraction to the total dust mass} 

PAHs are planar molecules ($\sim$1nm) made of aromatic cycles of carbon and hydrogen and are thought to be responsible for the strong emission features observed 
in the near- to mid-IR \citep{Leger1984}. The main features are centered at 3.3, 6.2, 7.7, 8.6 and 11.3 \mic. \citet{Draine_Li_2007} and \citet[][bare silicate and graphite 
grain models]{Zubko2004} found that 4.6$\%$ of the total dust mass in the Milky Way could reside in PAHs. Using \spitz/IRS spectra, \citet{Compiegne2011} obtain a 
larger f$_{\rm PAH}$ (7.7$\%$) for the same environment. It has been shown that the PAH fraction also varies significantly depending on the intensity of the radiation 
field and with the metallicity of the environment \citep[][among others]{Engelbracht2008, Galliano_Dwek_Chanial_2008}.
Our fitting procedure can help us assess these variations on local scales. Figure~\ref{SED_results} (bottom right panel) shows the distribution of the PAH-to-total dust 
mass fraction (in $\%$) across the N11 complex. We find a median f$_{\rm PAH}$ across the complex of $\sim$4$\%$, thus close to the \citet{Draine_Li_2007} and the 
\citet{Zubko2004} studies\footnote{A lower value of f$_{\rm PAH}$ = 3.3$\%$ is obtained when the region is modeled as one single ISM element.}. However, the fraction 
varies significantly across the complex. Lower fractions ($<$1\%) are for instance estimated for regions with high radiation field intensities while higher fractions ($>$6-12\%) 
are observed in more diffuse regions of the complex. This is similar to what has been observed in the LMC N158-N159-N160 complex \citep{Galametz2013a} and consistent 
with the expected destruction of PAH molecules through photodissociation processes scaled with the radiation field hardness and intensity \citep{Madden2005}. \\

\noindent {\it Caveats -} Laboratory experiments have shown that the various PAH bending modes (C-C, C-H etc.) vary with the PAH charge: neutral PAHs preferably emit 
around 11\mic\ while ionized PAHs preferably emit around 8 \mic. Because of our lack of constraint in the mid-IR spectrum, we fixed the fraction of ionized PAHs f$_{\rm PAH+}$ 
to 0.5. This means that {\it i)} the f$_{\rm PAH}$ we derive naturally scales, by model construction, with 8 \mic-to-L$_{\rm IR}$ luminosity ratio (correlation coefficient r=0.91) 
and {\it ii)} the neutral PAHs scale with the ionized PAHs. Yet, by studying local f$_{\rm PAH}$ in the SMC \citep[12+log(O/H) $\sim$8.0;][]{Kurt1998}, \citet{Sandstrom2012} 
find that PAHs in the SMC tend to be smaller and more neutral than in more metal-rich environments. This could also be the case in the LMC, affecting the PAH mass. 
Mid-IR spectra would be necessary to properly quantify the ionized-to-neutral PAH fraction in N11. On a side note, \citet{Jones2013} \citep[see also][]{Jones2014} recently 
proposed nanometer-sized aromatic hydrogenated amorphous carbon grains (a-C(:H)) in lieu of free flying PAHs to explain the mid-IR diffuse interstellar bands we 
observe in the Galaxy. A more systematic comparison of the two hypotheses would enable us to test their ability to reproduce the observations of different environments.

\subsection{Correlations between parameters}

Figure~\ref{ModelsResults_correlations} presents correlation plots between the dust temperature (Temp), the mean starlight heating intensity ($<$U$>$), the PAH fraction 
f$_{\rm PAH}$, the IR luminosity L$_{\rm IR}$ and the dust surface density $\Sigma$$_{\rm dust}$. We are using ISM elements of 14\arcsec, which means that 
our neighboring pixels are not independent. However, a quick test using 42\arcsec\ pixels shows that the correlations observed in Figure~\ref{ModelsResults_correlations} 
are not affected by our choice of pixel size. 

The top left panel shows the strong relation between the dust temperature 
and the mean starlight heating intensity across the N11 complex. In thermal equilibrium conditions, the energy absorbed by a dust grain is equal to that re-emitted. 
This leads to a direct link between the temperature of the grain T, its emissivity $\beta$ and the intensity of the surrounding radiation field U, with U $\propto$ T$^{(4+\beta)}$. 
By model construction, the submillimeter effective emissivity of our ``AC model" is $\beta$ = 1.7 \citep[see][]{Galliano2011}, so $<$U$>$ is proportional to T$^{5.7}$. 
We fit our ISM elements, fixing $\beta$ to 1.7, and derive the relation $<$U$>$ = (T/18.6$^{\pm0.02}$)$^{5.7}$, thus a normalizing equilibrium dust temperature of 
18.6K (dashed line in Fig.~\ref{ModelsResults_correlations}, top panel). Fitting our ISM elements with no {\it a priori} on $\beta$ leads to the relation: $<$U$>$ = 
(T/18.9$^{\pm0.05}$)$^{6.4^{\pm0.10}}$ (solid line). Uncertainties in the fit are calculated using a jackknife technique: we apply the fit to 1/10 ISM elements randomly 
selected and derive the parameters. We then repeat this procedure 1000 times to derive a final median and standard deviation per parameter. The small uncertainties 
highlight the very tight correlation between these two parameters. The predicted relation between $<$U$>$, T and $\beta$ is different from that fitted to the ISM elements
modeled. This discrepancy is driven by ISM elements with log $<$U$>$ $<$ 0.5, i.e. the `diffuse' ISM of N11. These elements show a very steep submm slope. They reach
effective $\beta$ higher than 2 when you let $\beta$ vary, values that are difficult to explain from our current knowledge about grain physics. Because we fix $\beta$ 
to 1.5 in our MBB fitting technique, our cold temperatures are higher than what would be derived with a higher index (i.e. $\beta$=1.7 or more). This translates 
into an increase of the fitting coefficient from 5.7 to 6.4.

Figure~\ref{ModelsResults_correlations} also shows the close relation between f$_{\rm PAH}$ and $<$U$>$. f$_{\rm PAH}$ reaches a constant fraction ($\sim$4\%) 
when $<$U$>$ is lower than $\sim$3. Above this threshold value, we observe a linear decrease of f$_{\rm PAH}$ with log $<$U$>$. The lower panels in 
Fig.~\ref{ModelsResults_correlations} also show the correlations with $\Sigma$$_{\rm dust}$. We do not observe strong variations of column density across the complex. 
Most of the variations of the IR power are rather driven by the variations in the radiation field intensity. Dense regions have a higher $<$U$>$ and a lower f$_{\rm PAH}$, 
as expected from regions with embedded star formation.

\subsection{Dissecting the components of the 870 \mic\ emission}

As explained in $\S$2.3, the data reduction of LABOCA data can lead to a filtering of faint extended emission. If our iterative data reduction procedure helps 
recover a significant amount of emission around the brighter structures, part of the very extended faint emission (well traced by the \hersc/SPIRE instrument 
for instance) is not recovered. Only the flux densities in regions with sufficient S/N (i.e. $>1.5$) can then be trusted. For these reasons, we decide not to 
include the 870 \mic\ data as a direct constraint in the dust modeling procedure. However, the LABOCA 870 \mic\ map traces the coldest phases of dust in 
the complex. By comparing the 870 \mic\ emission predicted by our dust models (fixed emissivity properties) with the observed 870 \mic\ emission, it can also help
us investigate potential variations in the grain emissivity in the submm regime. We decompose the various contributors to the 870 \mic\ emission in this 
dedicated section. Part of the emission at 870 \mic\ in this particularly massive star forming region is linked with non-dust contributions, namely $^{12}$CO(3-2) 
emission line falling in the wide LABOCA passband (at 345GHz) and thermal bremsstrahlung emission produced by free electrons in the ionized gas. 
We will quantify how much of the measured 870 \mic\ surface brightness can reasonably be attributed to thermal dust emission, free-free continuum and 
CO(3-2). This study will allow us to investigate the presence (or not) of any submm emission not explained by these ``standard'' components (the so-called `submm excess').

\begin{figure*}
    \centering
    \begin{tabular}{m{8.5cm}m{9cm}}
       \includegraphics[width=8.2cm]{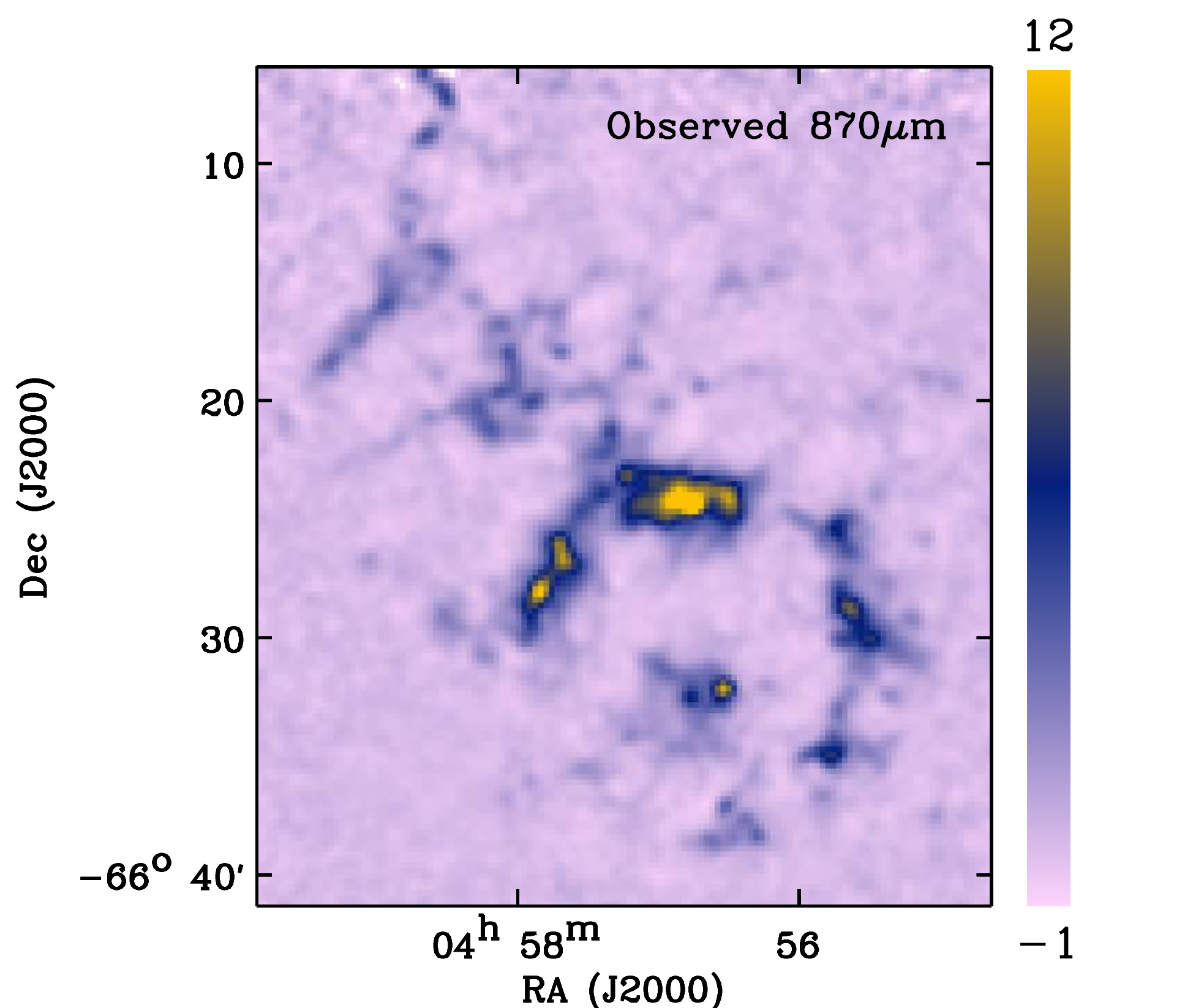} &
       \includegraphics[width=8.2cm]{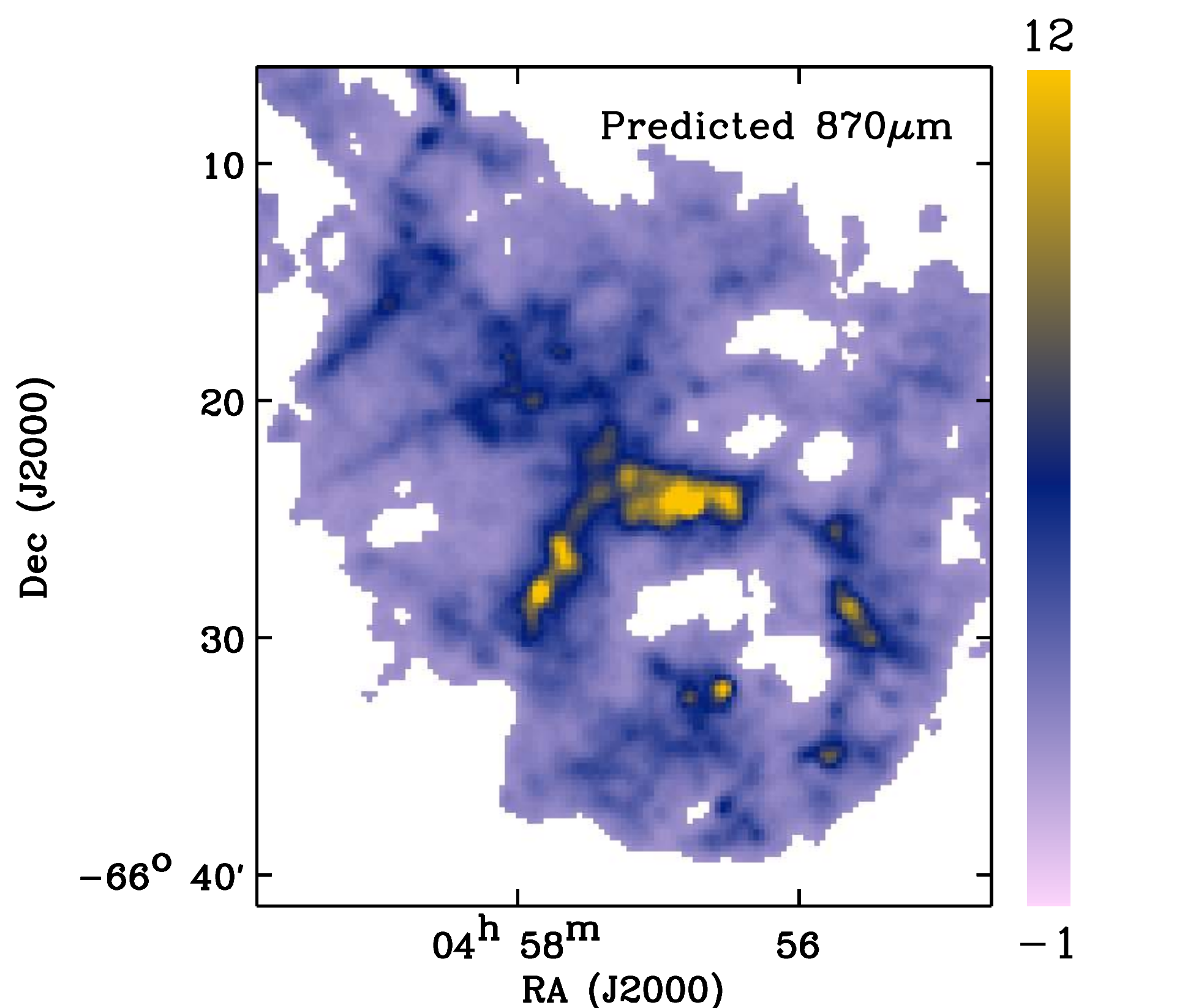} \\
       \vspace{-3pt}
        \includegraphics[width=8.2cm]{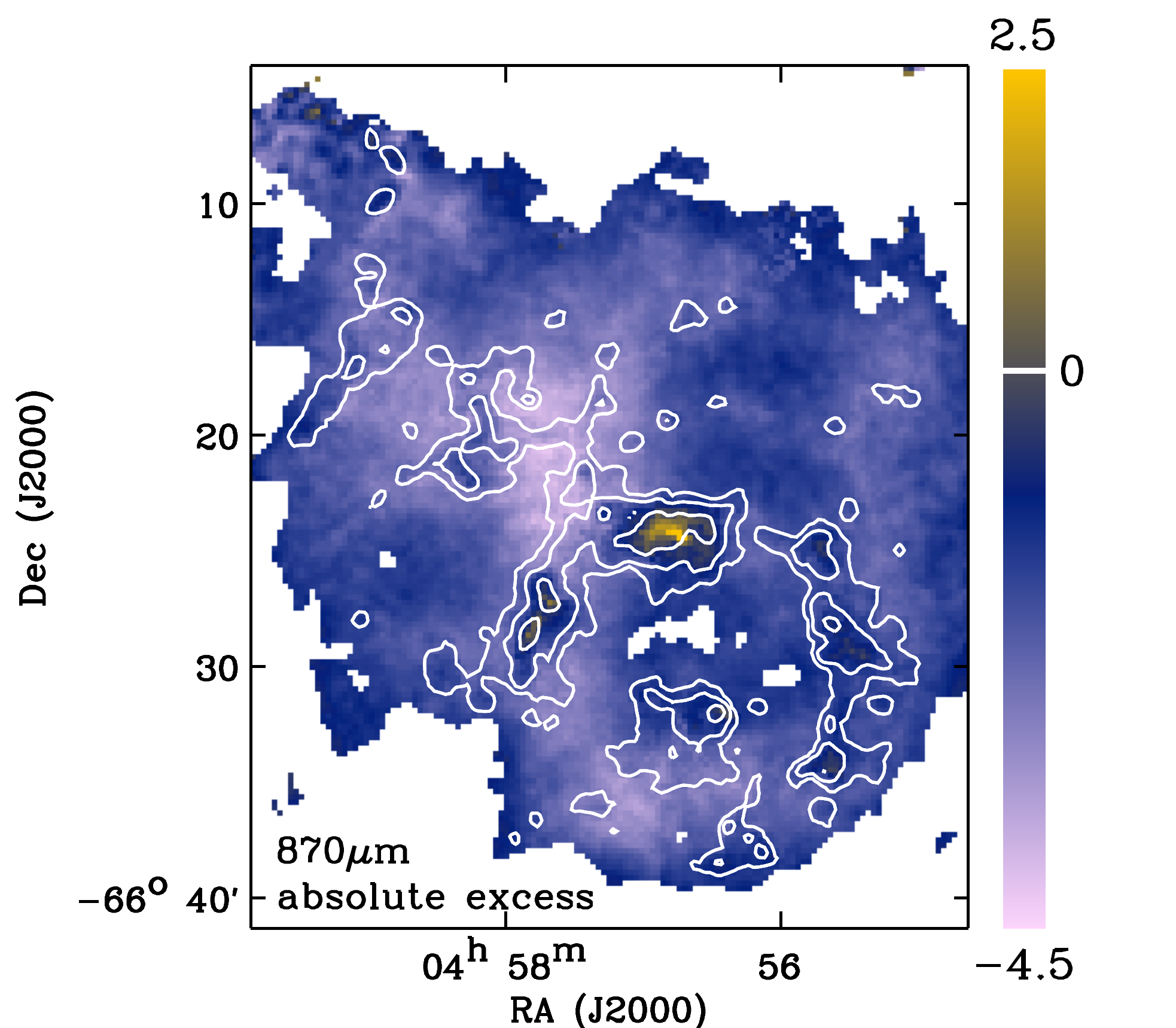} & 
       \vspace{-3pt}
       \includegraphics[width=8.2cm]{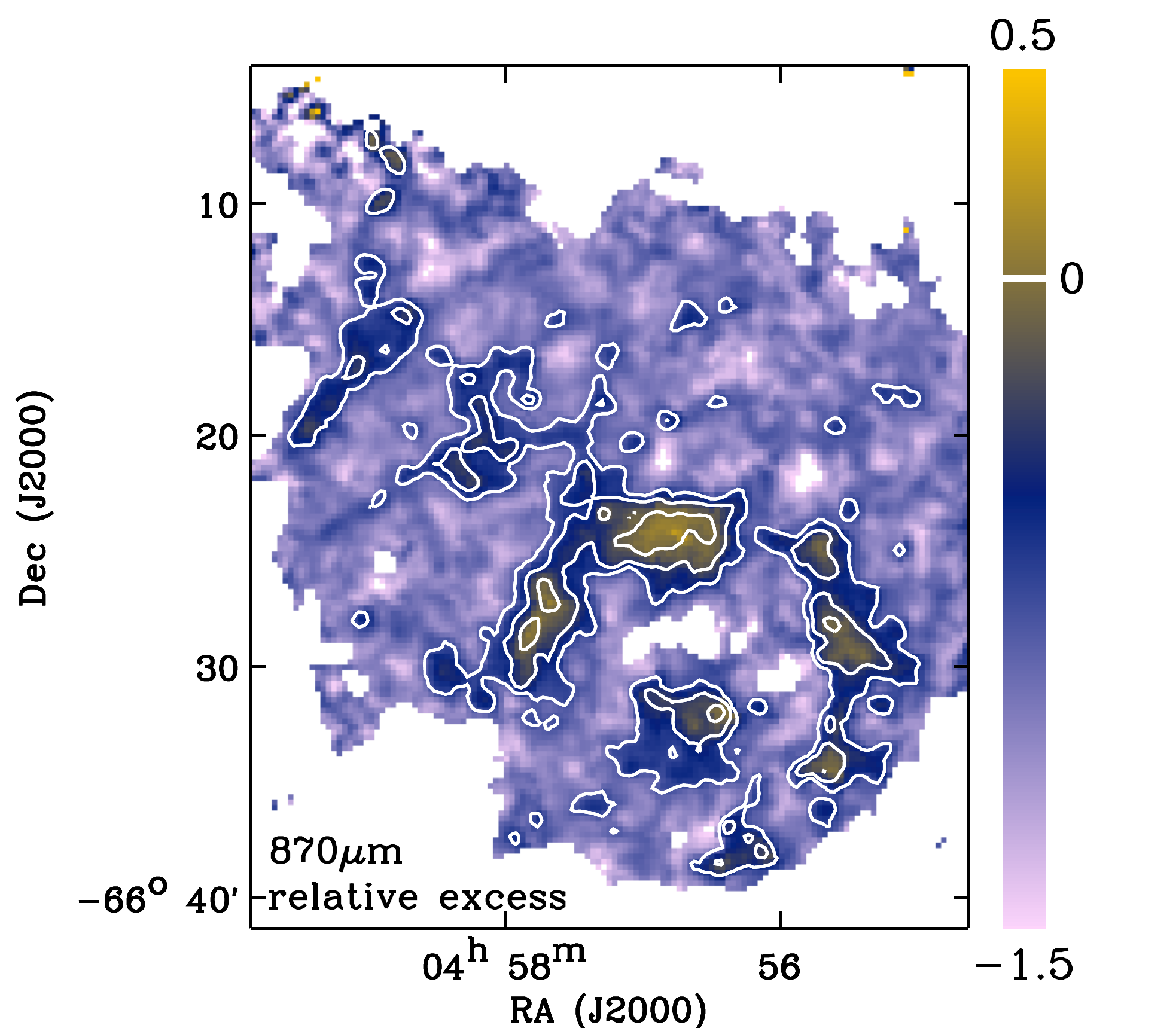} \\              
       \vspace{-3pt}
      \includegraphics[width=8.2cm]{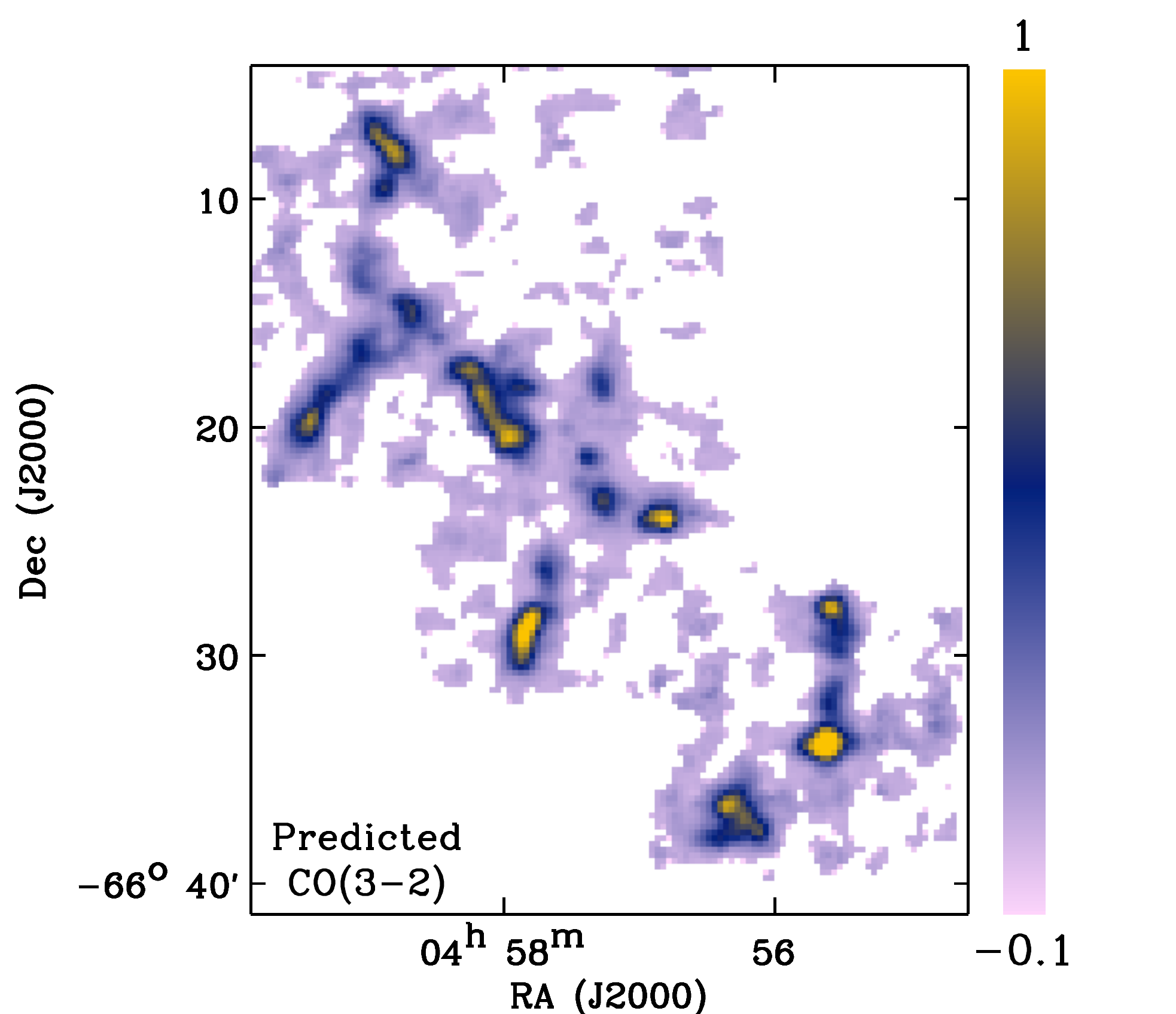}  &
       \vspace{-3pt}
      \includegraphics[width=8.2cm]{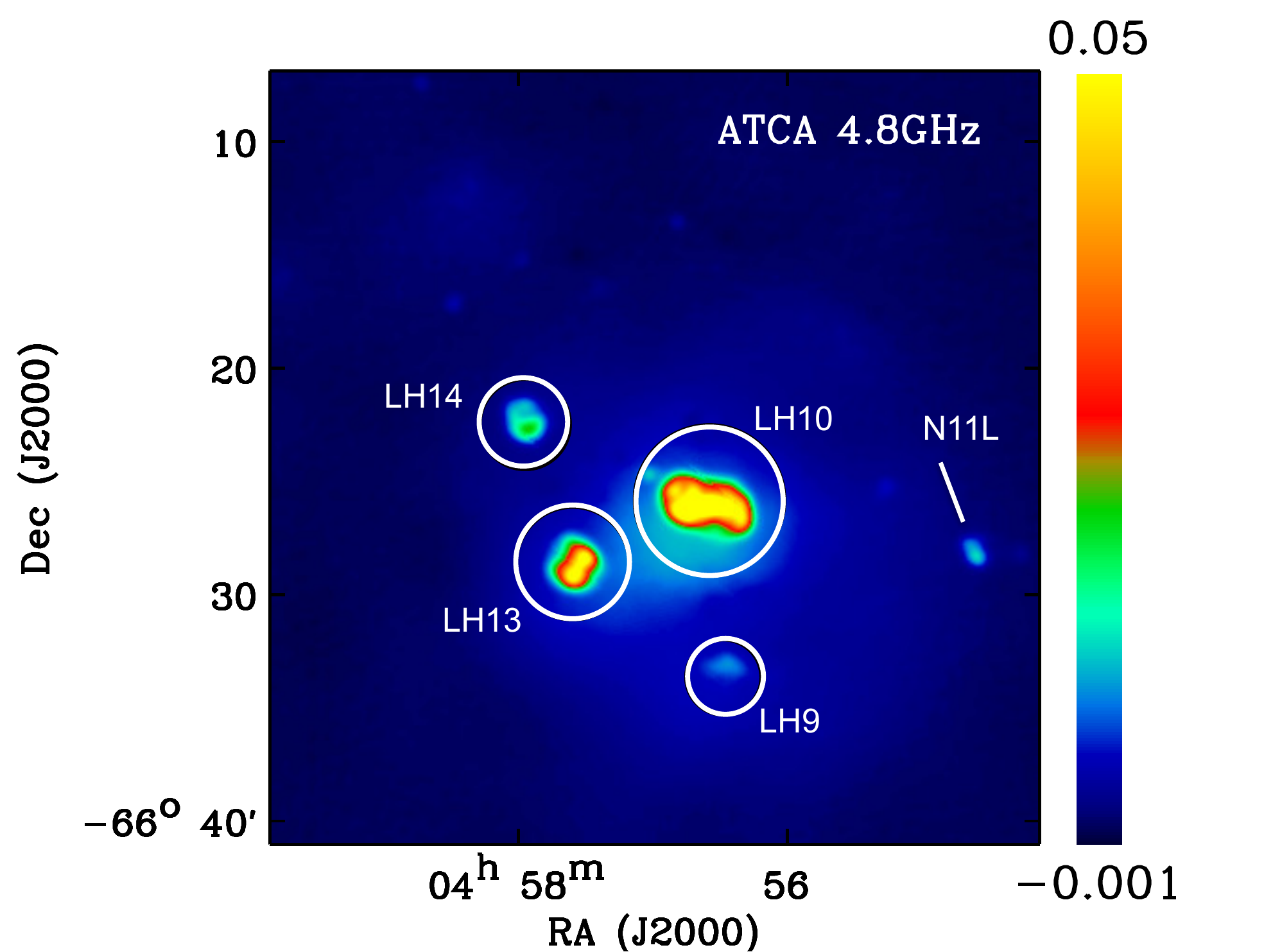}  \\  
          \end{tabular}
    \caption{     
    {\it Top left:} 870 \mic\ flux density in MJy~sr$^{-1}$ observed with LABOCA (resolution of SPIRE 500\mic, i.e. 36\arcsec). 
    {\it Top right:} 870 \mic\ flux density in MJy~sr$^{-1}$ predicted by the AC model, i.e. pure thermal dust emission (same resolution). 
    {\it Middle left:} 870 \mic\ absolute excess in MJy~sr$^{-1}$ defined as (observed - modeled flux). The modeled flux is that predicted 
    by the AC model (so pure thermal dust emission). LABOCA signal-to-noise contours (S/N = 1.5, 5, 15) are overlaid. 
    {\it Middle right:} 870 \mic\ relative excess defined as ((observed - modeled flux) / modeled flux). 
    {\it Bottom left:} CO(3-2) pseudo-continuum flux densities in MJy~sr$^{-1}$ (resolution: 45\arcsec).
    {\it Bottom right:} ATCA radio map at 4.8 GHz \citep[from][]{Dickel2005} in Jy~beam$^{-1}$ (original resolution). 
    The positions of the 4 main OB associations in the complex as well as the SNR N11L are indicated. The circles indicate the positions 
    and sizes of the photometric apertures used to derive the fluxes provided in Table~\ref{Regions_Fluxes}.
     }
    \label{N11_Excess_Betafree}
\end{figure*}

 \begin{table*}
\caption{IR-to-submm photometry of the OB associations in N11 and non-dust contribution to the 870\mic\ fluxes}
\label{Regions_Fluxes}
 \centering
 \begin{tabular}{ >{\centering\arraybackslash}m{2cm}  >{\centering\arraybackslash}m{1.5cm}  >{\centering\arraybackslash}m{2cm}  >{\centering\arraybackslash}m{2cm}  >{\centering\arraybackslash}m{2cm}  >{\centering\arraybackslash}m{2cm} }
 \hline 
\hline
&&&&&\\

&& LH9 / N11F & LH10 / N11B & LH13 / N11C & LH14 / N11E \\

&&&&&\\

\hline
     
&&&&&\\

 Center $^{a}$   & (RA)		& 04$^h$56$^m$35$^s$   & 04$^h$56$^m$48$^s$  & 04$^h$57$^m$46$^s$  & 04$^h$58$^m$12$^s$  \\
~~~~~~~~~~~~ & (DEC)	 	& -66$^{\circ}$32\arcmin04\arcsec\  & -66$^{\circ}$24\arcmin18\arcsec  & -66$^{\circ}$27\arcmin36\arcsec  & -66$^{\circ}$21\arcmin38\arcsec\ \\

 Radius &  (arcsec)		& 100 & 200 & 150 & 120 \\  

&&&&&\\

\hline
     
&&&&&\\

 8 \mic\     & (Jy) 	&  2.7 {\small $\pm$ 0.3} 		&14.9 {\small $\pm$ 1.5} 		& 7.2 {\small $\pm$ 0.7} 		& 3.2 {\small $\pm$ 0.3} 		\\
 24 \mic\   & (Jy) 	&   7.5 {\small $\pm$ 0.7} 		& 84.2 {\small $\pm$ 8.4} 		& 28.0 {\small $\pm$ 2.8} 		& 7.7 {\small $\pm$ 0.8} 		\\
 70 \mic\   & (Jy) 	&  101.4 {\small $\pm$ 10.1} 	& 815.6 {\small $\pm$ 81.6} 	& 312.1 {\small $\pm$ 31.2} 	& 110.2 {\small $\pm$ 11.0} 	\\
 100 \mic\ & (Jy) 	&    153.6 {\small $\pm$ 15.4} 	& 1220.3 {\small $\pm$ 122.0} & 507.0 {\small $\pm$ 50.7} 	& 190.8 {\small $\pm$ 19.1} 	\\
 160 \mic\ & (Jy) 	&  162.8 {\small $\pm$ 6.5} 	& 968.4 {\small $\pm$ 38.7} 	& 456.0 {\small $\pm$ 18.2} 	& 193.8 {\small $\pm$ 7.8} 	\\
 250 \mic\ & (Jy) 	&  76.9 {\small $\pm$ 5.4} 	& 416.7 {\small $\pm$ 29.2} 	& 212.6 {\small $\pm$ 14.9} 	& 102.4 {\small $\pm$ 7.2}	\\
 350 \mic\ & (Jy) 	&   38.1 {\small $\pm$ 3.8} 	& 191.4 {\small $\pm$ 19.1} 	& 102.5 {\small $\pm$ 10.3} 	& 50.5 {\small $\pm$ 5.1} 		\\
 500 \mic\ & (Jy) 	&  16.1 {\small $\pm$ 1.6} 	& 77.1 {\small $\pm$ 7.7} 		& 42.3 {\small $\pm$ 4.2} 		& 21.3 {\small $\pm$ 2.1} 		\\ 
 870 \mic\ $^{b}$ & (Jy) 	&   4.2 {\small $\pm$ 0.3} 		& 12.2 {\small $\pm$ 0.9} 		& 7.1 {\small $\pm$ 0.5} 		& 3.1 {\small $\pm$ 0.2} 		\\
              
&&&&&\\

\hline

&&&&&\\

f$_{\rm 870, free-free}$     & (\%) & 7.4 & 10.4 & 9.1 & 6.3 \\
f$_{\rm 870, CO}$		 & (\%) & $<$6 & $<$4 & $<$20 &  - \\
     
&&&&&\\

 \hline
\end{tabular}
\vspace{5pt}
\begin{list}{}{}
\item[$^a$] The photometric apertures are shown in Fig.~\ref{N11_Excess_Betafree} (bottom right panel).
\item[$^b$] Observed flux density not corrected for the non-dust contribution.
\end{list}
 \end{table*} 

\subsubsection{Thermal dust emission at 870 \mic\ and excess maps} 

From our local SED modeling (the physically motivated ``AC model"), we derive a prediction of the pure thermal contribution to the 870 \mic\ observations. The 
observed and predicted maps at 870 \mic\ are shown in Fig.~\ref{N11_Excess_Betafree} (top panels). Their resolution is our working resolution of 36\arcsec. 
To compare the two maps, we compute the absolute differences between observations and model predictions
defined as (observed flux at 870 \mic\ - modeled flux at 870 \mic) and the relative differences between observations and model predictions defined as (observed 
flux at 870 \mic\ - modeled flux at 870 \mic) / (modeled flux at 870 \mic). The maps (that we call the absolute and relative excess maps) are shown in Fig.~\ref{N11_Excess_Betafree} 
(middle panels). We overlay the contours of LABOCA S/N to highlight regions where the LABOCA emission is higher than a 1.5-$\sigma$ threshold. The image shows 
that most of the structures below our S/N threshold correspond to regions where the SED model over-predicts the observed 870 \mic\ (negative difference). As 
previously suggested, part of the diffuse emission might be filtered out during the data reduction in these faint regions. In regions above our S/N criterion, the 
distribution of the relative excess seems to follow the structure of the complex. The observed emission at 870 \mic\ is close to the predicted value (weak emission 
in excess) on average and the relative excess reaches $\sim$20$\%$ at most in the center of N11B. We will now try to quantify the non-dust contribution to the 
870 \mic\ flux that could partly account for this excess.

\subsubsection{CO(3-2) line contribution} 

A $^{12}$CO(1-0) mapping of LMC giant molecular clouds (GMCs) has been performed using the Australia Telescope National Facility Mopra Telescope as part of the 
Magellanic Mopra Assessment (MAGMA\footnote{Data can be retrieved at http://mmwave.astro.illinois.edu/magma/}; 45\arcsec\ resolution) project \citep{Wong2011}. 
We are using these observations to estimate the $^{12}$CO(3-2) line contribution to the 870 \mic\ flux. To do the conversion, we need to assume a brightness 
temperature ratio R$_{\rm 3-2,1-0}$. In LMC GMCs, this ratio ranges between 0.3 and 1.4 (average of the clumps: 0.7), with high values ($>$1.0) being associated 
with strong H$\alpha$ fluxes \citep{Minamidani2008}. They find an average brightness temperature ratio of about 0.9 in the N159 star forming complex. The dust 
temperature in N11 being lower than that in N159 as shown in $\S$4.2, the R$_{\rm 3-2,1-0}$ is probably lower than this value. We used the average R$_{\rm 3-2,1-0}$ 
ratio of 0.7 derived in \citet{Minamidani2008} to convert the CO(1-0) map into a CO(3-2) map. We use the formula from \citet{Drabek2012} to convert our CO line 
intensities (K km s$^{-1}$) to pseudo-continuum fluxes (mJy~beam$^{-1}$):

\begin{equation}
\frac{C}{(mJy~beam^{-1})(K~km~s^{-1})^{-1}} = \frac{2k\nu^3}{c^3}\frac{g_{\rm \nu}(line)}{\int g_{\rm \nu}~d\nu} \Omega_{\rm B}
\end{equation}

\noindent where k is the Boltzmann constant, $\nu$ is the frequency, $\Omega$$_{\rm B}$ is the telescope beam area, g$_{\rm \nu}$(line) is the transmission at the 
frequency of the CO(3-2) line and $\int$ g$_{\rm \nu}$ d$\nu$ is the transmission integrated across the full frequency range. g$_{\rm \nu}$(345GHz)/$\int$g$_{\rm \nu}$ d$\nu$ 
is $\sim$0.017 for LABOCA (private communication; G. Siringo, ESO/MPIfR; 2007). The derived CO(3-2) map is shown in Fig.~\ref{N11_Excess_Betafree} (bottom left). 
We convolve the 870 \mic\ map to the resolution of the CO(3-2) map (Gaussian kernel) to compare the two maps. We estimate a contribution of $\sim$15-20$\%$ in 
LH13, $<$6$\%$ in LH10 and LH9 and from 4 to 12$\%$ in the west ring and in the northern elongated structure detected 
by LABOCA. We note that LH14 is not fully covered in the public MAGMA map we are using. These values are reported in Table~\ref{Regions_Fluxes}.

\subsubsection{Free-free contribution} 

Figure~\ref{N11_Ha_I4_S250} shows the H$\alpha$ distribution (whose distribution should be co-spatial with that of the free-free component across the complex) 
while Fig.~\ref{N11_Excess_Betafree} (bottom right) presents a mosaicked image of the 4.8 GHz radio continuum emission taken with the Australia Telescope 
Compact Array (ATCA). The radio (ATCA) maps at 4.8 and 8.6 GHz (resolution of 33\arcsec\ and 20\arcsec\ respectively) obtained from \citet{Dickel2005} are 
used to model the free-free emission where radio emission is detected. 
We first convolve the two maps to the working resolution of 36\arcsec\ using a Gaussian kernel. We then estimate the 870 \mic, 6.25 and 3.5 cm flux densities in 
the 4 OB associations of the complex (LH9, LH10, LH13 and LH14) shown in Fig.~\ref{N11_Excess_Betafree} (middle). The circles indicate the positions and sizes 
of the photometric apertures. 
The 8 to 870 \mic\ flux densities in the 4 individual regions are provided in Table~\ref{Regions_Fluxes}. We use the two ATCA constraints to extrapolate the free-free 
emission in the 870 \mic\ band, assuming that the free-free flux density is proportional to $\nu$$^{-0.1}$. We estimate free-free contributions of 10.4, 9.1, 7.4 and 
6.3$\%$ in LH10, LH13, LH9 and LH14 respectively. These values are reported in Table~\ref{Regions_Fluxes}.

\vspace{5pt}
\noindent {\it Potential synchrotron contamination -} In this analysis, we assume that the radio emission across the complex is dominated by free-free emission. 
However, radio continuum observations can trace both thermal emission from H\,{\sc ii} regions and synchrotron emission. Polarized synchrotron emission can, 
for instance, be produced in supernova remnants (SNRs). This is the case in particular for the SNR N11L detected in the ATCA observations and indicated in 
Fig.~\ref{N11_Excess_Betafree}. This SNR is however located outside the regions where the 870 \mic\ excess peaks. Synchrotron radiation can also be an 
artificial source of X-rays in the ISM. Using X-ray observations of the N11 superbubble from the Suzaku observatory, \citet{Maddox2009} detected non-thermal 
X-ray emission around the OB association LH9. However, the photon index of the required non-thermal power-law component is too hard to be explained by 
a synchrotron origin. Our hypothesis of negligible contamination of the 870 \mic\ emission in N11 by synchrotron emission thus seems reasonable.

\subsubsection{Conclusion}

The excess above the pure thermal dust emission we observe in $\S$4.5.1 is weak and can be, within uncertainties, accounted for by the various non-dust 
contributions (CO line emission and free-free emission) we estimated. We conclude that the 870 \mic\ can fully be reproduced by these 3 standard components.

\section{Comparison between the dust and the gas tracers}

In this section, we use the dust surface density map we generate to study the relation between the dust reservoir and 
the gas tracers. We use H\,{\sc i} and CO observations to derive an atomic and a molecular surface density map of N11. 
By studying the variations of the `observed' GDR across the complex, we will be able to explore the influence of each of the 
assumptions we made.

\subsection{The gas reservoirs in N11}

\subsubsection{Atomic gas}

\begin{figure*}
    \centering
    \begin{tabular}{m{8cm}m{8cm}}
    \vspace{-10pt}
      \includegraphics[height=8cm]{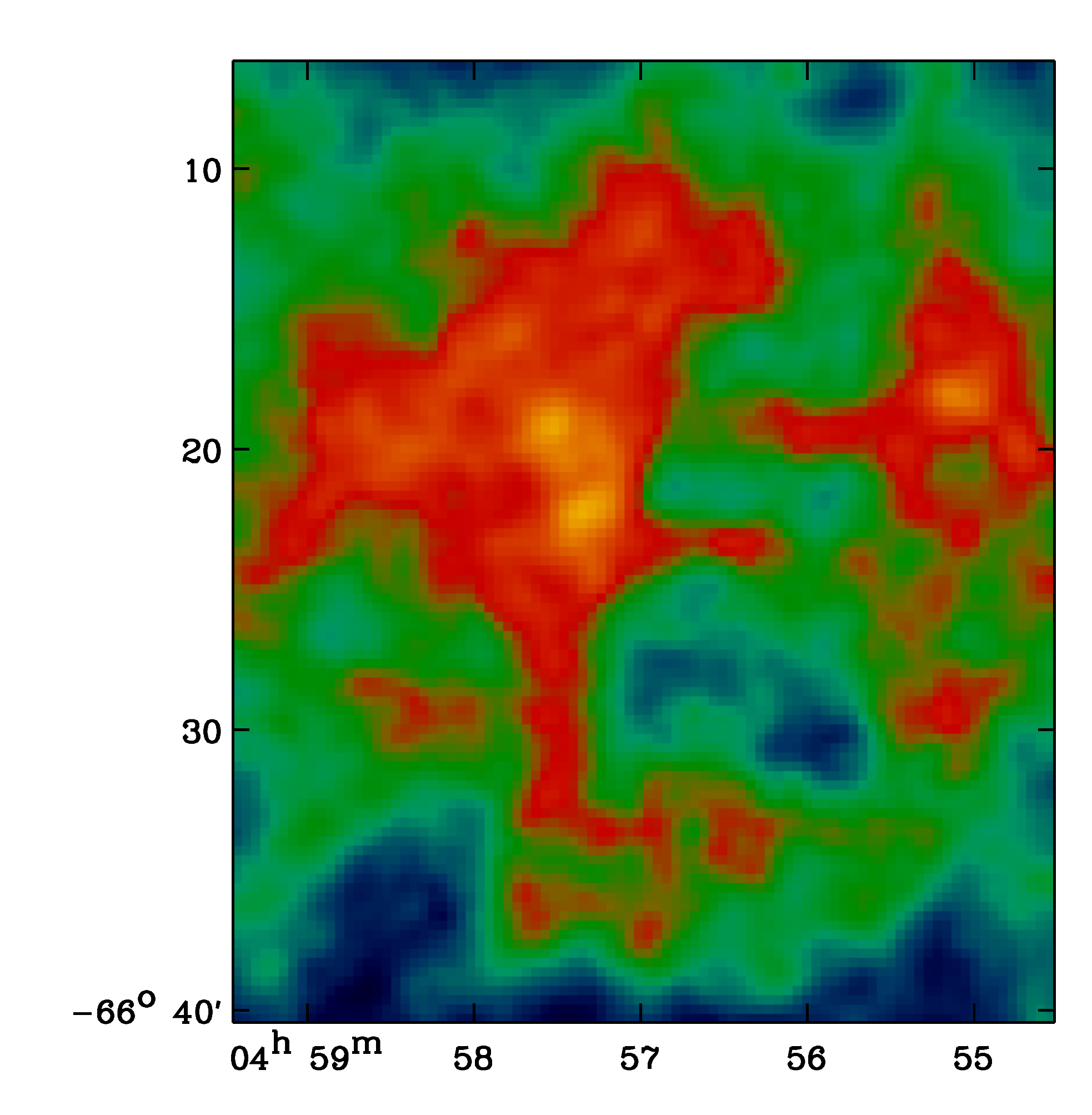} & \includegraphics[height=8cm]{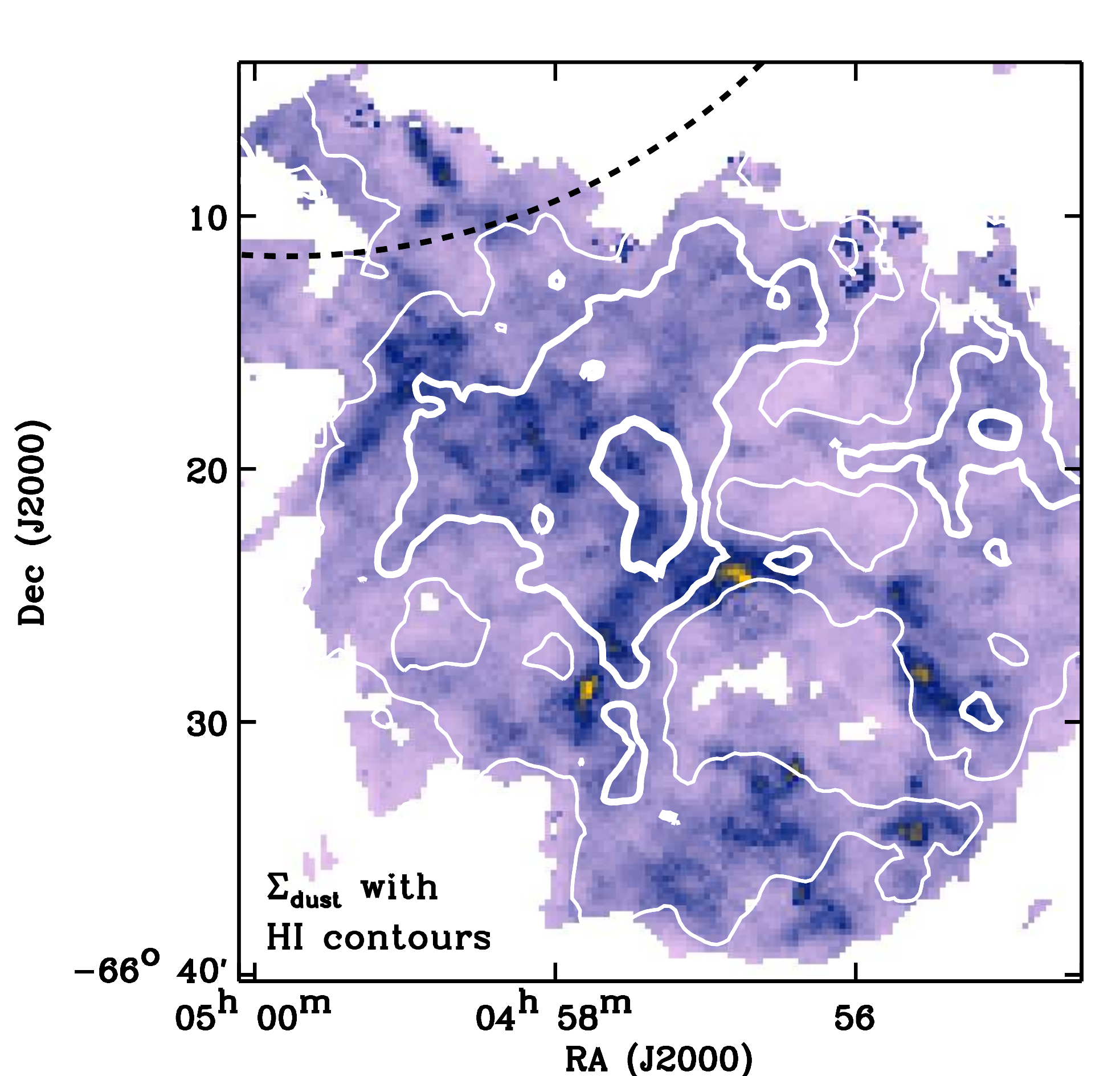}  \\
      \includegraphics[height=8cm]{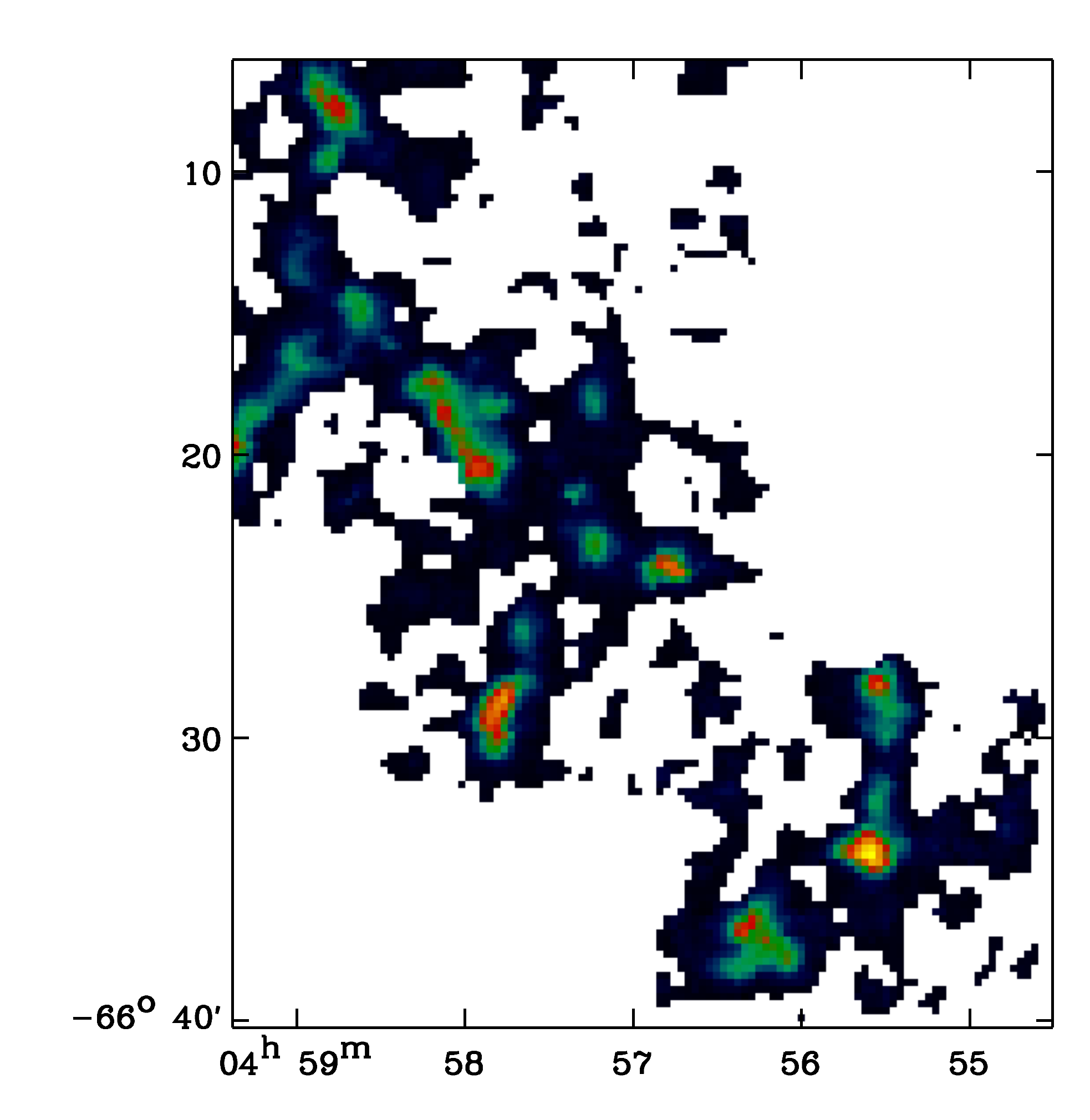} & \includegraphics[height=8cm]{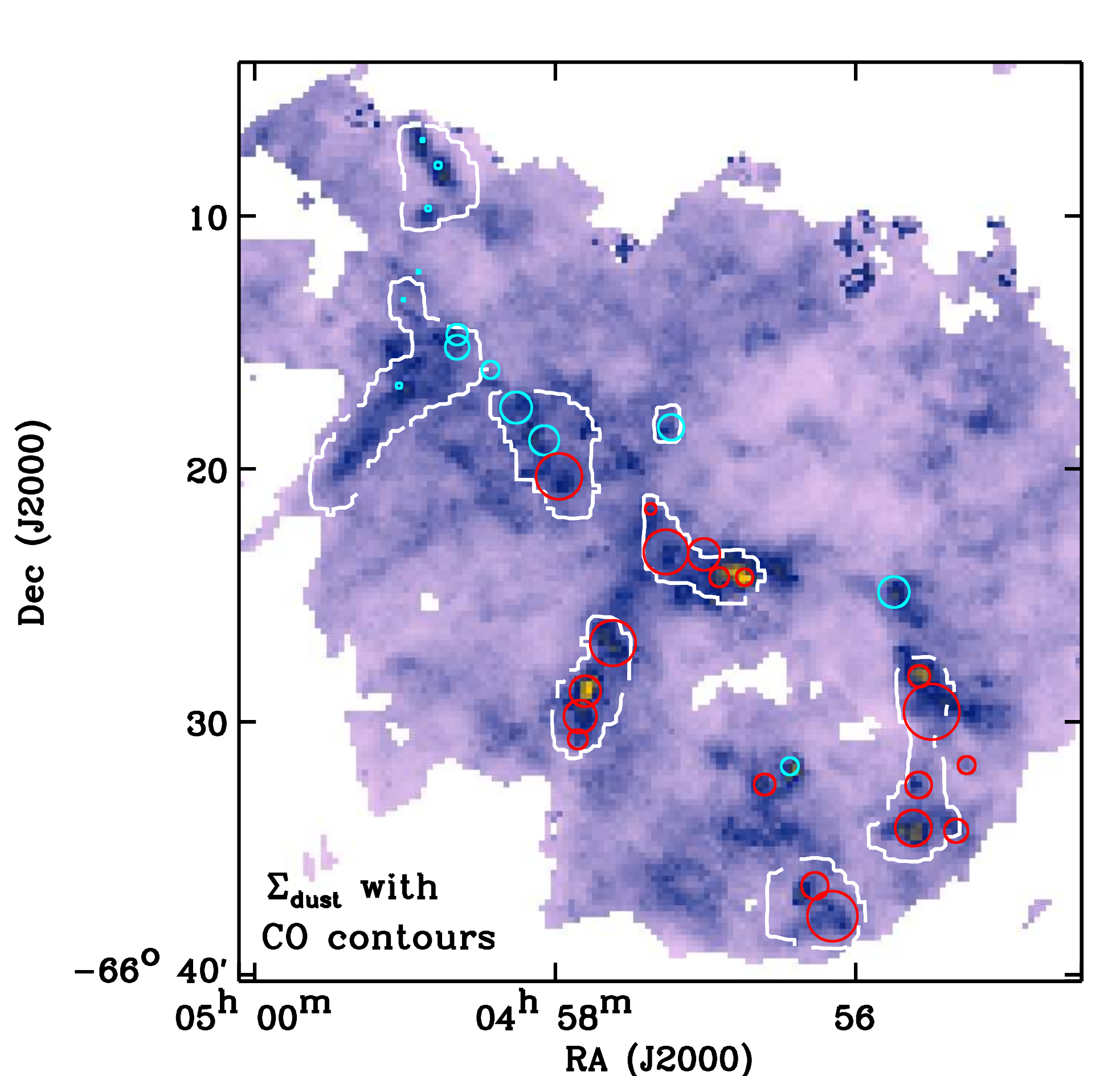}\\
          \end{tabular}
    \caption{Relation between dust and gas in N11. 
    {\it Top left:} H\,{\sc i} map of N11.
    {\it Top right:} $\Sigma$$_{\rm dust}$ map of the complex (units of \msun~pc$^{-2}$) with H\,{\sc i} contours overlaid. The levels are 3$\times$10$^{21}$, 
    4$\times$10$^{21}$ and 5$\times$10$^{21}$ atoms~cm$^{-2}$ (from thinner to thicker lines). The dashed circle indicates the location of the supergiant shell 
    located north to the N11 star forming region.
    {\it Bottom left:} MAGMA CO map of N11.
    {\it Bottom right:} $\Sigma$$_{\rm dust}$ map of N11 with a MAGMA CO contour at 1.4 K~km~s$^{-1}$ (the survey has a 3-$\sigma$ sensitivity limit of 1.2 K~km~s$^{-1}$). 
    Molecular clouds identified by \citet{Herrera2013} are shown in red. Additional molecular clouds previously identified by \citet{Israel2003_1} are overlaid in cyan. 
    Both studies have identified molecular clouds based on SEST observations of N11. The radii of the circles indicate the geometrical radii of the clouds fitted to the SEST observations at these positions.}
    \label{HI_CO_Dust}
\end{figure*}

\citet[][]{Kim2003} produced a high-resolution H\,{\sc i} data cube of the LMC by combining their ATCA data with observations of the Parkes 64-m radio 
telescope from \citet{Staveley2003}. The final cube has a velocity resolution of 1.649 km s$^{-1}$ and a spatial resolution of 1\arcmin. The H\,{\sc i} 
intensity map of the N11 complex was obtained by integrating the data cube over the 190 $<$ v$_{\rm hel}$ $<$ 386 km s$^{-1}$ velocity range. This 
excludes the galactic contamination caused by the low galactic latitude of the LMC (b $\sim$ -34$^{\circ}$). We transform this integrated H\,{\sc i} 
intensity map to atoms cm$^{-2}$ units assuming that the 21cm line is optically thin across the complex. We discuss potential consequences of this 
assumption in $\S$5.3.3. We finally fit the distribution of pixels outside the LMC by a Gaussian and subtract the peak value (as a background estimate) 
from the data. The H\,{\sc i} contours (in atoms cm$^{-2}$) are overlaid on the dust surface density $\Sigma$$_{\rm dust}$ map in Fig.~\ref{HI_CO_Dust}. 
N11 is located at the south of a supergiant shell ($\sim$30\arcmin\ in radius; see Fig.~\ref{HI_CO_Dust}) that is expanding at a velocity of 15 km~s$^{-1}$ 
\citep[see][]{Kim2003}. A significant reservoir of atomic gas is detected in the complex, in particular in the north-eastern structure delimiting the supergiant 
shell rim. Compared to the N158-N159-N160 region where the major peaks of the FIR emission systematically reside in H\,{\sc i} holes \citep{Galametz2013a}, 
the H\,{\sc i} distribution broadly follows the FIR (and dust) distribution in N11. However, the peaks of the H\,{\sc i} distribution are not co-spatial with the 
peaks in the dust mass distribution. The ionized cavity around LH9 is H\,{\sc i} free.

\subsubsection{Molecular gas} 

The CO(1-0) line emission is widely used as an indirect tracer of the H$_{\rm 2}$ abundance. The 1.4 K~km~s$^{-1}$ contour of the MAGMA CO(1-0) 
map presented previously in the paper is overlaid on the $\Sigma$$_{dust}$ map in  Fig.~\ref{HI_CO_Dust} (bottom). \citet{Israel2003_1} also 
mapped some of the N11 clouds in $^{12}$CO(1-0) and 
$^{12}$CO(2-1) emission lines as part of the ESO-SEST key program (FWHM=45\arcsec\ and 23\arcsec\ respectively). \citet{Herrera2013} present a follow-up 
study using fully sampled maps. Both studies provide catalogues of the physical properties of individual molecular clouds (overlaid in Fig.~\ref{HI_CO_Dust}) 
across the N11 complex. We observe that N11 is composed of many individual molecular clumps that account for a significant part of the 
CO emission detected in the complex. Most of the peaks in CO correspond to peaks in the dust mass map. The main H\,{\sc i} peak resides in-between two 
CO complexes. This H\,{\sc i} / H$_{\rm 2}$ interface is often observed in the LMC. If molecular clouds are thought to primarily form from global gravitational 
instabilities, \citet{Dawson2013} have shown that up to 25\% of the molecular mass residing in LMC supergiant shells could be a direct consequence of stellar 
feedback such as accumulation, shock compression mechanisms or ionizing radiation. The H\,{\sc i} / H$_{\rm 2}$ interface we observe in N11 could partly 
result from the same process at smaller scales (the N11 ring is 4 times smaller in size than the supergiant shell located 0.8$^{\circ}$ north of N11). \\

{\it Choice of the X$_{\rm CO}$ factor~~-~~}
In order to build a map of the H$_{\rm 2}$ column density, we need to convert the intensities of the MAGMA CO(1-0) map into masses. The `standard' 
X$_{\rm CO}$ factor in the Solar neighborhood is $\sim$2 $\times$ 10$^{20}$~cm$^{-2}$~(K~km~s$^{-1}$)$^{-1}$ \citep{Scoville1987,Solomon1987} 
but has a strong dependence on the ISM physical conditions, especially with metallicity \citep{Bolatto2013}. 
Because of the lower dust content in low-metallicity objects, the UV photons penetrate deeper into the molecular clouds, leading to a drop in the optical depth 
and a photo-dissociation of the CO molecule (so less CO emission to trace the same H$_{\rm 2}$ mass). The X$_{\rm CO}$ factor is thus usually higher in 
low-metallicity environments. Indirectly using dust measurements to trace the gas reservoirs, \citet{Leroy2011} derived a X$_{\rm CO}$ factor of 
3 $\times$ 10$^{20}$~cm$^{-2}$~(K~km~s$^{-1}$)$^{-1}$ for the LMC. X$_{\rm CO}$ factors were also estimated from the NANTEN survey of nearly 300 GMCs 
over the whole LMC \citep{Fukui1999, Mizuno2001}. They find an average value of $\sim$9 $\times$ 10$^{20}$~cm$^{-2}$~(K~km~s$^{-1}$)$^{-1}$. 
The estimate was then refined to be 7 $\times$ 10$^{20}$~cm$^{-2}$~(K~km~s$^{-1}$)$^{-1}$ by improving the rms noise level by a factor of 
two \citet{Fukui2008}. By targeting more specifically the CO clumps of N11, X$_{\rm CO}$ was estimated to be $\sim$5 $\times$ 
10$^{20}$~cm$^{-2}$~(K~km~s$^{-1}$)$^{-1}$ in \citet{Israel2003_1} and $\sim$8.8 $\times$ 10$^{20}$ $\alpha$$_{\rm vir}$$^{-1}$ 
in \citet{Herrera2013}, with $\alpha$$_{\rm vir}$ the virial parameter corresponding to the ratio of total kinetic energy to gravitational energy. In 
their analysis, \citet{Herrera2013} suggest to use $\alpha$$_{\rm vir}$$\sim$2 (this leads to a X$_{\rm CO}$ factor of 4.4 $\times$ 
10$^{20}$~cm$^{-2}$~(K~km~s$^{-1}$)$^{-1}$). In our analysis, we decide to use the statistically robust average from the whole MAGMA 
GMC sample obtained by \citet{Hughes2010}, i.e. 4.7 $\times$ 10$^{20}$~cm$^{-2}$~(K~km~s$^{-1}$)$^{-1}$. This assumes that $\alpha$$_{\rm vir}$=1. 
We will call this value the MAGMA X$_{\rm CO}$ factor. We note that in the case of an $\alpha$$_{\rm vir}$ equal to 2, the X$_{\rm CO}$ value 
will be close to the Galactic X$_{\rm CO}$ (4.7 / 2 = 2.35). As a comparison, we will thus also present the results obtained when a standard 
Galactic X$_{\rm CO}$ factor is used. We discuss the consequences of these choices further in $\S$5.3.1.

\begin{figure*}
    \centering
    \begin{tabular}{m{10cm}m{9cm}}
\includegraphics[height=7.5cm]{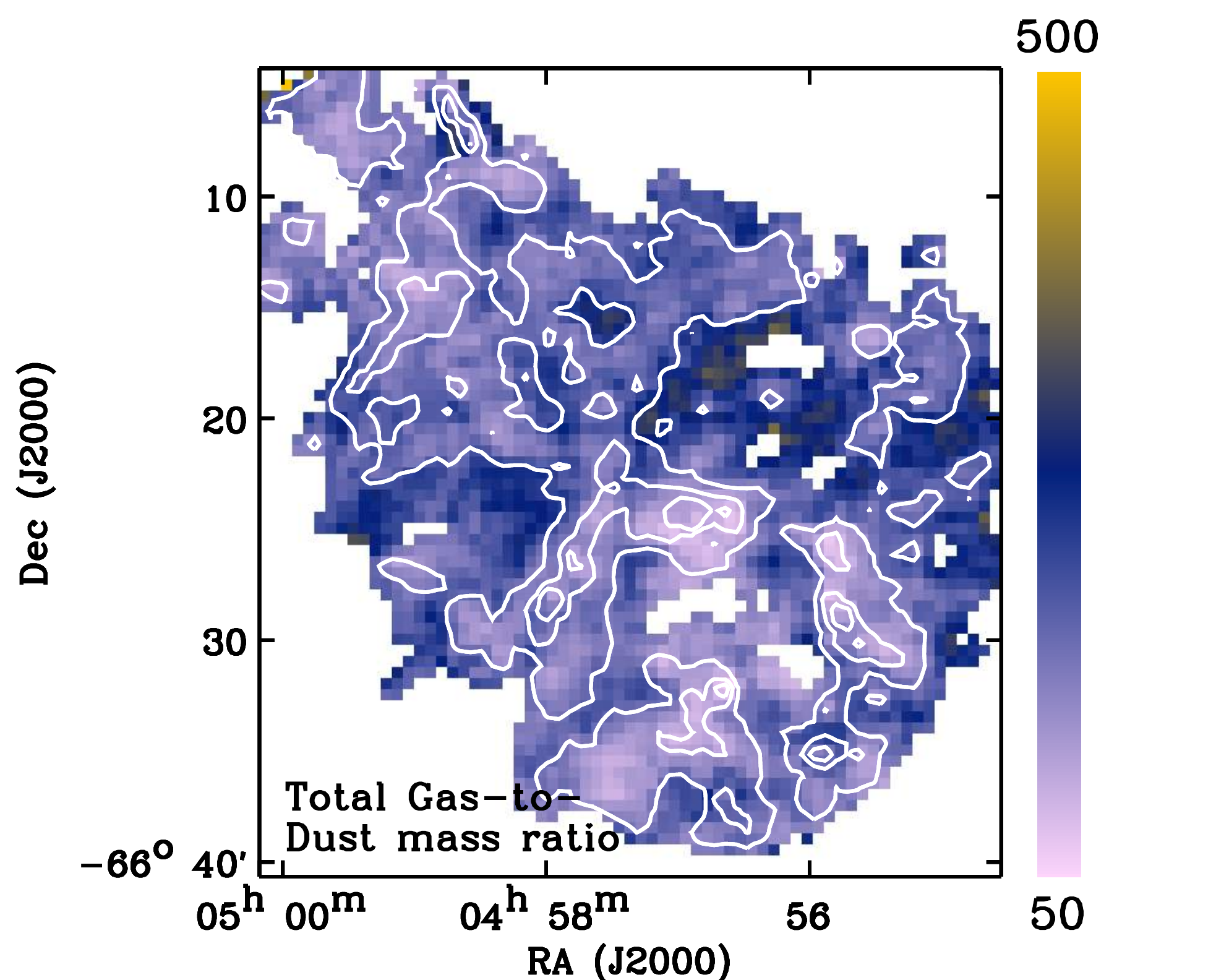} &
\hspace{-40pt}    
\includegraphics[height=6cm]{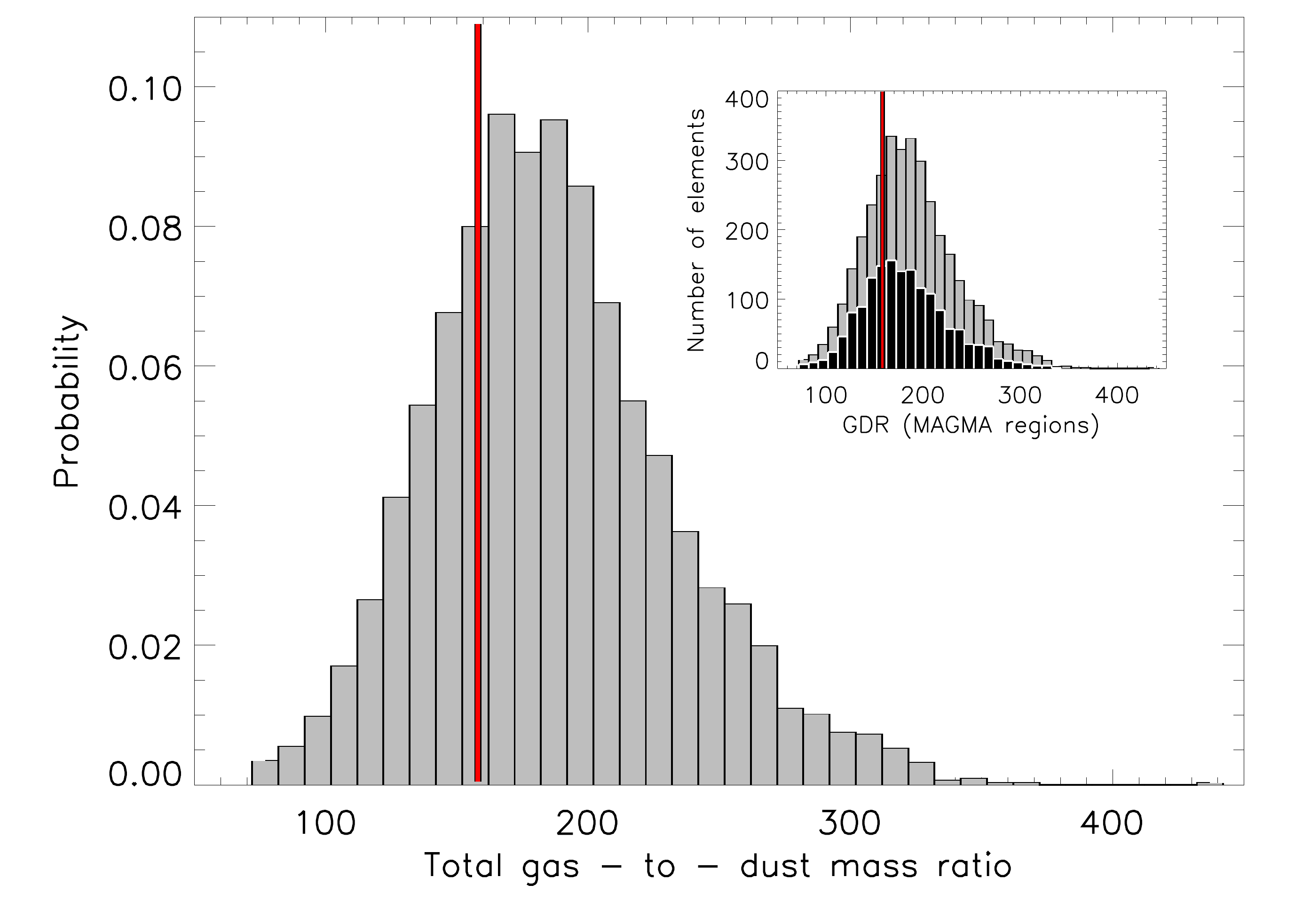} \\
          \end{tabular}
    \caption{
    Total gas-to-dust mass ratio map derived at a 1\arcmin\ resolution. $\Sigma$$_{\rm dust}$ contours at 0.2, 0.35 and 0.5 \msun~pc$^{-2}$ are overlaid (same resolution). 
    The total gas is equal to 1.36 M$_{\rm HI}$ + M$_{\rm H2}$. The molecular mass is estimated from CO(1-0) observations 
    and using an X$_{\rm CO}$ factor of 4.7 $\times$ 10$^{20}$~cm$^{-2}$~(K~km~s$^{-1}$)$^{-1}$ \citep{Hughes2010}. 
    The corresponding probability distribution of the GDR is shown on the right panel. The vertical red line indicates the Galactic total GDR value \citep[i.e. 158;][]{Zubko2004}. 
    The inset shows the histogram of ISM elements not normalized. Regions for which we have CO coverage (thus molecular information) are overlaid in darkgrey.}
    \label{GD}
\end{figure*}

\begin{figure*}
\centering
\hspace{-20pt}
\begin{tabular}{cc}
 \includegraphics[width=10cm]{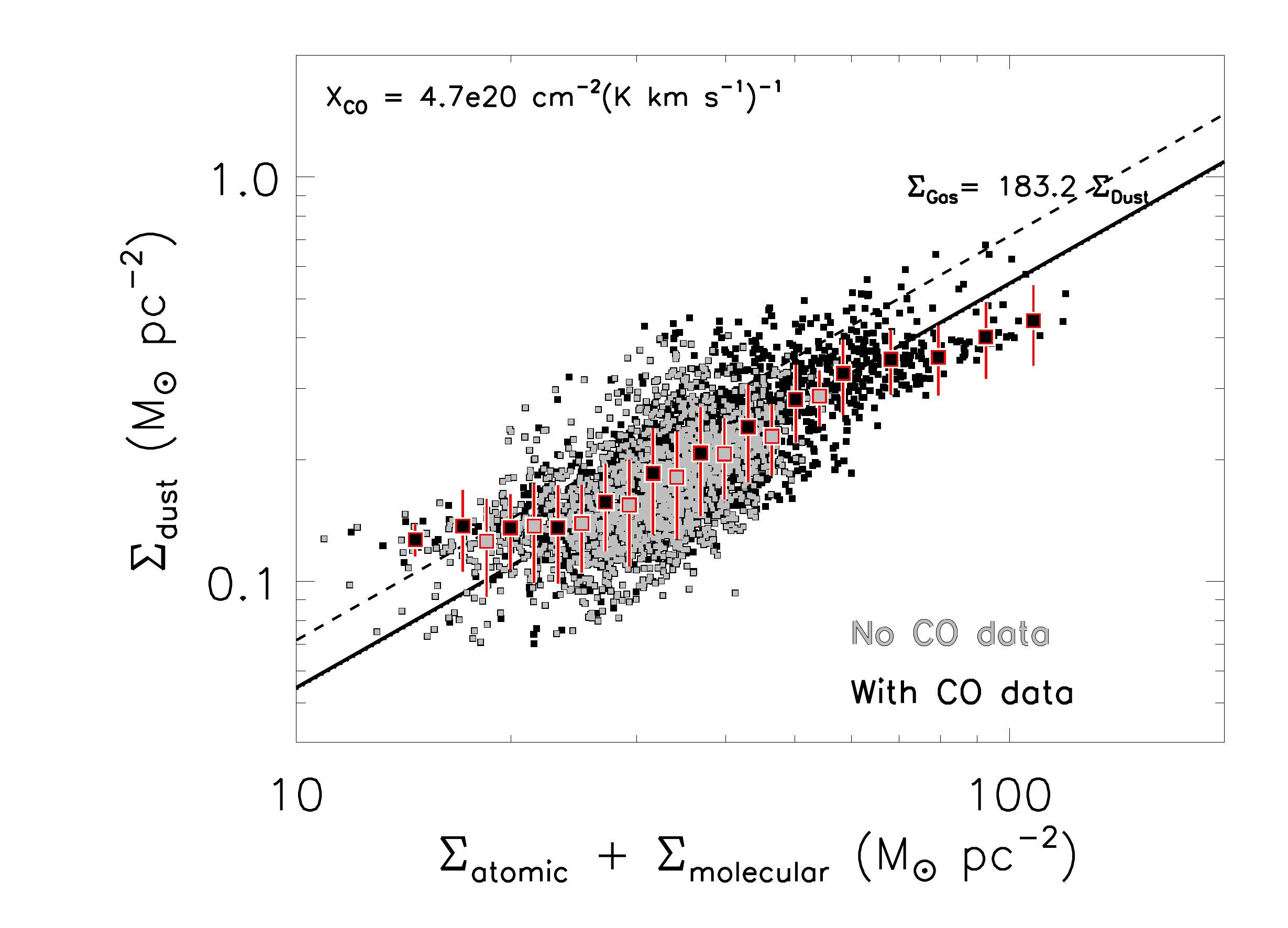} &
\hspace{-80pt}  \includegraphics[width=10cm]{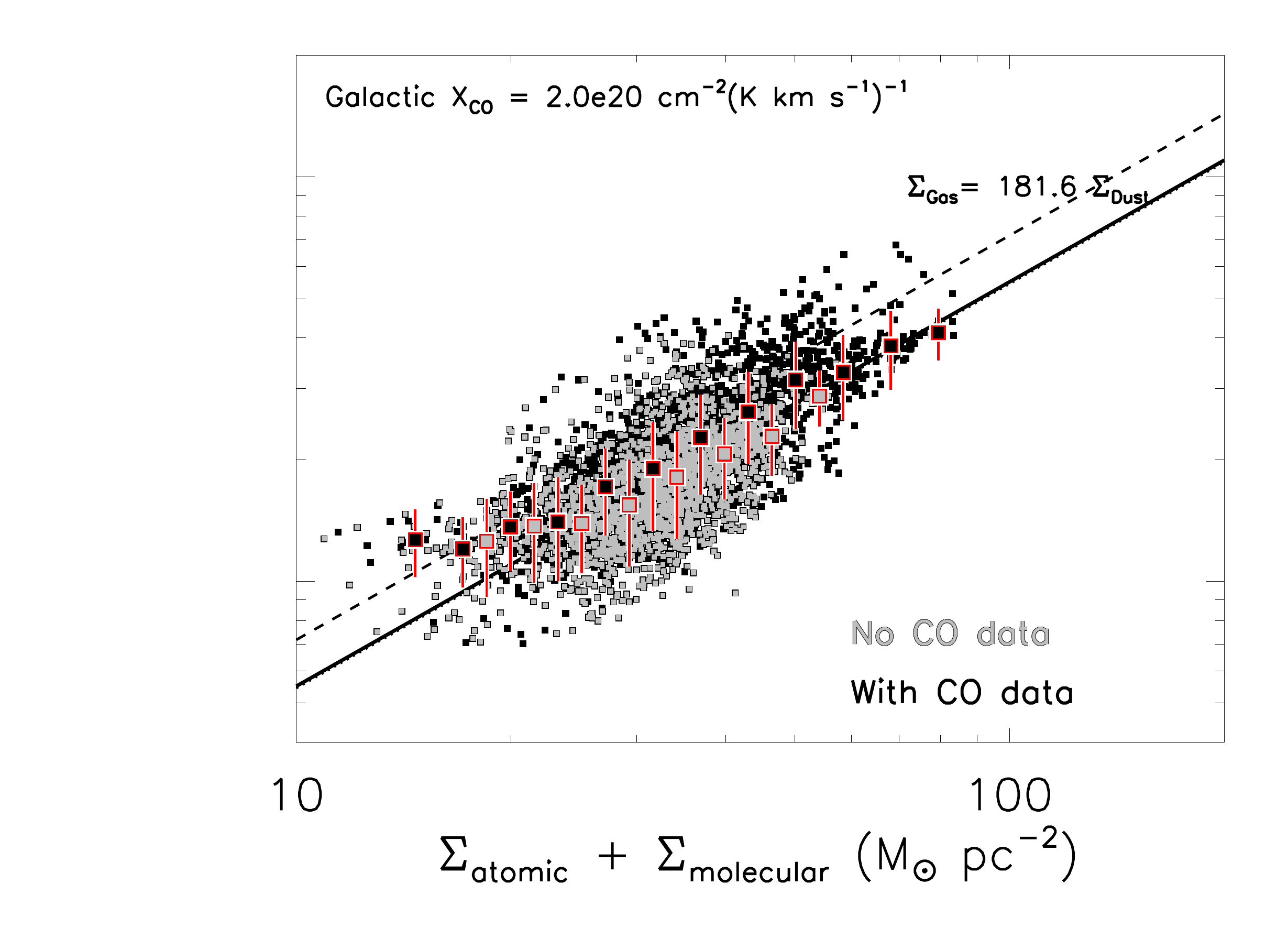} \\
 \end{tabular} 
 \caption{ 
 Dust surface density as a function of the total gas surface density. The molecular mass is estimated using an X$_{\rm CO}$ 
factor of 4.7 $\times$ 10$^{20}$~cm$^{-2}$~(K~km~s$^{-1}$)$^{-1}$ on the left panel and a Galactic X$_{\rm CO}$ factor 
of 2.0 $\times$ 10$^{20}$~cm$^{-2}$~(K~km~s$^{-1}$)$^{-1}$ on the right panel. 
On both plots, black points are ISM elements where MAGMA CO data is available and grey points where MAGMA CO data 
is not available. The larger squares overlaid over the individual ISM elements indicate the averaged $\Sigma$$_{\rm dust}$ 
per bins of $\Sigma$$_{\rm gas}$ (we choose regular bins in logarithmic scale). The linear fit to the pixel-by-pixel relation 
(in the shape of $\Sigma$$_{\rm gas}$ = GDR $\times$ $\Sigma$$_{\rm dust}$) is indicated for the whole sample with the 
black solid line, for ISM elements with $\Sigma$$_{\rm dust}$ $<$ 0.2\msun pc$^{-2}$ with the dotted line and for ISM 
elements with $\Sigma$$_{\rm dust}$ $>$ 0.2 \msun pc$^{-2}$ with the dashed line.}
 \label{PlotGD}
\end{figure*}

\subsubsection{Surface density maps of the gas}

The MAGMA CO map at a 1\arcmin\ resolution (that of the H\,{\sc i} map) is already provided on the MAGMA website. We regrid this map and the 
H\,{\sc i} map to a final pixel grid of half the resolution of the H\,{\sc i} map, i.e. 7 pc at the distance of the LMC. We derive the H\,{\sc i} surface 
density map ($\Sigma$$_{\rm HI}$) and the H$_{\rm 2}$ surface density map ($\Sigma$$_{\rm H2}$) by dividing the local H\,{\sc i} 
masses (M$_{\rm HI}$) and the local H$_{\rm 2}$ masses (M$_{\rm H2}$) by the area of our reference pixel (final units: \msun~pc$^{-2}$). 
We multiply the $\Sigma$$_{\rm HI}$ map by 1.36 to take the presence of helium into account. The MAGMA X$_{\rm CO}$ factor 
being derived from virial masses, our $\Sigma$$_{\rm H2}$ map includes all the material contributing to the dynamical mass of the 
cloud, thus already includes helium (this contribution is added however in the Galactic X$_{\rm CO}$ case). The two final maps will 
be respectively referred to as the atomic and molecular gas surface density maps $\Sigma$$_{\rm atomic}$ and $\Sigma$$_{\rm mol, CO}$, their sum
as $\Sigma$$_{gas}$.
Assuming that the region follows the \citet{Kennicutt1998} relation ($\Sigma$$_{\rm SFR}$ = 2.5 $\times$ 10$^{-4}$ ($\Sigma$$_{\rm gas}$)$^{1.4}$), 
we can use the $\Sigma$$_{gas}$ map to derive local estimates of the SFRs. For the regions detected at a 2-$\sigma$ level in the \hersc\ 
bands, the SFR ranges from 4.5 $\times$ 10$^{-3}$ to 2.4 $\times$ 10$^{-1}$ \msun~kpc$^{-2}$~yr$^{-1}$, with an average $\Sigma$$_{\rm SFR}$ 
of 4.4 $\times$ 10$^{-2}$ \msun~kpc$^{-2}$~yr$^{-1}$ across the whole complex. Using YSO candidates in the N11 region, \citet{Carlson2012} 
estimated the SFR in the region to be between 1.8 and 8.8 $\times$ 10$^{-2}$  \msun~kpc$^{-2}$~yr$^{-1}$ (depending on the timescale selected 
for the Stage\,{\sc i} formation, i.e. for embedded sources). The value we estimate from the gas mass is consistent with that SFR range.

\subsection{Relations between dust and gas surface densities}

We convolve the $\Sigma$$_{\rm dust}$ map (resolution: 36\arcsec) to the resolution of the $\Sigma$$_{gas}$ map 
using Gaussian kernels (same pixel grid). We then derive an `observed' total GDR map of N11. The map obtained 
using the MAGMA X$_{\rm CO}$ factor to derive the molecular gas is shown in Figure~\ref{GD}. The corresponding 
probability distribution is shown in the right panel. The vertical line indicates the Galactic GDR value derived by 
\citet{Zubko2004} (i.e. 158). The inset compares this distribution with that restricted to the regions covered by the 
MAGMA public release. The range of $\Sigma$$_{\rm gas}$ we are probing covers about an order of magnitude. 
The distribution of GDR broadly follows the dust distribution, with highest values of the GDR observed in the most 
diffuse regions and lowest values detected toward the H\,{\sc ii} regions. H\,{\sc i} dominates the local gas masses in 
many regions of the complex. 

\vspace{5pt}
Figure~\ref{PlotGD} shows the relation between the dust surface density and the atomic+molecular gas surface density (MAGMA X$_{\rm CO}$ case 
on the left panel and Galactic X$_{\rm CO}$ case on the right panel). Black points distinguish ISM elements where MAGMA 
CO data is available in the publicly released map. They represent about 45$\%$ of our ISM elements. Grey points indicate elements for which the CO information is 
not available. They are identical in both panels. Some of these `grey' elements might have CO emission (but are simply not covered). 
This is probably the case for regions that possess a non-negligible dust surface density at the top of the ``grey cloud" of points.
Additional molecular gas would shift these elements and tighten the relation between $\Sigma$$_{\rm dust}$ and $\Sigma$$_{\rm gas}$. 
The larger squares indicate the averaged $\Sigma$$_{\rm dust}$ per bins of $\Sigma$$_{\rm gas}$ (we choose 
regular bins in logarithmic scale). The error bars indicate the scatter within these $\Sigma$$_{\rm gas}$ bins. In the $\Sigma$$_{\rm dust}$ range we are 
studying here (0.1 $\le$ $\Sigma$$_{\rm dust}$ $\le$ 0.5 \msun pc$^{-2}$), the relation between $\Sigma$$_{\rm dust}$ and $\Sigma$$_{\rm gas}$ 
seems to be linear. In the case of the MAGMA X$_{\rm CO}$ factor, we observe a flattening of the relation above $\Sigma$$_{\rm gas}$=60 \msun 
pc$^{-2}$. The flattening resides within the error bars in the Galactic X$_{\rm CO}$ case. 

\vspace{5pt}
We take the uncertainties on the individual $\Sigma$$_{\rm dust}$ into account to derive the linear scaling coefficients 
linking the dust surface densities to the gas surface densities (so to derive the error-weighted averaged GDR of the sample). 
The thick line in the plots of Fig.~\ref{PlotGD} indicates these relations. We see that the GDR of the whole region is close 
to 180 in both X$_{\rm CO}$ cases. How does this compare to the expected GDR in the LMC \citep[12+log(O/H)=8.3-8.4; see][]{Russell1990}? 
We can predict this value using the formula of \citet{RemyRuyer2014} for a broken power-law and a X$_{\rm CO,\mathbb{Z}}$ case (i.e. X$_{\rm CO}$ $\sim$ Z$^{-2}$). 
This leads to a GDR of 350, thus $\sim$50\% more than the global ratio we find. To probe the variations of GDR with the dust surface density, 
we cut our $\Sigma$$_{\rm dust}$ range in two intervals, estimating the GDR for $\Sigma$$_{\rm dust}$ $<$ 0.2 \msun pc$^{-2}$ (dotted line) 
and $\Sigma$$_{\rm dust}$ $>$ 0.2 \msun pc$^{-2}$ (dashed line). The values are 186{\small $\pm$12} and 140{\small $\pm$33} respectively 
in the case of MAGMA X$_{\rm CO}$ and 184{\small $\pm$12} and 140{\small $\pm$30} respectively in the case of a Galactic X$_{\rm CO}$). 
The GDR thus decreases with the dust surface densities in the N11 complex. This decrease was previously observed in a strip of the LMC 
in \citet{Galliano2011} or in the recent study of \citet{RomanDuval2014}.

\subsection{Discussion on the low `observed' GDR}

Our analysis of GDR in N11 lead to values lower than those expected for an environment such as the LMC. We recall that several assumptions 
have been made to derive this `observed' GDR map: we assume that {\it i)} the X$_{\rm CO}$ factor is constant across the complex, {\it ii)} the CO traces the 
full molecular gas reservoir, {\it iii)} the H\,{\sc i} line is optically thin in the region and {\it iv)} the dust composition does not vary across the complex. 
In the following section, we discuss the impact of these hypotheses on the derived GDR and analyse the possible origin of its decrease with the dust 
surface density.

\subsubsection{Underestimation or variations in the X$_{\rm CO}$ factor}

The CO molecule can be highly photo-dissociated in the outer regions of molecular clouds while H$_{\rm 2}$ is shielded by dust or self-shields 
from UV photodissociation. As mentioned in $\S$5.1.2, this effect often translates into higher X$_{\rm CO}$ factors between the observed CO 
intensities and the H$_{2}$ abundance they trace in porous media submitted to strong radiation fields such as N11. In this analysis, we conservatively choose 
the MAGMA factor derived for GMCs across the LMC, factor that is already 2.4 times higher than the Galactic X$_{\rm CO}$ factor. However, given the intense 
radiation fields arising from the multiple OB associations in N11, the X$_{\rm CO}$ factor could be above the mean MAGMA value that is driven by less 
energetic environments. \citet{Israel1997} for instance suggests a X$_{\rm CO}$ factor of 6 $\times$ 10$^{20}$~cm$^{-2}$~(K~km~s$^{-1}$)$^{-1}$ 
in the northeast filament of N11 and of up to 2.1 $\times$ 10$^{21}$~cm$^{-2}$~(K~km~s$^{-1}$)$^{-1}$ in the N11 ring itself. 

\begin{figure}
    \centering
    \vspace{-10pt}
     \includegraphics[width=9cm]{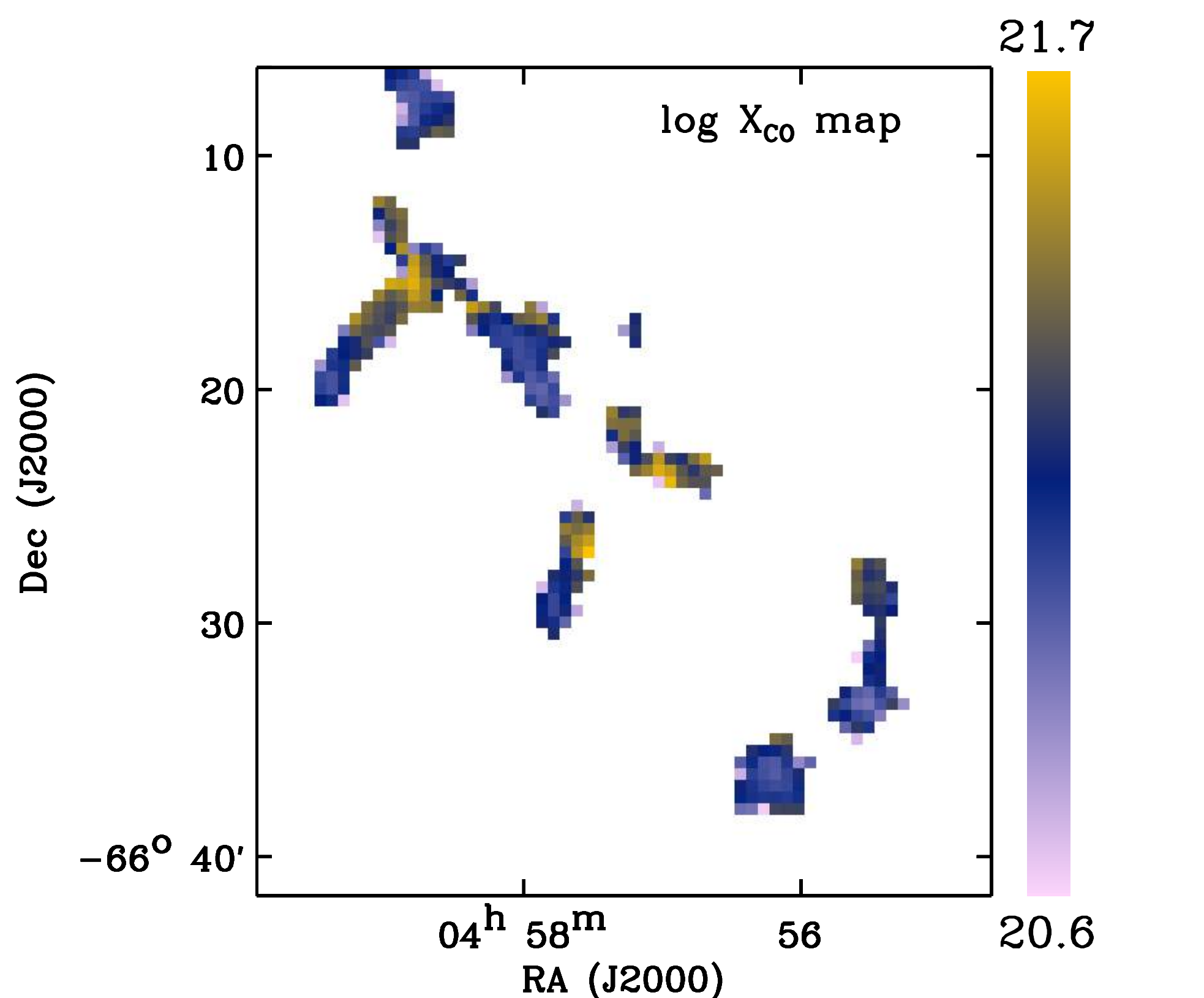}  
    \caption{Local variations of the X$_{\rm CO}$ factor estimated using a constant GDR = 350 across the complex. Units are in cm$^{-2}$~(K~km~s$^{-1}$)$^{-1}$ (log scale). 
    We limit the study to ISM elements above the 3-$\sigma$ sensitivity limit of the MAGMA CO(1-0) map. }
    \label{N11_XCO}
\end{figure}

Let's assume a constant GDR of 350 across the region. Which conversion factors would then be required to reach this value? X$_{\rm CO}$ would 
be equal to (350 $\times$ $\Sigma$$_{\rm dust}$ - $\Sigma$$_{\rm HI}$) / I$_{\rm CO}$. Figure~\ref{N11_XCO} shows the map of the 
X$_{\rm CO}$ factors we obtain. Because low signal-to-noise pixels around the edge of the CO-bright clouds are very dependent on the baselines 
and the signal identification method used to generate the MAGMA CO map, we choose to limit this study to regions above the 3-$\sigma$ sensitivity limit of 
the MAGMA survey. We observe the highest X$_{\rm CO}$ values near the bright OB associations LH10 and LH13 and in the 
northeastern filament. The derived X$_{\rm CO}$ factor can vary by an order of magnitude: it ranges between 5.2 $\times$ 10$^{20}$~cm$^{-2}$~(K~km~s$^{-1}$)$^{-1}$ 
and 4.9 $\times$ 10$^{21}$~cm$^{-2}$~(K~km~s$^{-1}$)$^{-1}$, with a mean value of 1.3 $\times$ 10$^{21}$~cm$^{-2}$~(K~km~s$^{-1}$)$^{-1}$. 
The minimum factor we obtain is consistent with our nominal (MAGMA) choice for X$_{\rm CO}$ but using a higher X$_{\rm CO}$ factor would indeed  increase
the local GDR values toward the expected LMC GDR in most of these regions. The maximum X$_{\rm CO}$ factor we find is twice the value estimated for the ring by 
\citet{Israel1997}. Even if possible, these very high X$_{\rm CO}$ factors are, nevertheless, expected in more extreme environments than the LMC, such as
the dwarf irregular galaxies NGC~6822 or the Small Magellanic Cloud \citep{Leroy2011}. A modification of the X$_{CO}$ factor alone is probably 
not sufficient to fully explain the low `observed' GDR.

\subsubsection{CO-dark gas}

Studies tracing gas through the dust emission or the gamma rays emission \citep[produced through cosmic ray collisions in the 
Galaxy; see][]{Grenier2005, Fermi2012} have highlighted the presence of H$_{\rm 2}$ that is not detectable 
through CO observations. This molecular phase, called the ``CO-dark" phase is particularly difficult to quantify 
and can not be related to the CO-emitting reservoirs through the usual X$_{\rm CO}$ factor. If we consider that 
this `missing molecular phase' is responsible for the low GDR observed in N11, we can quantify its abundance 
by doing $\Sigma$$_{dark~gas}$ = 350 $\times$ $\Sigma$$_{\rm dust}$ - $\Sigma$$_{\rm HI}$ - $\Sigma$$_{\rm mol, CO}$,
with $\Sigma$$_{dark~gas}$ the surface density of the molecular dark gas not traced by CO.
Before continuing with this analysis, we need to take into account the fact that in the more quiescent regions of the complex, part of 
the missing gas mass could be already linked with our lack of CO constraints due to the limited coverage of the public MAGMA map. If we 
assume a constant GDR of 350 throughout the complex, $\sim$ 8.6 $\times$ 10$^{5}$ \msun\ are missing from the total gas budget 
in the regions with no CO data (see Fig.~\ref{N11_Excess_Betafree} bottom left).
We use the good correlation between $\Sigma$$_{dust}$ and $\Sigma$$_{\rm mol, CO}$ in the regions covered in CO
(Spearman coefficient r=0.5 and $\Sigma$$_{dust}$ = 0.16{\small $\pm$0.007} $\Sigma$$_{\rm mol, CO}$$^{\rm 0.19\pm0.02}$) to estimate the 
CO-emitting molecular gas in the regions not covered in CO. Using this completed $\Sigma$$_{\rm mol, CO}$ map, we can then derive the mass
of the dark gas in the region. We find a fraction of the dark gas to the total gas mass equal to 55-60\%. The fraction of the dark gas to the 
total molecular mass (f$_{DG}$ = M$_{dark~gas}$ / (M$_{dark~gas}$+M$_{\rm mol, CO}$)) is equal to 70-80\%, with larger values outside the
dust peaks of N11. By theoretically model the dark component, \citet{Wolfire2010} predict that f$_{DG}$ would be relatively invariant with the 
incident UV radiation field strength ($\sim$0.3 in the Galaxy for instance) but that this value could increase with {\it i)} a decreasing visual extinction
and {\it ii)} a decreasing metallicity. If our results are consistent with these trends, the very high f$_{DG}$ fraction we obtain is pushing the
models to the limits. This suggests that a hidden reservoir of CO-dark gas is probably not the unique explanation to the low `observed' GDR throughout N11.

We note that further observations would be needed to correctly estimate the CO-faint phase in the complex. The
CO-dark reservoirs could also be quantified using the [C\,{\sc ii}] 157 \mic\ line as suggested by \citet{Madden1997}. Several studies 
have indeed shown that the [C\,{\sc ii}] emission is more extended than that of CO \citep{Israel2011,Lebouteiller2012}.

\subsubsection{Optically thick HI}

The H\,{\sc i} column density N$_{\rm HI}$ can be estimated from the measured H\,{\sc i} brightness temperature T$_{\rm B}$ and the optical depth $\tau$ using equation 3-38 from \citet{Spitzer1978}:

\begin{equation}
N_{HI} = 1.823\times10^{18} \int{T_B \frac{\tau}{1-e^{-\tau}} d\varv}
\end{equation}

\noindent where T$_{\rm B}$ is the measured brightness temperature (K), $\tau$ is the optical depth and $\varv$ is the velocity. In this analysis, we assume 
that the 21cm line is optically thin across the complex in order to derive the local H\,{\sc i} masses. This reduces the equation to N$_{\rm HI,thin}$ = 1.82 
$\times$ 10$^{18}$ $\int$ T$_{\rm B}$ $\tau$ d$\varv$. However, because the H\,{\sc i} optical depth strongly depends on N$_{\rm HI}$, the assumption 
of optically thin H\,{\sc i} starts to be questionable for large column densities. Recent studies have proposed optically thick H\,{\sc i} envelope around CO 
clouds to explain the large scatter in the relation between the H\,{\sc i} velocity integrated intensity and the submm dust optical depth \citep[see][for instance]{Fukui2014,Fukui2015}. 
In their study of the Perseus molecular cloud, \citet{Lee2012} estimated the optical depth effects to be responsible of an underestimation of the H\,{\sc i} 
mass by a factor of 1.2-2. Using Eq. 3-37 from \citet{Spitzer1978} and a single spin temperature of T$_{\rm s}$ = 60K, we estimate that we would need 
an H\,{\sc i} column density of about 10$^{20}$ cm$^{-2}$ per km/s to reach $\tau$ = 1. If we consider a H\,{\sc i} line width of $\sim$14km/s \citep[mean 
value of the H\,{\sc i} line width toward the position of the GMCs;][]{Fukui2009}, this leads to a N$_{\rm HI}$ threshold of 1.4 $\times$ 10$^{21}$ cm$^{-2}$ 
above which optical depth effects could be expected, which is mostly the case in the N11 complex we are studying here. Moreover, several surveys 
(\citet{Dickey1994} and \citet{Marx-Zimmer2000}) targeting absorption features in compact radio continuum sources toward the LMC have shown that 
its cold atomic gas is probably colder than that of the Milky Way (down to T$_{\rm s}$$\sim$30 K). The theoretical N$_{\rm HI}$ threshold we determine 
could thus be even lower in N11. However, \citet{Marx-Zimmer2000} also find that if a few regions such as 30 Doradus and the eastern H\,{\sc i} boundary 
are optically thick, most of the clouds they target in other regions have low values of  $\tau$. Unfortunately none of them were in the direction of N11. 
If present, optically thick H\,{\sc i} would lead to an underestimation of the gas masses at high column densities. Higher atomic gas masses in N11 would 
bring the GDR closer to that expected for a low-metallicity galaxy like the LMC. However, the pixel-by-pixel effect is difficult to quantify. In the N11 complex, 
the peak in $\Sigma$$_{\rm dust}$ and $\Sigma$$_{\rm HI}$ are not co-spatial. As one can see in Fig.~\ref{HI_CO_Dust}, $\Sigma$$_{\rm HI}$ is rather 
constant around the N11 shell where the $\Sigma$$_{\rm dust}$ peaks can be found. These regions are going to be less affected by the `optically thick' 
correction than the northeast filament where the H\,{\sc i} distribution peaks. This suggests that the `H\,{\sc i} optical depth' effect could only partly explain 
the low `observed' GDR we find but is not sufficient to explain its progressive decrease with the dust surface densities we obtain.

\subsubsection{Dust emissivity variations}

\begin{figure}
    \centering
    \vspace{-15pt}
     \includegraphics[width=9cm]{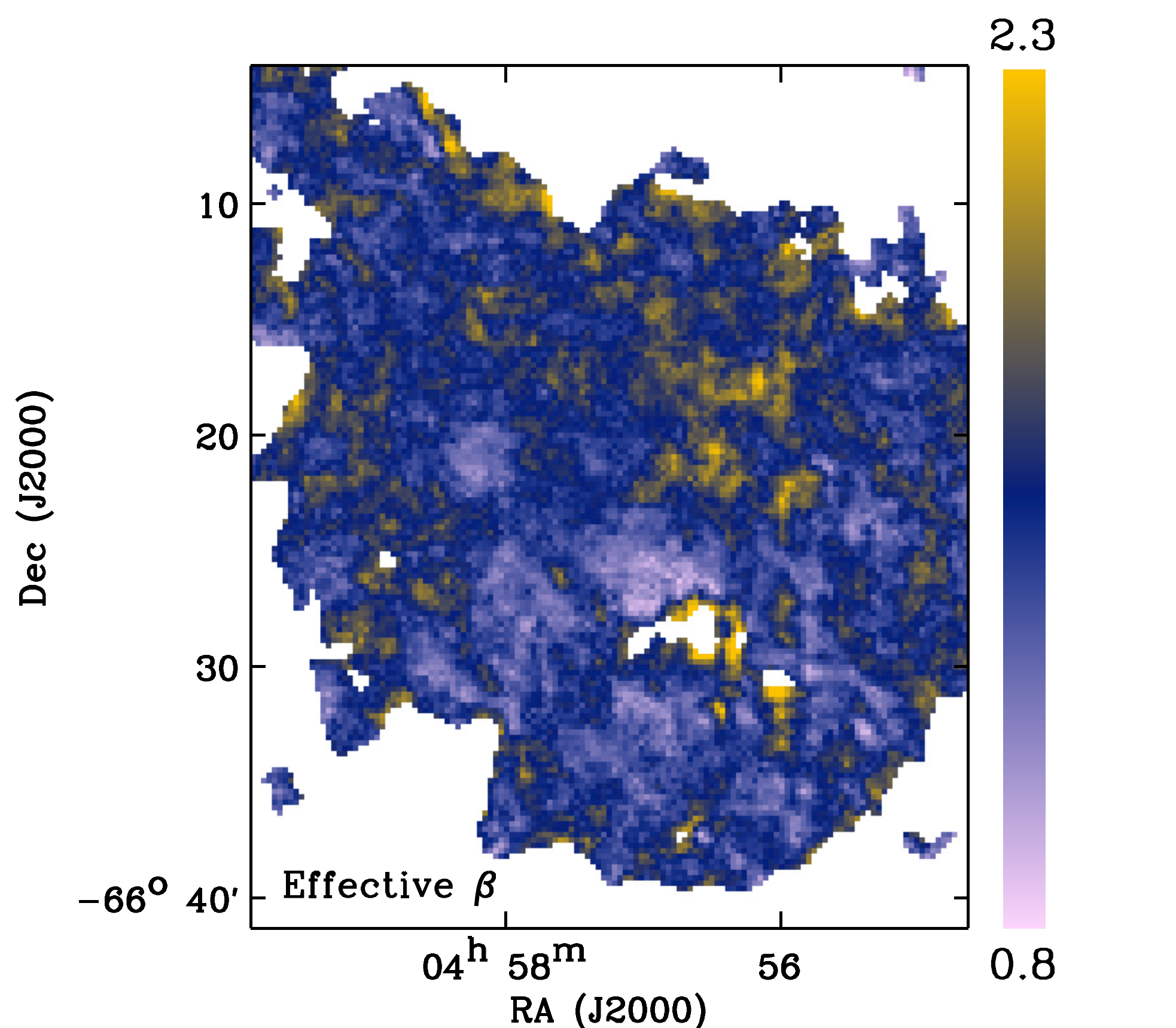}  
    \caption{Map of the effective emissivity index of the cold dust grains derived from a two-temperature modeling of the N11 complex. The cold emissivity index 
    $\beta$ is thus considered as a free parameter. }
    \label{N11_Emissivity}
\end{figure}

The three previous sections were trying to find an explanation of the low `observed' GDR by invoking a modification of the hypotheses made on the gas phase. 
This section is exploring the hypotheses of a dust abundance variation and their effect on the GDR on local scales. To properly understand the reprocessing of 
grains in the ISM and quantify the dust abundance variations, one would need to access the intrinsic emissivity of the dust grains. In our local MBB modeling, we 
fixed the effective emissivity index (i.e. the apparent emissivity index) of the cold dust component to 1.5 in order to minimize degeneracies between the dust 
temperature and $\beta$. Leaving both parameters free is, however, a commonly used technique to probe potential variations in the dust grain emissivity
\citep[see][; among many others]{PlanckCollaboration2014_GalacticPlane,Tabatabaei2014,Grossi2015}. So what do we obtain when we let the grain 
emissivity vary? Figure~\ref{N11_Emissivity} shows the map of the effective emissivity index derived from a two-temperature 
modeling of the complex if $\beta$ is used as a free parameter. The median value of the emissivity index $\beta$ is 1.52{\small $\pm$0.19} across the complex 
if we restrict the analysis to ISM elements with a 2-$\sigma$ detection in the \hersc\ bands. We note that this median value is consistent with the value we chose 
to fix in our SED modeling procedure ($\beta$=1.5). Nevertheless, we do observe systematically lower values of $\beta$ (so a flattening of the submm slope) in 
dense regions compared to the diffuse medium\footnote{Note that the trend is similar if we restrict the study to ISM elements with a 3-$\sigma$ at 870\mic\
and include this 870\mic\ constraint in the fitting procedure.}. Part of the explanation could be linked to the fact that the large ISM elements we are 
studying in this paper contain dust populations with a large range of temperatures. Temperature mixing effects usually lead to shallower observed 
submm SEDs \citep{Shetty2009}, with the effective emissivity index we observe being, in fact, a 
non-trivial combination of the intrinsic emissivity index of each dust population. Our SED modeling procedure includes a radiation field intensity range (thus a 
temperature range) and is able to fit the data up to 500 \mic, suggesting that our assumptions (silicate + AC grains) could be sufficient to explain the submm 
\hersc\ fluxes we observe and that flattening of the effective emissivity can be fully explained by temperature gradients in the H\,{\sc ii} regions of the complex. 
The study of the excess at 870 \mic\ ($\S$4.5) also showed that the residuals we observe at 870 \mic\ compared to our local SED models (thus compared to 
`pure emission from the cold grains') could be linked with non-dust contribution from the CO(3-2) line or the free-free emission.

Does this necessarily mean that there is no variation of the dust emissivity index of the grains across the region? The variation of GDR with the dust surface density 
could suggest otherwise. Indeed, our modeling procedure assumes the same dust grain properties whatever the density in the ISM element we are looking at. 
However, \citet{RomanDuval2014} suggested that dust accretion and coagulation processes could be happening in dense phases such as the ones probed in 
this analysis. If this is the case, this would lead to an overestimation of the dust surface density in environments where the accretion and coagulation processes 
could occur. A variation of the emissivity index (to lower indices in that particular case) would decrease our estimated dust masses.

\subsubsection{Conclusion}

As explained in \citet{RomanDuval2014}, it is very difficult to disentangle between the various hypothesis we explore at our working resolution, i.e. the presence 
of CO dark-gas or real GDR variations with environment, H\,{\sc i} optical depth effect or potential variations of the dust abundances. The explanation is probably 
a combination of all these various effects that affect the GDR in the same direction. Higher resolution observations of the molecular clouds themselves (at much 
smaller scales than the giant complexes we are mapping here), more specifically a mapping of the submm dust continuum, a sampling of CO spectral line energy 
distributions, but also observations of other tracers of the dense gas such as HCN or HCO+ with ALMA would be necessary to unambiguously probe the dense 
clouds and constrain the GDR on small scales in the LMC.

\section{A Principal Component Analysis of the IR/submm dataset}

In this paper, we have analyzed the local characteristics of dust grains across the star forming region N11. However, the properties we derive are dependent on 
the assumptions made on the dust composition in the SED modeling technique we have selected. In this last section, the goal would be to explore the capacity 
of a data processing method based a Principal Component Analysis (PCA) to decompose the various dust populations contributing to the local SEDs in N11 
with no {\it a priori} on the dust populations.  

\subsection{Method}

The Karthunen-Lo\`{e}ve transform, more familiarly called Principal Component Analysis (hereafter PCA), is the orthogonal projection of a set of data into a new 
system of coordinates. This technique has been used on many astronomical objects and for multiple purposes: analysis of spectral cubes \citep{Mekarnia2004}, 
reduction of dimensionality for SED libraries \citep{HanHan2014}, among many others. 
Our SED modeling of the N11 complex has enabled us to build a library of spectra across the complex. If we take into account pixels with a 1-$\sigma$ detection 
in the \hersc\ bands, this leaves us with 21392 individual dust SEDs. Our goal is to apply a PC decomposition to this multispectral data cube. Using the fully 
modeled SEDs in lieu of the individual bands provides a much better constraint of the SED shape in each resolution element and will decrease the biases linked 
with noise in the data. The first step of our analysis is to compute the covariance matrix of the multispectral data cube (we use the IDL function {\it CORRELATE}). 
Since we are mostly interested in the dust SED here, we keep the spectral coverage from 8 \mic\ to 1000\mic. Each SED in our database is sampled with 117 
wavelength points between these two wavelengths. We use the logarithm of our local models to perform the PCA decomposition. From the covariance matrix, 
we can calculate the eigenvectors and eigenvalues (we use the IDL function {\it EIGENQL}, with the covariance option). This directly provides us with an ensemble 
of vectors that constitute the main building blocks of our local SEDs. The total number of eigenvectors derived in a PCA is equal to the number of spectral elements 
used for the analysis. We obtain 117 independent eigenvectors. 
We will only analyze the 6 first eigenvectors in the following analysis. The last step of the analysis is to reconstruct the principal basis images, rotating by the 
eigenvectors. As we decided to apply a non-standardized PCA (no mean-centering), we expect the first component to mainly correspond to a `mean' SED of 
the ISM elements while the following components will reflect the deviations from this main SED. 

\begin{figure}
    \centering
    \hspace{-10pt} \includegraphics[width=8.5cm]{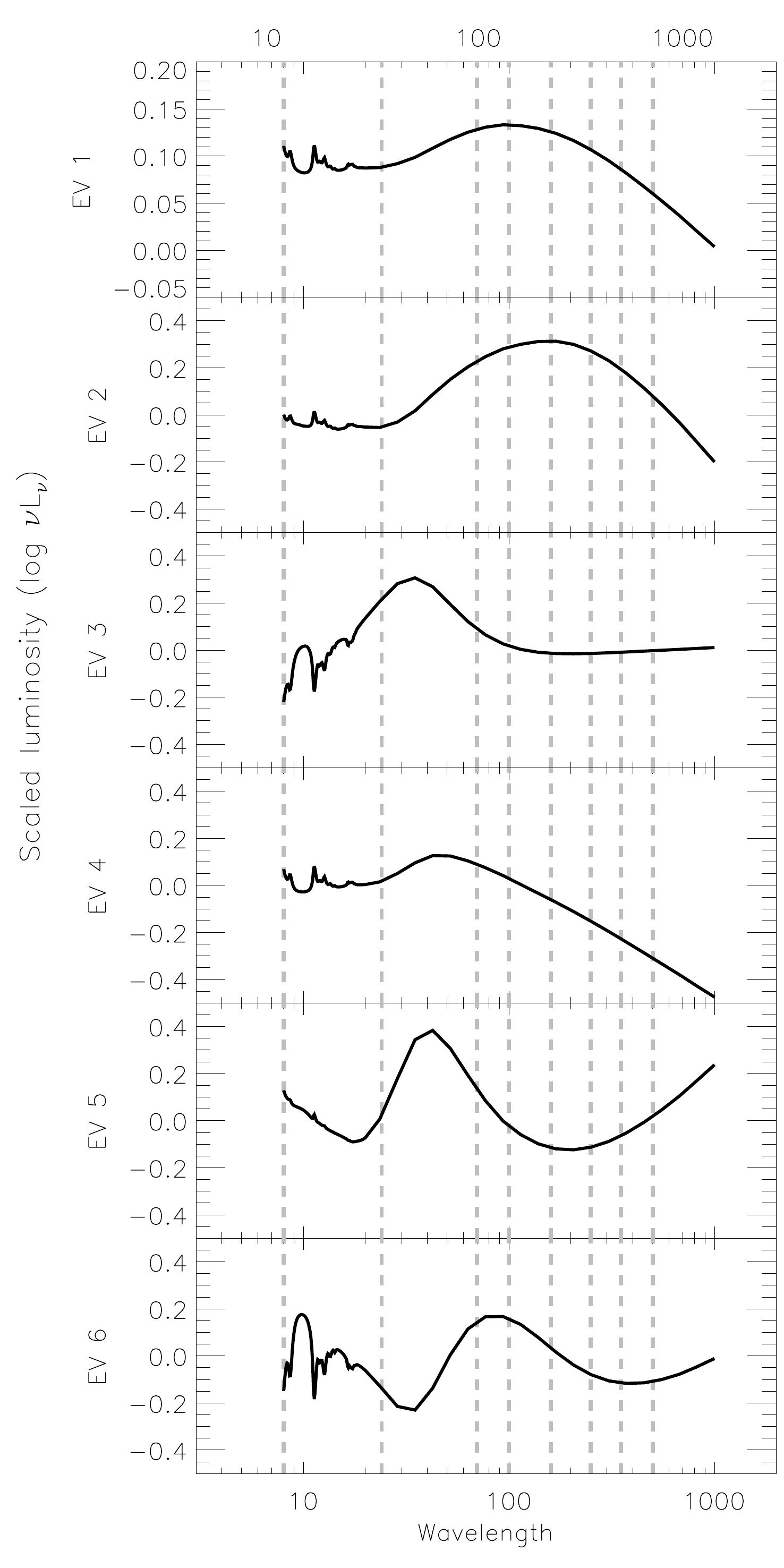}     \\   
    \caption{The 6 main eigenvectors. They respectively contain 98.8, 0.53, 0.42, 0.17, 0.06 and 0.02$\%$ of the total variance. The 8, 24, 70, 100, 160, 250, 350 
    and 500 \mic\ wavelengths are overlaid as references but the full SED models (sampled with 117 elements) have been used for the decomposition.}
    \label{PCA_images_all2}
\end{figure}

\begin{figure*}
    \centering    
  \includegraphics[width=18cm]{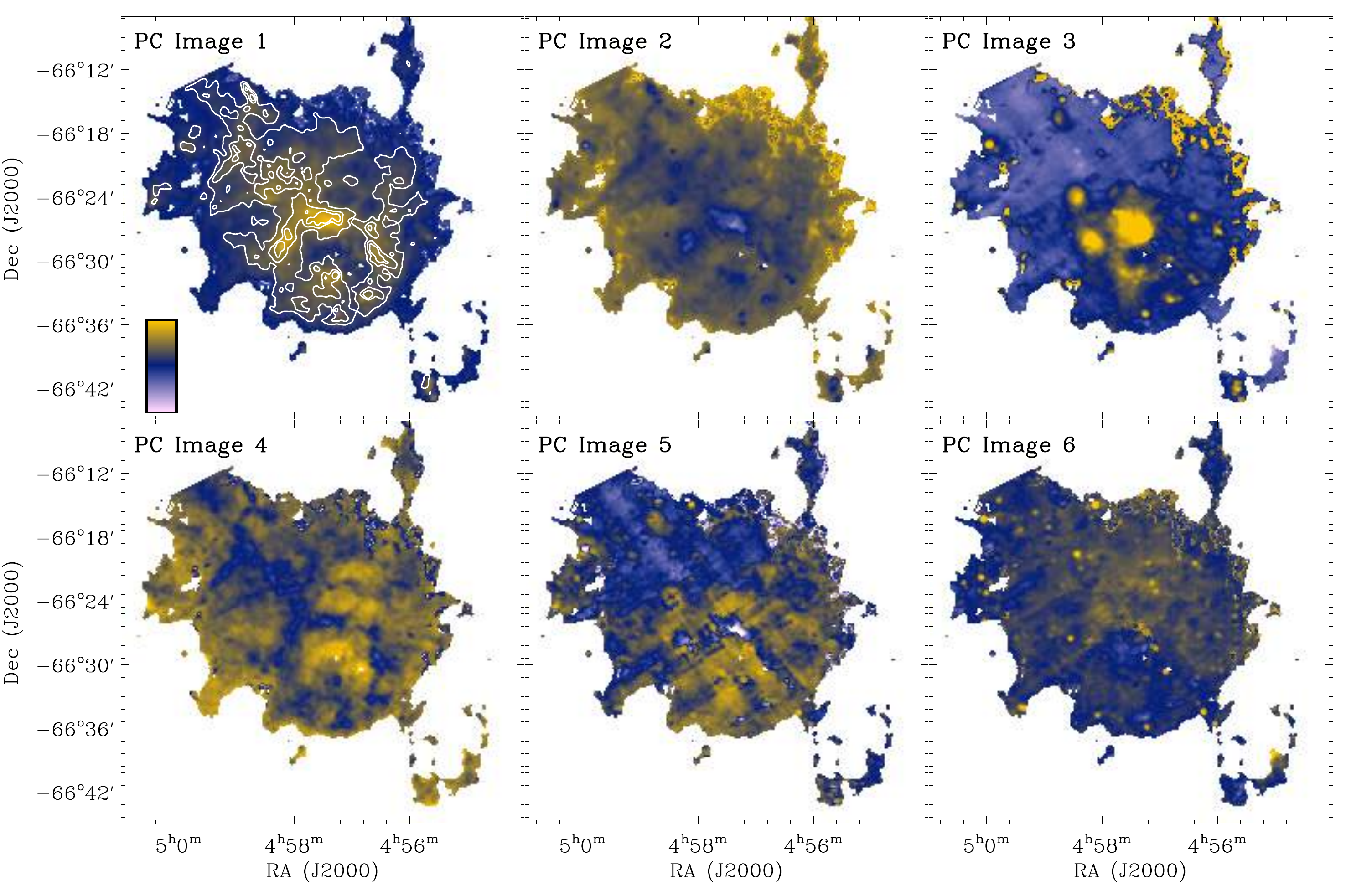}     \\  
    \caption{The 6 corresponding component images of the PCA. $\Sigma$$_{\rm dust}$ contours are overlaid on the first top panel.}
    \label{PCA_images_all}
\end{figure*}

\begin{figure*}
    \centering 
    \begin{tabular}{cc} 
  \includegraphics[width=9cm]{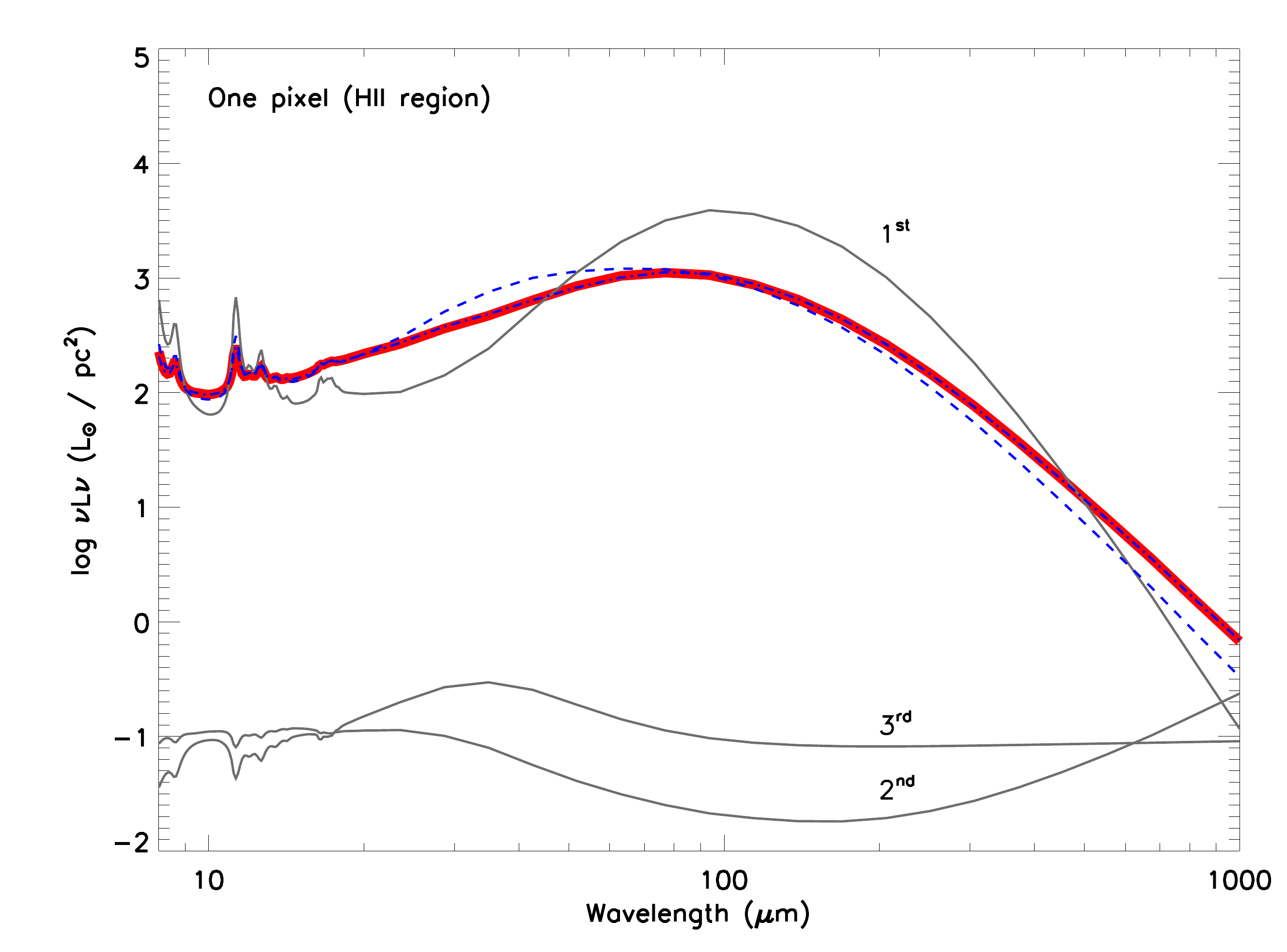} &
  \hspace{-15pt}
  \includegraphics[width=9cm]{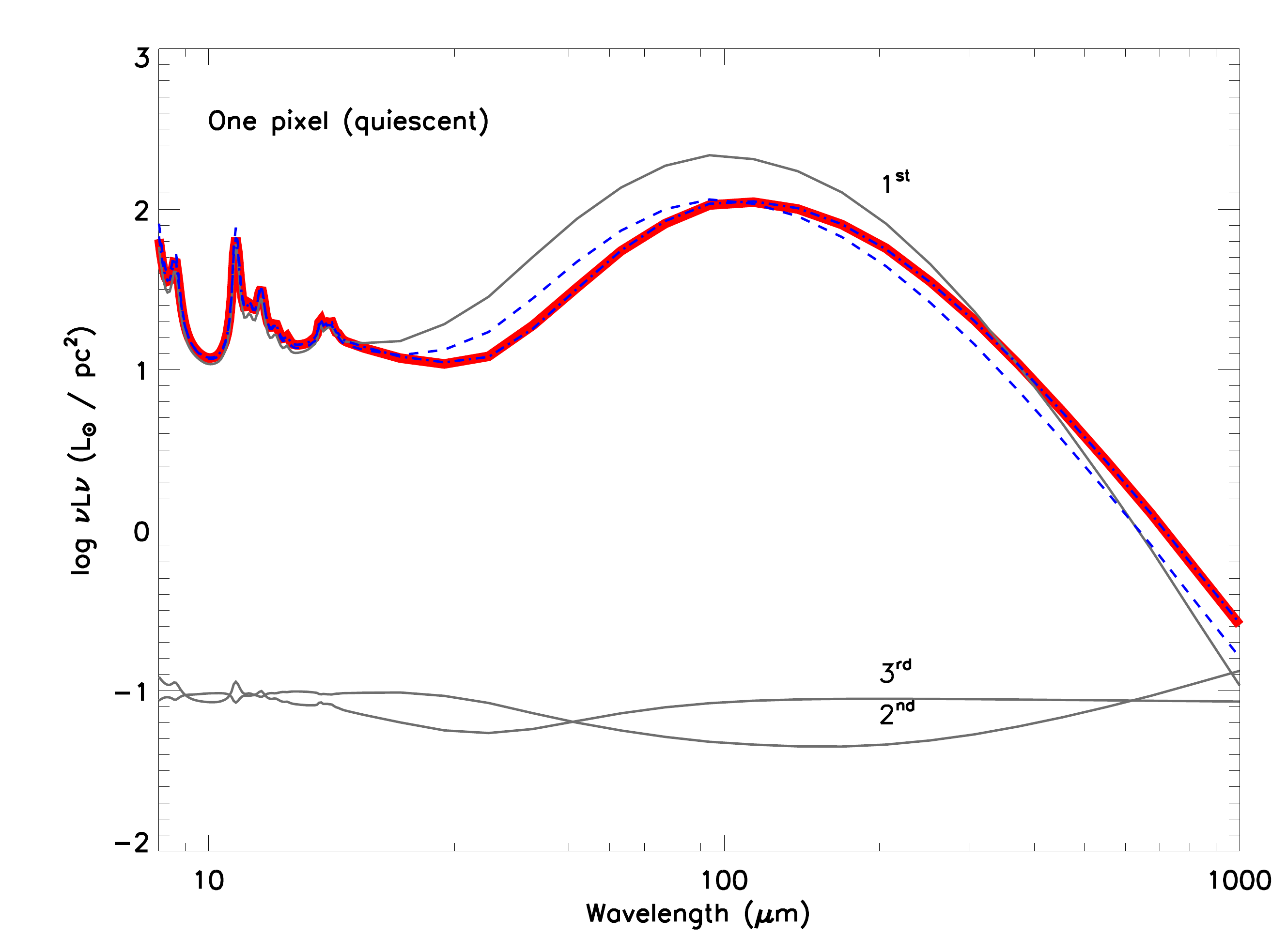}     \\  
  \end{tabular}
    \caption{Decomposition of two local SEDs (same pixels as in Fig.~\ref{N11_GlobalSED}) using our PCA eigenvectors. The red line shows the 8 to 1000\mic\ 
    modeled SED and 3 black lines the contribution from the 3 first components (see labels). The dashed line indicates the SED reconstructed from the 3 first 
    components while the dashed-dotted line indicates the SED reconstructed from the 6 first components (those are indistinguishable from the modeled SED).}
    \label{PCA_decomposition}
\end{figure*}

\subsection{Decomposition and interpretation}

We find that the first component of the decomposition dominates and contributes to 98.8$\%$ of the total variance in the observed variables (the higher the 
variance along a given `axis', the more representative the axis). The contributions from the second, third and fourth component rapidly decrease (0.53, 0.42 
and 0.17$\%$ respectively). We show the 6 main components/vectors of the PCA in Fig.~\ref{PCA_images_all2} and the reconstructed PC images in 
Fig.~\ref{PCA_images_all} (color scale from low/purple to high/yellow in the top left panel). An examination of each component allows us to interpret the 
results of the decomposition process. We observe that the first component (see the vectors in Fig.~\ref{PCA_images_all2}) peaks around 100 \mic: the first 
basis image (see the PC images in Fig.~\ref{PCA_images_all}) is, as expected, linked with the bulk of emission arising from the warm/cool regions of the 
N11 complex and resemble the average SED of the complex. 
The second component of the PCA peaks around 160 \mic: the second basis image seems to correspond to the contribution of the coldest regions mostly 
arising outside the N11 ring and filament. 
The third PCA component peaks between 30 and 40 \mic: the third basis image is thus dominated by the contribution of the hottest regions in the LH14, LH10, 
LH13 and LH9. We note that in this component, the PAHs are inverted and negative, which is consistent with the fact that PAHs are partly destroyed in H\,{\sc ii} 
regions.

The interpretation of the following PC images is less straightforward. The fourth PCA component peaks around 50 \mic\ and smoothly decreases toward longer 
wavelengths: the fourth basis image could thus be related to the hardness of the radiation fields, as it highlights regions where small grains (VSG or PAHs) seems 
to be excited. This could explain why the distribution of PC4 in Fig.~\ref{PCA_images_all} seems to be anti-correlated with the distribution of the dust surface density, 
the radiation being usually harder in regions of lower optical depth.
The fifth component is associated with instrumental noise effects in the MIPS observations (particularly recognizable stripping pattern of the MIPS 70 \mic\ map). 
Note that even if we do not include the direct observed 70 \mic\ map in the PCA, this information seems to be contained in the modeled SEDs. This map thus 
highlights the non-negligible uncertainties on the dust SED modeling in that part of the spectrum and the power of the PCA approach to derive these small fluctuations. 
Finally, the sixth basis image is difficult to associate to a specific dust phase. Part of the component could be associated with the old stellar population as peaks in the 
PC Image 6 are visible toward old stars emitting in the mid-IR. We note that the stellar contribution was not removed from the SED spectra used in this PCA analysis in 
order to observe its influence on the final SED library. 

Figure~\ref{PCA_decomposition} shows the PC decomposition for two local SEDs (we select the same bright and quiescent pixels as in Fig.~\ref{N11_GlobalSED}). 
The 3 black lines highlight the contribution from the 3 first components (see labels) to the local modeled SED (red line). We overlay the SED reconstructed from the 3 
first components with the dashed line and the SED reconstructed from the 6 first components with the dashed-dotted line. This last SED and the reference SED are 
nearly indistinguishable for both regions. The shape of the first PC component is really similar to the SED shape in the quiescent region. This is expected from the 
fact that quiescent regions dominate in the complex we are studying. For the `warm' pixel, the contribution from the following components is crucial to account from 
the warm dust populations and the flattening of the SED shape. \\

Contrary to many approaches to decompose the local dust SED in nearby objects, the PCA approach has no a priori assumption included in the procedure rather 
than the information contained in the individual spectra. Our results show quite convincingly that even without a SED modeling procedure, we are able to recover 
a spatial structure of the different dust populations that has a coherence compared to what we expect for the region. The decomposition of the SED in various dust 
components (hot, warm, cold) sounds really promising. We will carry out further studies on the full LMC or on a larger sample of galaxies, for instance the KINGFISH 
survey of nearby galaxies, in order to understand how unique this decomposition is or how it evolves from one environment to another.


\section{Conclusions}

We combine \spitz\ and \hersc\ data from 3.6 to 500 \mic\ in order to perform a local dust SED modeling of the N11 massive star forming complex in the Large Magellanic 
Cloud. We use the amorphous carbon model of \citet{Galliano2011} to derive maps of various grain properties, such as the dust mass or the PAH fraction as well as 
physical conditions of the ISM (e.g the interstellar radiation field intensity) on scales smaller than 10pc. By performing a careful study of the residuals to the model at 
all the observed wavelengths, we find non-negligible residuals in the PACS bands, especially at 160 \mic. Part of these local residuals could originate from the strong 
[C\,{\sc ii}] line emission. We also model the data with a more standard two-modified blackbody model to derive local average temperatures and compare them with 
the radiation field intensity derived from the \citet{Galliano2011}. This model also helps us investigate potential variations in the dust emissivity index across N11.

We observe strong local SED variations. We find colder temperatures (on average) than in the LMC N158-N159-N160 star forming complex previously studied by 
\citet{Galametz2013a}, with a median temperature of 20.5K. We find a total dust mass of 3.3$\times$10$^4$ \msun\ in the regions we model. This is a factor of 2.5 
times lower than the dust mass derived if we would have used standard graphite to model the carbon dust and a factor of 2.5 higher than the dust masses derived 
by \citet{Gordon2014} using a BEMBB. The PAH fraction strongly decreases in our H\,{\sc ii} regions where they are probably destroyed by the strong radiation fields. 
We compare our model predictions at 870 \mic\ with observations obtained with the APEX/LABOCA instrument at the same wavelength. We evaluate the potential 
non dust contribution to the 870 \mic\ (CO line, free-free emission) emission. Our local dust SED models provide a satisfying fit of the Ôdust-onlyÕ emission at 870 \mic.

We relate the dust surface densities to the atomic and molecular surface densities and investigate the variations of the total gas-to-dust ratio across the region. We 
obtain a low GDR compared to those expected for a low-metallicity environment like the LMC. We also observe a decrease of GDR with increasing dust surface 
densities. Several effects could be responsible for the variations in the `observed' gas-to-dust mass ratios and we explore the influence of each of the assumptions 
we made to derive the ratio map. Potential explanations are a variation of the X$_{CO}$ factor due to the intense radiation fields arising from the H\,{\sc ii} regions 
in N11, the presence of a dark molecular gas component not traced by the available CO observations, the presence of optically thick H\,{\sc i} or a variation of the 
dust abundances in the densest regions of N11. The complete explanation is probably a combination of all these effects that affect the GDR in the same direction.

In the last section, we perform a Principal Component Analysis of the data cube. The first three components allow us to isolate three various dust phases (diffuse 
medium, cold dust and warm dust) with a spatial structure that is consistent with what we expect for the region. These tests will be extended in future studies to 
analyze the potential of the PCA method in disentangling the various dust reservoirs for a wider variety of environments.


\section*{Acknowledgments}
We would like to first thank the referee for his/her careful reading of this paper and useful suggestions.
We would also like to thank Karl Gordon for providing us with the reprocessed Herschel maps and the MegaSAGE consortium for our motivating collaboration and meetings. 
We acknowledge financial support from the NASA Herschel Science Center, JPL contracts \# 1381522, \# 1381650 \& \# 1350371. Meixner acknowledges support from 
NASA grant, NNX14AN06G, for this work.
This publication is based on data acquired with the {\it Herschel Space Observatory}. The \hersc/PACS instrument has been developed by MPE (Germany); UVIE (Austria); 
KU Leuven, CSL, IMEC (Belgium); CEA, LAM (France); MPIA (Germany); INAF-IFSI/OAA/OAP/OAT, LENS, SISSA (Italy); IAC (Spain). This development has been supported 
by BMVIT (Austria), ESA-PRODEX (Belgium), CEA/CNES (France), DLR (Germany), ASI/INAF (Italy), and CICYT/MCYT (Spain). 
The \hersc/SPIRE has been developed by a consortium of institutes led by Cardiff Univ. (UK) and including: Univ. Lethbridge (Canada); NAOC (China); CEA, LAM (France); 
IFSI, Univ. Padua (Italy); IAC (Spain); Stockholm Observatory (Sweden); Imperial College London, RAL, UCL-MSSL, UKATC, Univ. Sussex (UK); and Caltech, JPL, NHSC, 
Univ. Colorado (USA). This development has been supported by national funding agencies: CSA (Canada); NAOC (China); CEA, CNES, CNRS (France); ASI (Italy); 
MCINN (Spain); SNSB (Sweden); STFC, UKSA (UK); and NASA (USA). 
This publication is also based on data acquired with the Atacama Pathfinder Experiment. APEX is a collaboration between the Max-Planck-Institut fur Radioastronomie, 
the European Southern Observatory, and the Onsala Space Observatory.


\bibliographystyle{mn2e}
\bibliography{/Users/maudgalametz/Documents/Work/Papers/mybiblio.bib}

\end{document}